\documentclass[twocolumn]{aastex62}

\usepackage[hyphenbreaks]{breakurl} 

\def\photoz{photo-\textit{z}}
\def\specz{spec-\textit{z}}

\def\classic{\textsc{Classic}}
\def\lephare{\texttt{LePhare}}
\def\eazy{\texttt{EAZY}}
\def\tractor{\texttt{The Tractor}}
\def\farmer{\textsc{The Farmer}}
\def\Euclid{\textit{Euclid}}

\def\zmin{z_{\rm phot}^{\rm min}}
\def\zmax{z_{\rm phot}^{\rm max}}

\graphicspath{{./}{figures/}}

\received{\today}
\revised{\today}
\accepted{\today}
\submitjournal{ApJS}

\shorttitle{COSMOS2020}
\shortauthors{Weaver et al.}

\usepackage[flushleft]{threeparttable} 
\usepackage{amsmath} 

\begin{document}

\title{COSMOS2020: \\
A panchromatic view of the Universe to $z\sim10$ from two complementary catalogs
}   
\correspondingauthor{John R. Weaver}
\email{john.weaver@nbi.ku.dk}

\author[0000-0003-1614-196X]{J. R. Weaver}
\affil{Cosmic Dawn Center (DAWN)}
\affil{Niels Bohr Institute, University of Copenhagen, Jagtvej 128, 2200 Copenhagen, Denmark}

\author{O. B. Kauffmann}
\affil{Aix Marseille Univ, CNRS, LAM, Laboratoire d'Astrophysique de Marseille, Marseille, France}

\author[0000-0002-7303-4397]{O. Ilbert}
\affil{Aix Marseille Univ, CNRS, LAM, Laboratoire d'Astrophysique de Marseille, Marseille, France}

\author[0000-0002-9489-7765]{H. J. McCracken}
\affil{Institut d'Astrophysique de Paris, UMR 7095, CNRS,
    and Sorbonne Universit\'e, 98 bis boulevard Arago, 75014 Paris, France}

\author{A. Moneti}
\affil{Institut d'Astrophysique de Paris, UMR 7095, CNRS,
    and Sorbonne Universit\'e, 98 bis boulevard Arago, 75014 Paris, France}

\author[0000-0003-3631-7176]{S. Toft}
\affil{Cosmic Dawn Center (DAWN)}
\affil{Niels Bohr Institute, University of Copenhagen, Jagtvej 128, 2200 Copenhagen, Denmark}

\author[0000-0003-2680-005X]{G. Brammer}
\affil{Cosmic Dawn Center (DAWN)}
\affil{Niels Bohr Institute, University of Copenhagen, Jagtvej 128, 2200 Copenhagen, Denmark}

\author[0000-0002-7087-0701]{M. Shuntov}
\affil{Institut d'Astrophysique de Paris, UMR 7095, CNRS,
    and Sorbonne Universit\'e, 98 bis boulevard Arago, 75014 Paris, France}

\author[0000-0002-2951-7519]{I. Davidzon}
\affil{Cosmic Dawn Center (DAWN)}
\affil{Niels Bohr Institute, University of Copenhagen, Jagtvej 128, 2200 Copenhagen, Denmark}

\author{B. C. Hsieh}
\affil{Institute of Astrophysics \& Astronomy, Academia Sinica, Taipei 10617, Taiwan}

\author{C. Laigle}
\affil{Institut d'Astrophysique de Paris, UMR 7095, CNRS,
    and Sorbonne Universit\'e, 98 bis boulevard Arago, 75014 Paris, France}

\author{A. Anastasiou} 
\affil{Cosmic Dawn Center (DAWN)}
\affil{Niels Bohr Institute, University of Copenhagen, Jagtvej 128, 2200 Copenhagen, Denmark}

\author[0000-0002-8896-6496]{C. K. Jespersen} 
\affil{Cosmic Dawn Center (DAWN)}
\affil{Niels Bohr Institute, University of Copenhagen, Jagtvej 128, 2200 Copenhagen, Denmark}

\author{J. Vinther}         
\affil{Cosmic Dawn Center (DAWN)}
\affil{Niels Bohr Institute, University of Copenhagen, Jagtvej 128, 2200 Copenhagen, Denmark}

\author{P. Capak}
\affil{Cosmic Dawn Center (DAWN)}

\author[0000-0002-0930-6466]{C. M. Casey}
\affil{The University of Texas at Austin 2515 Speedway Blvd Stop C1400 Austin, TX 78712 USA}

\author[0000-0003-0639-025X]{C. J. R. McPartland}
\affil{Cosmic Dawn Center (DAWN)}
\affil{Institute for Astronomy, University of Hawaii, 2680 Woodlawn Drive, Honolulu, HI 96822, USA}
\affil{Department of Physics and Astronomy, University of California, Riverside, 900 University Avenue, Riverside, CA 92521, USA}
\affil{Niels Bohr Institute, University of Copenhagen, Jagtvej 128, 2200 Copenhagen, Denmark}

\author[0000-0002-2281-2785]{B. Milvang-Jensen}
\affil{Cosmic Dawn Center (DAWN)}
\affil{Niels Bohr Institute, University of Copenhagen, Jagtvej 128, 2200 Copenhagen, Denmark}
 
\author[0000-0001-5846-4404]{B. Mobasher}
\affil{Department of Physics and Astronomy, University of California, Riverside, 900 University Avenue, Riverside, CA 92521, USA}

\author{D. B. Sanders}
\affil{Institute for Astronomy, University of Hawaii, 2680 Woodlawn Drive, Honolulu, HI 96822, USA}

\author[0000-0001-5680-2326]{L. Zalesky}
\affil{Institute for Astronomy, University of Hawaii, 2680 Woodlawn Drive, Honolulu, HI 96822, USA}


\author{S. Arnouts}
\affil{Aix Marseille Univ, CNRS, LAM, Laboratoire d'Astrophysique de Marseille, Marseille, France}

\author{H. Aussel}
\affil{AIM UMR 7158, CEA, CNRS, Université Paris-Saclay, Université Paris Diderot, Sorbonne Paris Cité, F-91191 Gif-sur-Yvette, France}

\author{J. S. Dunlop}
\affil{Institute for Astronomy, University of Edinburgh, Royal Observatory, Edinburgh, EH9 3HJ, UK}

\author[0000-0002-9382-9832]{A. Faisst}
\affil{Infrared Processing and Analysis Center, California Institute of Technology, 1200 E. California Blvd, Pasadena, CA, 91125, USA}

\author[0000-0001-8221-8406]{M. Franx}
\affil{Leiden Observatory, Leiden University, P.O.Box 9513, NL-2300 AA Leiden, The Netherlands}

\author[0000-0001-6278-032X]{L. J. Furtak}
\affil{Institut d'Astrophysique de Paris, UMR 7095, CNRS,
    and Sorbonne Universit\'e, 98 bis boulevard Arago, 75014 Paris, France}

\author[0000-0002-8149-8298]{J. P. U. Fynbo}
\affil{Cosmic Dawn Center (DAWN)}
\affil{Niels Bohr Institute, University of Copenhagen, Jagtvej 128, 2200 Copenhagen, Denmark}

\author[0000-0003-4196-5960]{K. M. L. Gould}
\affil{Cosmic Dawn Center (DAWN)}
\affil{Niels Bohr Institute, University of Copenhagen, Jagtvej 128, 2200 Copenhagen, Denmark}

\author[0000-0002-2554-1837]{T. R. Greve}
\affil{Cosmic Dawn Center (DAWN)}
\affil{
DTU-Space, Technical University of Denmark, Elektrovej 327, DK2800 Kgs. Lyngby, Denmark
}
\affil{Department of Physics and Astronomy, University College London, Gower Place, WC1E 6BT London, UK}

\author[0000-0001-8221-8406]{S. Gwyn}
\affil{Herzberg Astronomy and Astrophysics, National Research Council of Canada, 5071 West Saanich Rd., Victoria, BC V9E 2E7, Canada}

\author[0000-0001-9187-3605]{J. S. Kartaltepe}
\affil{School of Physics and Astronomy, Institute of Technology, 84 Lomb Memorial Drive, Rochester, NY 14623, USA}

\author[0000-0001-9044-1747]{D. Kashino}
\affil{Department of Physics, ETH Z{\"u}rich, Wolfgang-Pauli-Strasse 27, 8093 Zurich, Switzerland}
\affil{Institute for Advanced Research, Nagoya University, Furocho, Chikusa, Nagoya, Aichi, 464-8602, Japan}

\author[0000-0002-6610-2048]{A. M. Koekemoer}
\affiliation{Space Telescope Science Institute, 3700 San Martin Dr., Baltimore, MD 21218, USA}

\author[0000-0002-5588-9156]{V. Kokorev}
\affil{Cosmic Dawn Center (DAWN)}
\affil{Niels Bohr Institute, University of Copenhagen, Jagtvej 128, 2200 Copenhagen, Denmark}
 
\author{O. Le F\`evre}
\affil{Aix Marseille Univ, CNRS, LAM, Laboratoire d'Astrophysique de Marseille, Marseille, France}

\author[0000-0002-6423-3597]{S. Lilly}
\affil{Department of Physics, ETH Z{\"u}rich, Wolfgang-Pauli-Strasse 27, 8093 Zurich, Switzerland}

\author{D. Masters}
\altaffiliation{NASA Postdoctoral Program (NPP) Fellow}
\affil{Jet Propulsion Laboratory, California Institute of Technology 4800 Oak Grove Drive Pasadena, CA 91109, USA}
\affil{Infrared Processing and Analysis Center, California Institute of Technology, 1200 E. California Blvd, Pasadena, CA, 91125, USA}

\author[0000-0002-4872-2294]{G. Magdis}
\affil{Cosmic Dawn Center (DAWN)}
\affil{
DTU-Space, Technical University of Denmark, Elektrovej 327, DK2800 Kgs. Lyngby, Denmark
}
\affil{Niels Bohr Institute, University of Copenhagen, Jagtvej 128, 2200 Copenhagen, Denmark}
\affil{Institute for Astronomy, Astrophysics, Space Applications and Remote Sensing, National Observatory of Athens, 15236
Athens, Greece}
 
\author[0000-0001-7166-6035]{V. Mehta}
\affil{Minnesota Institute for Astrophysics, University of Minnesota, 116 Church St SE, Minneapolis, MN 55455, USA}

\author{Y. Peng}
\affil{Kavli Institute for Astronomy and Astrophysics, Peking University, 5 Yiheyuan Road, Beijing 100871, China}

\author{D. A. Riechers}
\affil{Department of Astronomy, Cornell University, Space Sciences Building, Ithaca, NY 14853, USA}

 \author[0000-0001-7116-9303]{M. Salvato}
\affil{Max-Planck-Institut f\"ur extraterrestrische Physik, Giessenbachstrasse, D-85748 Garching, Germany}

\author{M. Sawicki}
\altaffiliation{Canada Research Chair}
\affil{Department of Astronomy \& Physics and the Institute for Computational Astrophysics, Saint Mary’s University, 923 Robie Street, Halifax, Nova Scotia, B3H 3C3, Canada}

\author[0000-0002-9136-8876]{C. Scarlata}
\affil{Minnesota Institute for Astrophysics, University of Minnesota, 116 Church St SE, Minneapolis, MN 55455, USA}

\author{N. Scoville}
\affil{California Institute of Technology, 1200 E. California Boulevard, Pasadena, CA, 91125, USA}

\author{R. Shirley}
\affil{Astronomy Centre, Department of Physics \& Astronomy, University of Southampton, Southampton, SO17 1BJ, UK}
\affil{Institute of Astronomy, University of Cambridge, Madingley Road, Cambridge CB3 0HA, UK}

\author[0000-0002-5460-6126]{A. Sneppen}
\affil{Cosmic Dawn Center (DAWN)}
\affil{Niels Bohr Institute, University of Copenhagen, Jagtvej 128, 2200 Copenhagen, Denmark}

\author{V. Smol\u{c}i\'c}
\affil{Department of Physics, Faculty of Science, University of Zagreb, Bijeni\u{c}ka cesta 32, 10000 Zagreb, Croatia}

\author[0000-0003-3780-6801]{C. Steinhardt}
\affil{Cosmic Dawn Center (DAWN)}
\affil{Niels Bohr Institute, University of Copenhagen, Jagtvej 128, 2200 Copenhagen, Denmark}

\author{D. Stern}
\affil{Jet Propulsion Laboratory, California Institute of Technology 4800 Oak Grove Drive Pasadena, CA 91109, USA}

\author{M. Tanaka}
\affil{National Astronomical Observatory of Japan, 2-21-1 Osawa, Mitaka, Tokyo 181-8588, Japan}
\affil{Department of Astronomical Science, The Graduate University for Advanced Studies, SOKENDAI, Mitaka, Tokyo, 181-8588, Japan}

\author{Y. Taniguchi}
\affil{The Open University of Japan, 2-11, Wakaba, Mihama-ku, Chiba 261-8586, Japan}

\author[0000-0002-7064-5424]{H. I. Teplitz}
\affil{Infrared Processing and Analysis Center, California Institute of Technology, 1200 E. California Blvd, Pasadena, CA, 91125, USA}

\author[0000-0002-6748-0577]{M. Vaccari}
\affil{Inter-university Institute for Data Intensive Astronomy, Department of
Physics and Astronomy, University of the Western Cape, Robert Sobukwe
Road, 7535 Bellville, Cape Town, South Africa}
\affil{INAF - Istituto di Radioastronomia, via Gobetti 101, 40129 Bologna, Italy}

\author[0000-0003-2588-1265]{W.-H. Wang}
\affil{Institute of Astrophysics \& Astronomy, Academia Sinica, Taipei 10617, Taiwan}

\author[0000-0002-2318-301X]{G. Zamorani}
\affil{Istituto Nazionale di Astrofisica - Osservatorio di Astrofisica e Scienza dello Spazio, via Gobetti 93/3, I-40129, Bologna, Italy}

\begin{abstract}

The Cosmic Evolution Survey (COSMOS) has become a cornerstone of extragalactic astronomy. Since the last public catalog in 2015, a wealth of new imaging and spectroscopic data has been collected in the COSMOS field. This paper describes the collection, processing, and analysis of this new imaging data to produce a new reference photometric redshift catalog. Source detection and multi-wavelength photometry is performed for 1.7 million sources across the $2\,\mathrm{deg}^{2}$ of the COSMOS field, $\sim$966\,000 of which are measured with all available broad-band data using both traditional aperture photometric methods and a new profile-fitting photometric extraction tool, \farmer{}, which we have developed. A detailed comparison of the two resulting photometric catalogs is presented. Photometric redshifts are computed for all sources in each catalog utilizing two independent photometric redshift codes. Finally, a comparison is made between the performance of the photometric methodologies and of the redshift codes to demonstrate an exceptional degree of self-consistency in the resulting photometric redshifts. The $i<21$ sources have sub-percent photometric redshift accuracy and even the faintest sources at $25<i<27$ reach a precision of $5\,\%$. Finally, these results are discussed in the context of previous, current, and future surveys in the COSMOS field. Compared to COSMOS2015, reaches the same photometric redshift precision at almost one magnitude deeper. Both photometric catalogs and their photometric redshift solutions and physical parameters will be made available through the usual astronomical archive systems (ESO Phase 3, IPAC IRSA, and CDS).

\end{abstract}

\keywords{catalogs --– galaxies: evolution –- galaxies: high-redshift –- galaxies: photometry -- methods: observational –- techniques: photometric}



\section{Introduction}
\label{sec:intro}

Photometric surveys are an essential component of modern astrophysics. The first surveys of the sky with photographic plates \citep{1888BuAsI...5..303B} permitted a quantitative understanding of our Universe; longer exposures on increasingly larger telescopes led to the first accurate understanding of the true size and scale of our Universe \citep{1934ApJ....79....8H}. Recent breakthroughs have been enabled by the advent of wide-field cameras capable of covering several square degrees at a time \cite[such as MegaCam,][]{2003SPIE.4841...72B}, coupled with wide-field spectroscopic instruments capable of collecting large numbers of spectroscopic redshifts like the Visible Multi-Object Spectrograph \citep[VIMOS;][]{2003SPIE.4841.1670L} and the Multi-Object Spectrograph For Infrared Exploration \citep[MOSFIRE;][]{2012SPIE.8446E..0JM}.

The launch of the Hubble Space Telescope (HST) led to the first Hubble Deep Field catalog \citep[HDF;][]{williams_hdf_96} which, although limited to an area of 7.5\,arcmin$^{2}$ in four optical bands to $\sim\,$28\,AB depth, revealed the morphological complexity of the distant Universe. This first step gave way to an explosion of data from similar surveys \citep[see][and references therein]{Madau_2014}. The installation of the Advanced Camera for Surveys (ACS) on HST led to a dramatic increase in the field-of-view and sensitivity of optical observations from space. This advancement laid the groundwork for the Great Observatories Origins Deep Survey \citep[GOODS;][]{Giavalisco_2004} which captured multi-band ACS observations over two 16$\times$10\,arcmin fields, totaling over 40 times more area than the original HDF. These observations provided groundbreaking insights into the nature of high-redshift galaxies, their rest-frame properties, and helped guide the development of methods to select different classes of objects. Although deep ground-based near-infrared imaging achieved notable successes \citep[e.g., FIRESurvey;][]{2003AJ....125.1107L}, the installation of the near-infrared camera WFC3 on HST in 2009 expanded our ability to probe the distant Universe. This allowed, for the first time, spatially-resolved measurements of rest-frame optical light at early cosmic times to depths unreachable from ground-based facilities, because of the high infrared sky background. The combined power of ACS and WFC3 yielded the deepest `blank-field' image of the Universe, the Hubble Ultra Deep Field \citep[HUDF;][]{Beckwith_2006, Ellis_2013, Illingworth13_hudf, Teplitz_2013}, observed over the course of a decade in 13 filters, some reaching depths $\sim\,$29.5$\,-\,$30\,AB. Together with ground-based spectroscopy, it was then possible to confirm some of the most distant galaxies which likely contributed to the reionization of the Universe \citep[e.g.,][]{Robertson13_reionzation, ishigaki18_UV_LF}. However, the transformative power of these forerunner observations was limited by their small area, complicating efforts to detect and characterize populations of rare high-redshift galaxies. To combat the effects of cosmic variance, the Cosmic Assembly Near-infrared Deep Extragalactic Legacy Survey \citep[CANDELS;][]{Grogin_2011, Koekemoer_2011} placed observations over five separate fields, covering $\sim\,$100 times more area than the HUDF with ACS and WFC3/IR in multiple filters to depths $\sim\,$28$\,-\,$29\,AB, which enabled precise measurements of the physical parameters of galaxies over cosmic time. Despite these significant advantages and the groundbreaking science they allowed, their individual areas proved still too small to fully combat cosmic variance to the extent required to probe large numbers of galaxies at high-redshift.

The Cosmic Evolution Survey \citep[COSMOS;][]{scoville_cosmic_2007} began in 2003 with a 1.7\,deg$^2$ mosaic with ACS over 583 HST orbits, reaching a $5\sigma$ depth of 27.2\,AB in the F814W band \citep{Scoville2007b, koekemoer_cosmos_2007}. This was the largest single allocation of HST orbits at the time and remains the largest contiguous area mapped with HST to date. Since then, the field has been covered with deep observations by virtually all major astronomical facilities which have consistently invested in extragalactic studies.

While various HST observations have been carried out with other bands in COSMOS, the programs completed to date generally cover no more than a few percent of the field. Ground-based broad- and narrow-band observations with Subaru Suprime-Cam were some of the first to be performed over the entire area in 2006, providing one of the largest imaging data sets available at that time \citep{capak_first_2007}. Mid-infrared observations of the entire COSMOS field were also taken using the Spitzer Space Telescope \citep{Sanders_2007}. 

The key to exploiting these multi-wavelength data sets has been `photometric redshift' estimation (hereafter \photoz), in which template spectral energy distributions (SEDs) are fit to photometry to estimate distances and physical parameters of galaxies \citep[see][for a review]{salvato19_photoz_review}. This has enabled the construction of large statistical samples of galaxies with well-characterized photometric redshifts calibrated to subsets of galaxies with accurate spectroscopic redshifts. COSMOS has been a benchmark testing ground for \photoz{} measurement techniques, due to its unrivaled multi-wavelength imaging data and thousands of measured spectroscopic redshifts.

Over the years, several COSMOS photometric catalogs have been publicly released \citep{capak_first_2007, ilbert_photoz_2009, ilbert_mass_2013, muzzin_public_2013, laigle_cosmos2015_2016}. Each of these releases followed new availability of progressively deeper data, such as the intermediate band Subaru/Suprime-Cam data \citep{taniguchi_subaru_2015} and the VISTA near-infrared coverage \citep{mccracken_ultravista_2012, milvangjensen_2013}. The most recent release, COSMOS2015 \citep{laigle_cosmos2015_2016}, contains half a million galaxies detected in the combined $zYJHK_s$ images from the Subaru and VISTA telescopes. Four ultra-deep stripes in VISTA and Spitzer, although non-uniform, cover a total area of 0.62\,deg$^{2}$ \citep[e.g.,][]{ashby_spitzer_2018}. The reported photometric redshifts reach a sub-percent precision at $i<22.5$. This methodology was applied also to the Subaru-XMM Deep Field \citep{Mehta2018}, the only other deep degree-scale field to feature similarly deep near- and mid-infrared coverage.

For more than a decade, the COSMOS field has occupied an outstanding position in the modern landscape of deep surveys, and has been relied upon to address fundamental scientific questions about our Universe. The 2\,deg$^2$ of COSMOS have been used to trace large-scale structure \citep{scoville13_cosmicweb, laigle18_filaments}, discover groups and clusters \citep[e.g.,][]{capak11_highz_cluster,2015ApJ...808L..33C,2016ApJ...826..130H,cucciati18_highz_cluster}, and link galaxies to their dark matter halos \citep[e.g.,][]{leauthaud_weak_2007, mccracken15_clustering, legrand19_hod}. The COSMOS \photoz{} distribution is used as reference to establish the true redshift distribution in redshift slices in the Dark Energy Survey \citep[DES;][]{troxel18_DES}, a crucial component when estimating cosmological parameters with weak lensing \citep[e.g.,][]{mandelbaum18_weaklensing}. COSMOS demonstrated feasibility of combining space-based shape measurements with ground-based photometric redshifts to map the spatial distribution of dark matter \citep{2007Natur.445..286M}, a method which will be used by the \Euclid{} mission \citep{EuclidRedBook}. COSMOS is already being used to prepare essential spectroscopic observations for the mission \citep{masters19_C3R2_2release} and to study biases in shape analyses.  COSMOS photometric data are being used to predict the quality of \Euclid{} \photoz{} (Duprez et al., in prep.), as well as the number of [O{\sc ii}] and H$\alpha$ emitters expected for future dark energy surveys \citep{saito20_emitters_COSMOS2015}. Hence, the photometric catalogs created in COSMOS continue to play a crucial role in cosmic shear surveys \citep{2006astro.ph..9591A}.

The combination of its depth in the visible and near-infrared, and the wide area covered, makes COSMOS ideal for identifying the largest statistical samples of the rarest, brightest, and most massive galaxies, such as ultra massive quiescent galaxies up to $z\sim4$ \citep[e.g.,][]{stockmann20_QGz2,Schreiber18_QGz4,Valentino20_QGz4}, as well as extremely luminous $z\sim5-6$ starbursts \citep[e.g.,][]{2010ApJ...720L.131R, 2014ApJ...796...84R, 2020ApJ...895...81R, 2018ApJ...861...43P, 2019ApJ...887...55C}, quasars \citep[e.g.,][]{2006ApJ...644..100P, 2016A&A...595A..13H}, and UV-bright star-forming galaxies at $6<z<10$ \citep[e.g.,][]{caputi15_highz_mass, Stefanon2019, bowler20_z8-10}. With rich multi-wavelength coverage at all accessible wavelengths from the X-ray \citep{civano16_chandra} to the radio \citep{smolcic17_radio}, an accurate picture of the galaxy stellar mass assembly was established with this data set, including numerous estimates of the galaxy stellar mass function \cite[e.g.,][]{ilbert_mass_2013, muzzin13_mass, Davidzon17_mass}, star formation rate density \citep[e.g.,][]{gruppioni13_LF_IR, novak17_sfrd}, mass and star formation rate relation \citep[][]{karim11_ssfr, rodighiero11_massSFR, Ilbert15_ssfr, lee15_ssfr, leslie20_ssfr}, and star formation quenching \citep[e.g.,][]{peng10_quenching}. A large number of follow-up programs have been conducted, including extensive spectroscopic coverage \citep[e.g.,][]{lilly07_zcosmos, lefevre15, vanderwel16_legac, hasinger18_deimos10k}, integral field spectroscopy \citep[e.g.,][]{foster-schreiber09_sins}, and ALMA observations \citep{scoville17_ALMA, lefevre19_alpine}.

This paper presents `COSMOS2020', the latest release of the COSMOS catalog. The principal additions comprise new ultra-deep optical data from the Hyper Suprime-Cam (HSC) Subaru Strategic Program (SSP) PDR2 \citep[SSP;][]{aihara_second_2019}, new Visible Infrared Survey Telescope for Astronomy (VISTA) data from DR4 reaching at least one magnitude deeper in the $K_s$ band over the full area, and the inclusion of all Spitzer IRAC data ever taken in COSMOS. Additionally, even deeper $u^*$ and new $u$ band imaging from the Canada-France-Hawaii Telescope (CHFT) program CLAUDS \citep{sawicki_cfht_2019} provides uniform, deep coverage over greater area than available in 2015. Legacy data sets (such as the Suprime-Cam imaging) have also been reprocessed. All imaging data is now aligned with Gaia DR1 \citep{gaia_collaboration_gaia_2016} for the optical and near-infrared data and DR2 \citep{gaia_collaboration_gaia_2018} for the U bands and IRAC data (see Moneti et al., submitted). This is reflected in band-to-band astrometric precision, which is comparably better than in \citet{laigle_cosmos2015_2016}. Taken together, these additions result in a doubling of the number of detected sources and an overall increase in photometric and astrometric homogeneity of the full data set. 

Previous COSMOS photometric catalogs were created with \texttt{SExtractor} \citep{bertin_sextractor:_1996}, wherein each image is first homogenized to a common `target' point-spread function (PSF). Fluxes are then extracted within circular apertures \citep{capak_first_2007, ilbert_photoz_2009, laigle_cosmos2015_2016}. While this approach is widely applied in the literature \citep[e.g.,][]{hildebrandt12_cfhtlens, sawicki1998}, other approaches avoid this homogenization process in order to preserve the original PSFs. The most common alternative involves using a model profile to estimate fluxes, with a wide variety of implementations and variations thereof \citep[e.g][]{Mobasher_1996, FS_HDF_1999, Labbe_2006, Hsieh_2012, Labbe_2015}. Of recent popularity are prior-based techniques \citep[e.g.,][]{desantis07_convphot, laidler07_tfit, merlin16_tphot} which use the highest resolution image as a prior, convolve it with the corresponding PSF kernel of the lower resolution images and utilize the normalization of the PSF convolved prior image to estimate the flux in the lower resolution images. Such an approach was instrumental to extract Spitzer/IRAC photometry in the CANDELS catalogs. Recently, \tractor{} \citep{lang_tractor_2016} has been developed to perform profile-fitting photometry. Instead of a prior cut from a high resolution image (e.g., HST), \tractor{} derives entirely parametric models from one or more images containing some degree of morphological information. This has two immediate advantages in that \tractor{} does not require a high resolution image from HST and can hence be readily and consistently applied to ground-based data sets, nor does it require that all the images are aligned on the same or integer-multiple pixel grid. Because the models are purely parametric, \tractor{} can provide shape measurements for resolved sources in addition to fluxes. \tractor{} has already been applied to several deep imaging surveys \citep[][]{Nyland17_servs_tractor, dey2019_decals}, the methods of which have greatly influenced this work. 

For COSMOS2020, two independent catalogs are created using different techniques. One is created using the same standard method as \citet{laigle_cosmos2015_2016} where aperture photometry is performed on PSF-homogenized images, with the exception of IRAC where PSF-fitting with the \texttt{IRACLEAN} software \citep{Hsieh_2012} is used. This is the \classic{} catalog. The other catalog is created with \farmer{} (Weaver et al., in prep.), a software package which generates a full multi-wavelength catalog utilizing \tractor{} to perform the modeling. In this sense, \farmer{} provides broadly reproducible source detection and photometry which \tractor{}, requiring a custom driving script, cannot do by itself. Detailed comparisons of both photometric catalogs and the quality of the \photoz{} derived from each of them are presented. By utilizing these two methods in tandem it is possible to evaluate the reliability of COSMOS2020. This work presents a detailed analysis of the advantages of each method and provide quantitative arguments which could guide photometric extraction choices for future photometric surveys. The most compelling advantage, however, lies not in discriminating between the catalogs but rather in using them constructively to evaluate the significance, accuracy, and precision of scientific results, a feature which has not yet been possible from a single COSMOS catalog release.

The paper is organized as follows. In Section~\ref{sec:data}, the imaging data set and the data reduction are presented. Section~\ref{sec:flux} describes the source extraction and photometry. The photometry from the two photometric catalogs are compared in Section~\ref{sec:comparison}. Section~\ref{sec:photoz} presents the photometric redshift measurements. In Section~\ref{sec:pp}, the physical parameters of the sources in the catalog are presented. Section~\ref{sec:conclu} presents our summary and conclusions. 

The two catalog files contain the position, extracted multi band photometry, matched ancillary photometry, area flags, derived photometric redshifts and physical parameters. Details of the catalog files including column names and descriptions will purposely not be presented in this paper, as at the time of writing the two catalog files have a combined 1,\,116 columns. Instead, reliable and up-to-date information corresponding to the particular catalog release version can be found in their accompanying \texttt{README} file and separate release documentation currently in preparation. More information can be found in Section~\ref{sec:release}.

The results presented in this paper adopt a standard $\Lambda$CDM cosmology with $H_0=70$\,km\,s$^{-1}$\,Mpc$^{-1}$, $\Omega_{\rm m,0}=0.3$ and $\Omega_{\Lambda,0}=0.7$. All magnitudes are expressed in the AB system \citep{oke_absolute_1974}, for which a flux $f_\nu$ in $\mu$Jy
($10^{-29}$~erg~cm$^{-2}$s$^{-1}$Hz$^{-1}$) corresponds to AB$_\nu=23.9-2.5\,\log_{10}(f_\nu/\mu{\rm Jy})$.

\section{Observations and Data Reduction}
\label{sec:data}

\subsection{Overview of included data}
The principal improvements in COSMOS2020 compared to previous catalogs are the significantly deeper optical and near-infrared images from ongoing Subaru-HSC and VISTA-VIRCAM surveys. In addition, this release contains the definitive reprocessing of all Spitzer data ever taken on COSMOS. `Legacy' or pre-existing data sets present in COSMOS2015 have been reprocessed to take advantage of improved astrometry from Gaia (the only exceptions being external ancillary data such as GALEX). All images are resampled to make final stacks with a $0\farcs15$ pixel scale. These stacks are aligned to the COSMOS tangent point, which has a right ascension and declination (J2000) of (10h00m27.92s  +02$^{\circ}12{\displaystyle '}03\farcs{}50$).

\begin{figure}[t]
	\centering
	\includegraphics[width=\hsize]{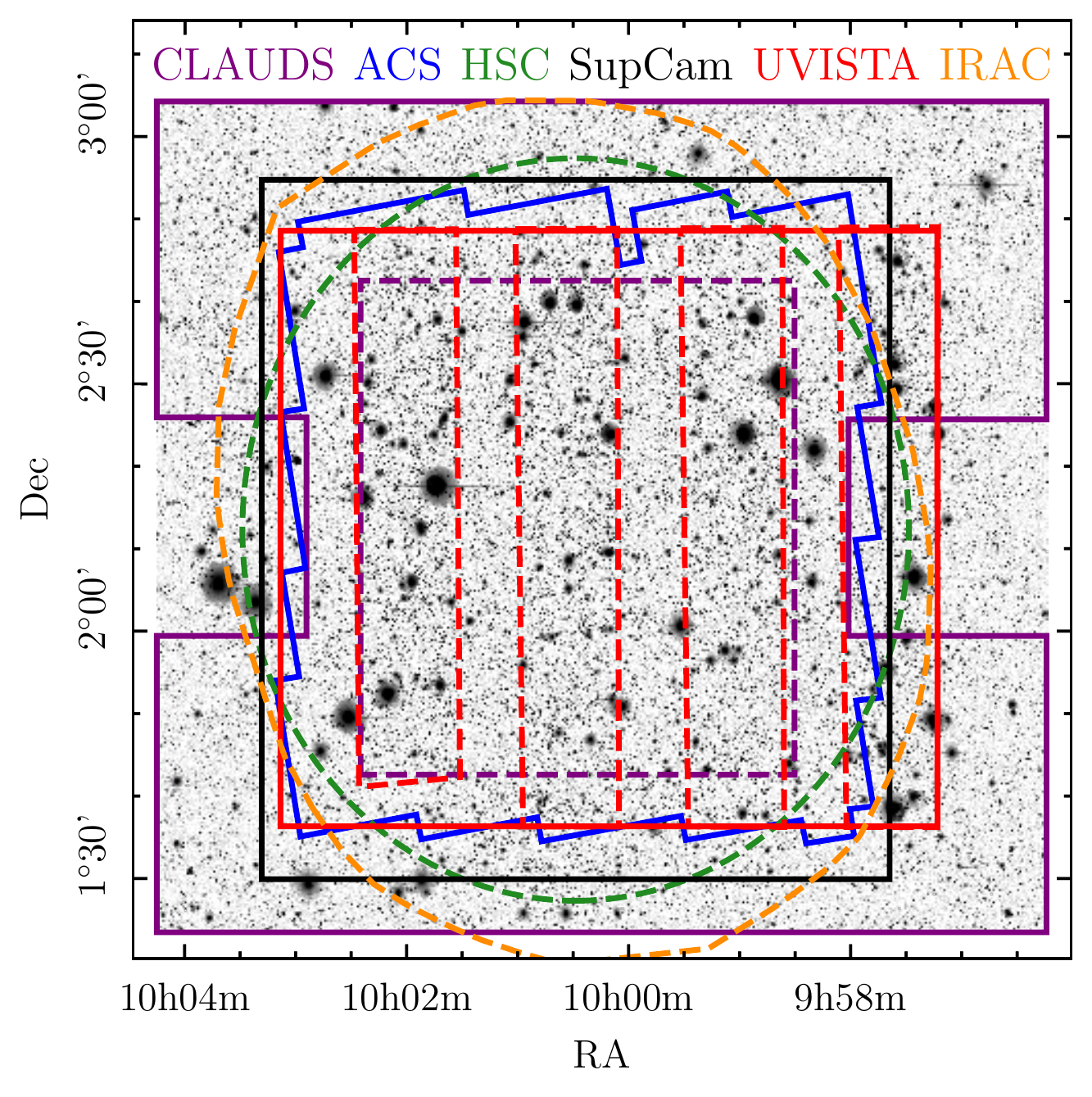}
	\caption{Schematic of the COSMOS field. The background image corresponds to the $izYJHK_s$ detection image. The solid lines represent survey limits, and the dashed lines indicate the deepest regions of the images. In the case of UltraVISTA, the dashed lines illustrate the `ultra-deep' stripes. In the case of CLAUDS, the solid line shows the limit of the $u$ band image and the dashed line shows the deepest region of the $u^*$ band image.}
	\label{fig:field}
\end{figure}

\begin{figure*}[t]
	\centering
	\includegraphics[width=\hsize]{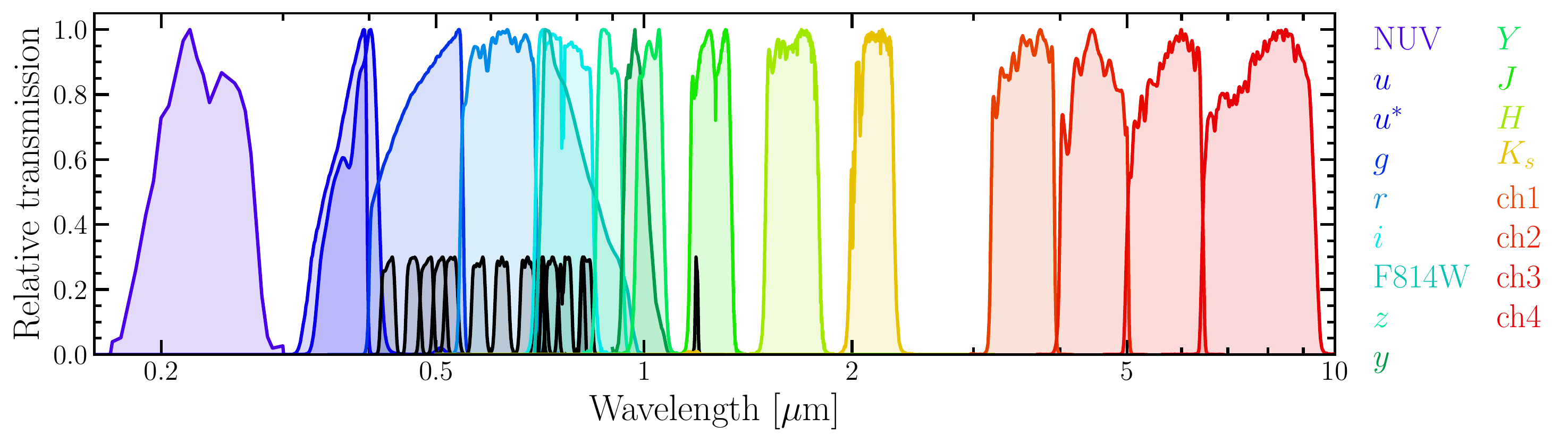}
	\caption{Relative transmission curves for the photometric bands used. The effect of atmosphere, telescope, camera optics, filter, and detector are included. The black curves represent medium and narrow-bands. The profiles are normalized to a peak transmission of 1.0 for the broad-bands, and to 0.3 for the medium and narrow-bands.}
	\label{fig:transmission}
\end{figure*}

\begin{figure}[t]
	\centering
	\includegraphics[width=\hsize]{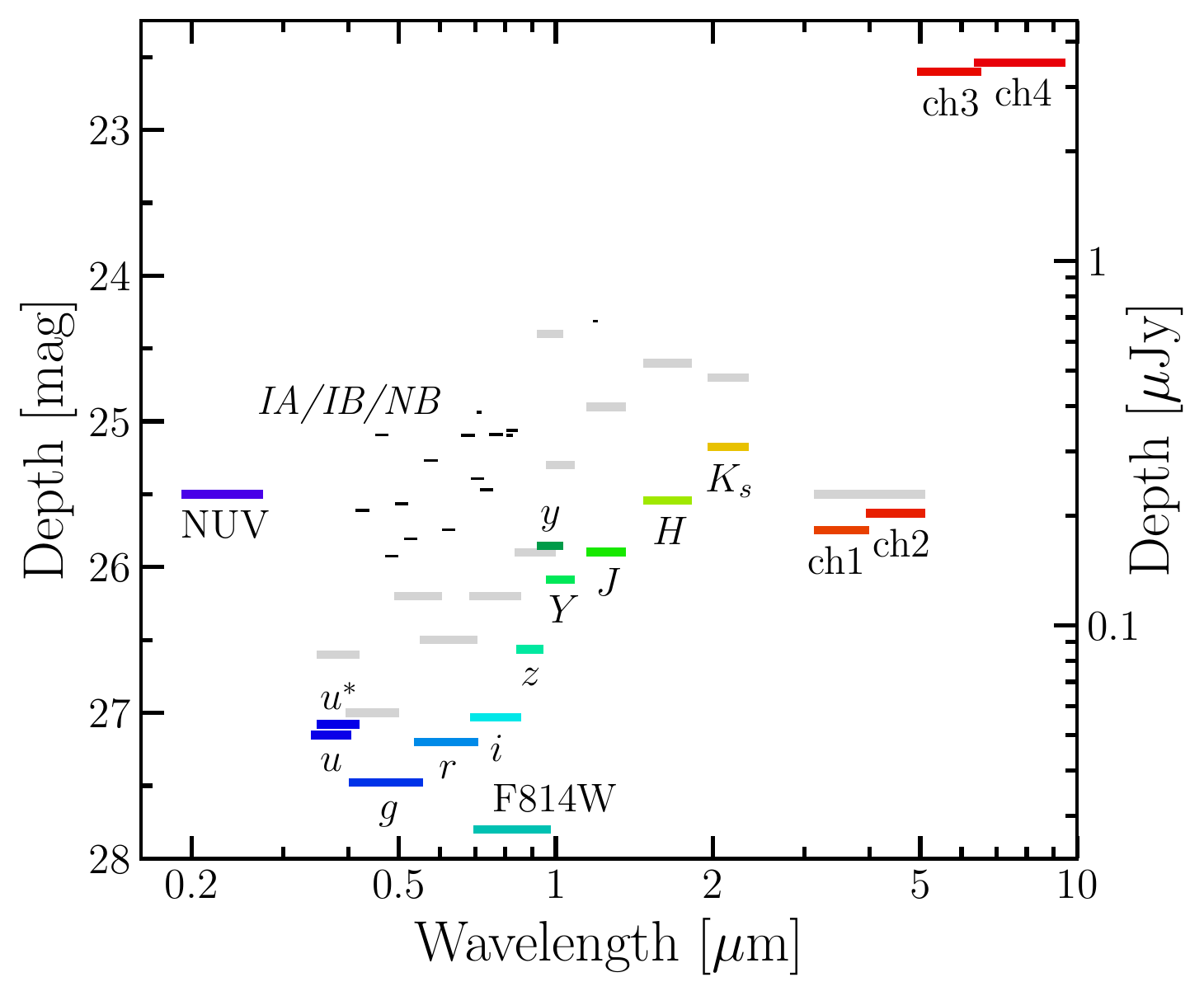}
	\caption{Depths at 3$\sigma$ measured in empty 3\arcsec{} diameter apertures in PSF-homogenized images, except for NUV and IRAC images. The NUV depth is from \citet{zamojski_deep_2007} and the F814W $3\sigma$ depth is derived from the $5\sigma$ value in \citet{koekemoer_cosmos_2007}. For the $Y$, $J$, $H$, $K_s$ bands, the depths in the ultra-deep regions are indicated. The length of each segment is the FWHM of the filter transmission curve. The thin black segments show the depths of the medium and narrow-bands. The grey segments indicate the depths of the images used in \citet{laigle_cosmos2015_2016} for comparison.}
	\label{fig:depth_comparison}
\end{figure}

Figure~\ref{fig:field} illustrates the footprint of the observations in the COSMOS field. Complete details of included data are listed in Table~\ref{tab:band_infos}. Image quality of the optical and near-infrared data, typically reported as the full-width at half-maximum (FWHM) of a Gaussian fit to the light profile, is excellent; with the exception of the Suprime-Cam $g^+$ band stack, FWHM values are all between 0\farcs6 and 1\farcs0. Figure~\ref{fig:transmission} shows the filter transmission curves. Figure~\ref{fig:depth_comparison} indicates the depths of the photometric data and provides a comparison with the COSMOS2015 depths. The depth computations are explained in Section~\ref{sec:aperturephoto} and follow largely the methods in \citet{laigle_cosmos2015_2016}. As in previous releases, in each band the image and the corresponding weight-map is resampled on the same tangent point using \texttt{SWarp} \citep{bertin_terapix_2002}. These images will be made publicly available through the COSMOS website at the NASA/IPAC Infrared Science Archive\footnote{\url{https://irsa.ipac.caltech.edu/Missions/cosmos.html}} (IRSA).

\begin{table}
\caption{UV-optical-IR data used in the catalogs}
\footnotesize
\setlength{\tabcolsep}{2pt}
\begin{threeparttable}

\begin{tabular*}{\hsize}{lccccc}
 \hline \hline
Instrument & Band & Central\tnote{a} & Width\tnote{b} & Depth\tnote{c} & Error Fact.\tnote{d} \\
/Telescope &  & $\lambda$ [\AA{}] & [\AA{}] & (2\arcsec/3\arcsec) & (2\arcsec/3\arcsec) \\
(Survey) &  &  &  & $\pm0.1$ & $\pm0.1$ \\
 \hline
GALEX & FUV & 1526 & 224 & 26.0\tnote{e} & - \\
 & NUV & 2307 & 791 & 26.0\tnote{e} & - \\
 \hline
MegaCam & $u$ & 3709 & 518 & 27.8/27.2 & 1.7/2.0 \\
/CFHT & $u^{*}$ & 3858 & 598 & 27.7/27.1 & 1.4/1.6 \\
 \hline
ACS/$HST$ & F814W & 8333 & 2511 & 27.8\tnote{f} & - \\
 \hline
HSC & $g$ & 4847 & 1383 & 28.1/27.5 & 1.4/1.8 \\
/Subaru & $r$ & 6219 & 1547 & 27.8/27.2 & 1.4/1.7 \\
HSC-SSP & $i$ & 7699 & 1471 & 27.6/27.0 & 1.5/1.9 \\
PDR2 & $z$ & 8894 & 766 & 27.2/26.6 & 1.4/1.7 \\
 & $y$ & 9761 & 786 & 26.5/25.9 & 1.4/1.7 \\
 \hline
Suprime-Cam & $B$ & 4488 & 892 & 27.8/27.1 & 1.5/1.8 \\
/Subaru & $g^{+}$ & 4804 & 1265 & 26.1/25.6 & 5.5/5.8 \\
 & $V$ & 5487 & 954 & 26.8/26.2 & 2.1/2.3 \\
 & $r^{+}$ & 6305 & 1376 & 27.1/26.5 & 1.6/1.9 \\
 & $i^{+}$ & 7693 & 1497 & 26.7/26.1 & 1.5/1.8 \\
 & $z^{+}$ & 8978 & 847 & 25.7/25.1 & 1.5/1.7 \\
 & $z^{++}$ & 9063 & 1335 & 26.3/25.7 & 2.3/2.6 \\
 & \textit{IB}$427$ & 4266 & 207 & 26.1/25.6 & 2.0/2.2 \\
 & \textit{IB}$464$ & 4635 & 218 & 25.6/25.1 & 3.1/3.3 \\
 & \textit{IA}$484$ & 4851 & 229 & 26.5/25.9 & 1.5/1.7 \\
 & \textit{IB}$505$ & 5064 & 231 & 26.1/25.6 & 1.6/1.8 \\
 & \textit{IA}$527$ & 5261 & 243 & 26.4/25.8 & 1.7/2.0 \\
 & \textit{IB}$574$ & 5766 & 273 & 25.8/25.3 & 2.4/2.5 \\
 & \textit{IA}$624$ & 6232 & 300 & 26.4/25.7 & 1.4/1.7 \\
 & \textit{IA}$679$ & 6780 & 336 & 25.6/25.1 & 2.5/2.7 \\
 & \textit{IB}$709$ & 7073 & 316 & 25.9/25.4 & 2.2/2.3 \\
 & \textit{IA}$738$ & 7361 & 324 & 26.1/25.5 & 1.5/1.7 \\
 & \textit{IA}$767$ & 7694 & 365 & 25.6/25.1 & 2.1/2.2 \\
 & \textit{IB}$827$ & 8243 & 343 & 25.6/25.1 & 2.4/2.6 \\
 & \textit{NB}$711$ & 7121 & 72 & 25.5/24.9 & 1.2/1.4 \\
 & \textit{NB}$816$ & 8150 & 120 & 25.6/25.1 & 2.3/2.5 \\
 \hline
VIRCAM & $Y^{\rm UD}$ & 10216 & 923 & 26.6/26.1 & 2.8/3.1 \\
/VISTA & $Y^{\rm Deep}$ &  &  & 25.3/24.8 & 2.7/2.8 \\
UltraVISTA & $J^{\rm UD}$ & 12525 & 1718 & 26.4/25.9 & 2.7/2.9 \\
DR4 & $J^{\rm Deep}$ &  &  & 25.2/24.7 & 2.5/2.7 \\
 & $H^{\rm UD}$ & 16466 & 2905 & 26.1/25.5 & 2.6/2.9 \\
 & $H^{\rm Deep}$ &  &  & 24.9/24.4 & 2.4/2.6 \\
 & $K_s^{\rm UD}$ & 21557 & 3074 & 25.7/25.2 & 2.4/2.6 \\
 & $K_s^{\rm Deep}$ &  &  & 25.3/24.8 & 2.4/2.6 \\
 & \textit{NB}$118$ & 11909 & 112 & 24.8/24.3 & 2.8/2.9 \\
 \hline
IRAC & ch1 & 35686 & 7443 & 26.4/25.7 & - \\
/\textit{Spitzer} & ch2 & 45067 & 10119 & 26.3/25.6 & - \\
 & ch3 & 57788 & 14082 & 23.2/22.6 & - \\
 & ch4 & 79958 & 28796 & 23.1/22.5 & - \\
 \hline
\end{tabular*}
\begin{tablenotes}
\item[a] Median of the transmission curve.
\item[b] Full width of the transmission curve at half maximum.
\item[c] $3\sigma$ depth computed on PSF-homogenized images (except for IRAC images) in empty apertures with the given diameter, averaged over the UltraVISTA area.
\item[d] Multiplicative correction factor for photometric flux uncertainties in the \classic{} catalog, averaged over the UltraVISTA area (see Section~\ref{sec:aperturephoto}).
\item[e] $3\sigma$ depth derived from the $5\sigma$ depth from \url{http://cesam.lam.fr/galex-emphot/}.
\item[f] $3\sigma$ depth derived from the $5\sigma$ depth in \citet{koekemoer_cosmos_2007}.
\end{tablenotes}
\end{threeparttable}

\label{tab:band_infos}
\end{table}

\subsection{$U$ band data}

Several programs have observed the COSMOS field in the $U$ band using the Canada-France-Hawaii telescope (CFHT) and the MegaCam instrument, the most efficient wide-field $U$ band instrument. For COSMOS2020, all archival MegaCam COSMOS $U$ data are recombined in addition to new data taken as part of the CFHT Large Area $U$ band Deep Survey\footnote{\url{https://www.ap.smu.ca/~sawicki/sawicki/CLAUDS.html}} (CLAUDS), which use a new bluer $u$ filter \citep{sawicki_cfht_2019} which lacks the red $\sim5000$\,\AA{} leakage present in the older and now retired $u^*$ filter. The methodology employed in the reprocessing is similar to that used by CLAUDS. For completeness, $u^*$ corresponds to the $u$ band used in \citet{laigle_cosmos2015_2016}. The depths\footnote{The reported $u^*$ band depth is deeper than COSMOS2015 because this work averages over the UltraVISTA layout, compared to the entire field in \citet{laigle_cosmos2015_2016}.} of the $u$ and the $u^*$ images are reported in Table~\ref{tab:band_infos}. The main motivations in reprocessing these data are to make deeper $U$ band images for the field, to make use of the new improved Gaia astrometric reference, and to resample each individual image onto the same COSMOS tangent point. 

Starting with the complete data set in both filters, these data were pre-processed by the \textit{Elixir} pipeline \citep{2004PASP..116..449M} at the CFHT before being ingested into the Canadian Astronomy Data Center, where the astrometric and photometric calibrations are recomputed using the image stacking pipeline \textit{MegaPipe} \citep{2008PASP..120..212G}. Images with sky fluxes above $\log_{10}(\rm{ADU/sec})> -0.1$ were rejected. The images were visually inspected and those with obvious flaws (bad tracking, bad seeing) were rejected. Several images were rejected during the calibration stage, having seeing worse than $1\farcs4$. In total, there were 649 $u^*$ band images and 500 $u$ band images. The median seeing of this final sample is $0\farcs9$. The two final stacked images were separately resampled onto the COSMOS tangent point and pixel scale and each were combined using a weighted 2.8 sigma clipping. The astrometric calibration used the Gaia DR2 reference catalog \citep{gaia_collaboration_gaia_2018}. The final images have an absolute astrometric uncertainty of $20$\,mas. The $u$ band calibration has been improved over earlier versions by carefully mapping the zeropoint variation across the mosaic for each observing run. Without this correction, the zeropoint could vary as much as 0.05\,mag across the field. After the correction, the variation is reduced to an estimated 0.005\,mag, a 10-fold improvement. This correction does not alter the average zeropoint. While the Sloan Digital Sky Survey (SDSS) is used as the photometric reference, it is not used as in-field standards to avoid propagating any local errors in the SDSS $u$ band calibration. Instead, zero-points are computed per night using all available images. Images taken on photometric nights were used to calibrate data taken in non-photometric conditions (see Section~3 of \citealt{sawicki_cfht_2019} for more details). In summary, both $u$ and $u^*$ images have equivalent average depths; however the newer $u$ images do not cover the entire COSMOS field but have two gaps at the left and right middle edges of the field (Figure~\ref{fig:field}). However, compared to the older $u^*$ data which is around 0.3\,mag deeper in the field center and substantially shallower outside of it, the newer $u$ data have uniform depth over the whole survey area.

\subsection{Optical data}

Wide-field optical data have played a key role in measuring COSMOS photometric redshifts. The commissioning of Subaru's $1.8$\,deg$^2$ Hyper Suprime-Cam (HSC; \citealt{2018PASJ...70S...1M}) instrument has enabled more efficient and much deeper broad-band photometric measurements over the entire COSMOS area. HSC/$y$ data were already included in \citet{laigle_cosmos2015_2016}. COSMOS2020 uses the second public data release (PDR2) of the HSC Subaru Strategic Program (HSC-SSP) comprising the $g,r,i,z,y$ bands \citep{aihara_second_2019}. 

The public stacks in COSMOS suffer from scattered light from the presence of bright stars in the field and the small dithers used. These are not removed at the image combination stage. Therefore, all the individual calibrated pre-warp CCD images (\textit{calexp} data) from the SSP public server are processed. These images were recombined with \texttt{SWarp} using \texttt{COMBINE\_TYPE} set to \texttt{CLIPPED} with a $2.8\sigma$ threshold (see \citealt{2014PASP..126..158G} for details). This removes a large fraction of the scattered light and satellite trails. As for the other data, images are centered on the COSMOS tangent point with a 0\farcs15 pixel scale. The Gaia DR1 astrometric solution computed by the HSC-SSP team agrees well with the solutions used here in other bands.

Finally, the Subaru Suprime-Cam data used in COSMOS2015 are retained for this work \citep{taniguchi_cosmic_2007,taniguchi_subaru_2015}, including 7 broad-bands ($B$, $g^+$, $V$, $r^+$, $i^+$, $z^+$, $z^{++}$), 12 medium-bands (\textit{IB}$427$, \textit{IB}$464$, \textit{IA}$484$, \textit{IB}$505$, \textit{IA}$527$, \textit{IB}$574$, \textit{IA}$624$, \textit{IA}$679$, \textit{IB}$709$, \textit{IA}$738$, \textit{IA}$767$, \textit{IB}$827$), and two narrow-bands (\textit{NB}$711$, \textit{NB}$816$). However, because the COSMOS2015 stacks had been computed with the old COSMOS astrometric reference, it was necessary to return to the individual images and recompute a new astrometric solution using Gaia DR1 with \texttt{Scamp} \citep{2006ASPC..351..112B}. The opportunity was taken to perform a tile-level PSF homogenization on the individual images. (see Section~\ref{sec:psf_homo}). 

\subsection{Near-infrared data}

The $YJHK_s$ broad-band and \textit{NB}$118$ narrow-band data from the fourth data release\footnote{\url{http://ultravista.org/release4/dr4_release.pdf}} (DR4) of the UltraVISTA survey \citep{mccracken_ultravista_2012, Moneti2019} are used. This release includes the images taken from December 2009 to June 2016 with the VIRCAM instrument on the VISTA telescope. Compared to DR2, the images are up to 0.8\,mag deeper in the ultra-deep stripes for the $J$ and $H$ bands, and 1\,mag in the deep stripes for the $K_s$ band, effectively homogenizing the $K_s$ depth across the full field. The additional \textit{NB}$118$ narrow-band image only covers the ultra-deep region. Characterization of the \textit{NB}$118$ filter is in \citet{milvangjensen_2013}. Only the publicly available stacks are used. These public stacks are aligned to the COSMOS tangent point described previously and have a $0\farcs15$ pixel scale. Gaia DR1 has been used to compute the astrometric solution.

\subsection{Mid-infrared data}

The infrared data comprise Spitzer/IRAC channel 1,2,3,4 images from the Cosmic Dawn Survey (Moneti et al., submitted). This consists of all IRAC data taken in the COSMOS field up to the end of the mission in January 2020. This includes the Spitzer Extended Deep Survey (SEDS; \citealt{ashby_seds_2013}), the Spitzer Large Area Survey with Hyper Suprime-Cam (SPLASH; \citealt{steinhardt_star_2014}), the Spitzer-Cosmic Assembly Deep Near-infrared Extragalactic Legacy Survey (S-CANDELS; \citealt{ashby_s-candels_2015}), and the Spitzer Matching Survey of the UltraVISTA ultra-deep Stripes survey (SMUVS; \citealt{ashby_spitzer_2018}). The resulting images have a 0\farcs6 pixel scale, and are resampled to the 0\farcs15 pixel scale of the optical and near-infrared images. The astrometric calibration used the Gaia DR2 reference. This work adopts the processed mosaics with stellar sources removed. A full listing of included programs and details of this processing are given in Moneti et al. (submitted). 

\subsection{X-ray, ultraviolet, and HST data}

The COSMOS2020 catalog provides basic measurements from ancillary datasets in COSMOS, including data unchanged from various source catalogs. Sources in COSMOS2020 are matched with ancillary photometric catalogs using positional cross-matching within a conservative radius of 0\farcs6 consistently for all ancillary catalogs, adopting only the most reliable sources, as described below. Measurements of the near-UV ($0.23$\,$\mu$m) and far-UV ($0.15$\,$\mu$m) are taken from the COSMOS GALEX catalog \citep{zamojski_deep_2007}, and X-ray photometry are taken from the \textit{Chandra} COSMOS Legacy survey \citep{civano_chandra_2016,marchesi_chandra_2016}. With the exception of the GALEX near-UV photometry of \citet{zamojski_deep_2007}, these ancillary data are not used in deriving \photoz{}, or physical parameters. Sources with significant X-ray detections are not used to assess \photoz{} performance, presented in Section~\ref{sec:photoz}. HST/ACS morphological measurements are used in identifying stellar contaminants. Summaries of the ancillary photometric datasets can be found in the \texttt{README} files accompanying the COSMOS2020 catalogs. Also included are column descriptions and corresponding reference literature where details of this ancillary data including their construction and caveats can be found. 

The \textit{HST}/ACS F814W high-resolution photometry from \citet{leauthaud_weak_2007} covering 1.64\,deg$^2$ of the COSMOS field are included for only unblended sources, as well as their morphological parameters. The ACS observations in the F475W and F606W bands cover about $5$\,\% of the field, so these are not included in the catalog. 

Unlike \citet{laigle_cosmos2015_2016}, far-infrared to millimeter photometry from the COSMOS Super-deblended catalog \citep{jin_super-deblended_2018} are not included as ancillary data in COSMOS2020. This is because the photometry was computed partly using a higher resolution prior catalog from COSMOS2015 and as such the identification of correct matches with COSMOS2020 is uncertain. Future work including \textit{Spitzer}/MIPS (24\,$\mu$m), \textit{Herschel}/PACS (100, 160\,$\mu$m) and SPIRE (250, 350, 500\,$\mu$m), JCMT/SCUBA2 (850\,$\mu$m), ASTE/AzTEC (1.1\,mm), IRAM/MAMBO (1.2\,mm) and VLA (1.4, 3\,GHz) photometry will be provided in an updated Super-deblended catalog using the COSMOS2020 positions as priors (Jin et al., in prep.).

\subsection{Masking}
\label{sec:masking}

Photometric extraction of sources can be significantly affected by the spurious flux of nearby bright stars, galaxies, and various other artifacts in the images. Thus, it is of interest to mark these sources. For this purpose, the COSMOS2020 catalogs provide flags for objects in the vicinity of bright stars, and for objects affected by various artifacts.

The bright-star masks from the HSC-SSP PDR2 \citep{Coupon_HSC_Masks_2018} are used to flag these sources. In particular, masks are taken from the Incremental Data Release 1 revised bright-star masks that uses Gaia DR2 as a reference star catalog, where stars brighter than $G=18$\,mag are masked. About 18\,\% of sources in the catalog are found within the masked regions in the vicinity of bright stars. Furthermore, artifacts in the Suprime-Cam images are masked using the same masks as in COSMOS2015. 

Masks indicating the area covered by the observations for the UltraVISTA deep and ultra-deep regions are provided as shown in Figure~\ref{fig:field}. Also included is a mask corresponding to coverage by Suprime-Cam. A conservative combined mask is prepared for sources within 1.27\,deg$^{2}$ which have coverage from HSC, UltraVISTA, and IRAC but which are not close to bright stars or large artifacts.

The most up-to-date descriptions of these masks and their respective flags can be found in the \texttt{README} files which accompany the catalogs.

\subsection{Astrometry}

\begin{figure}[t]
	\centering
	\includegraphics[width=\hsize]{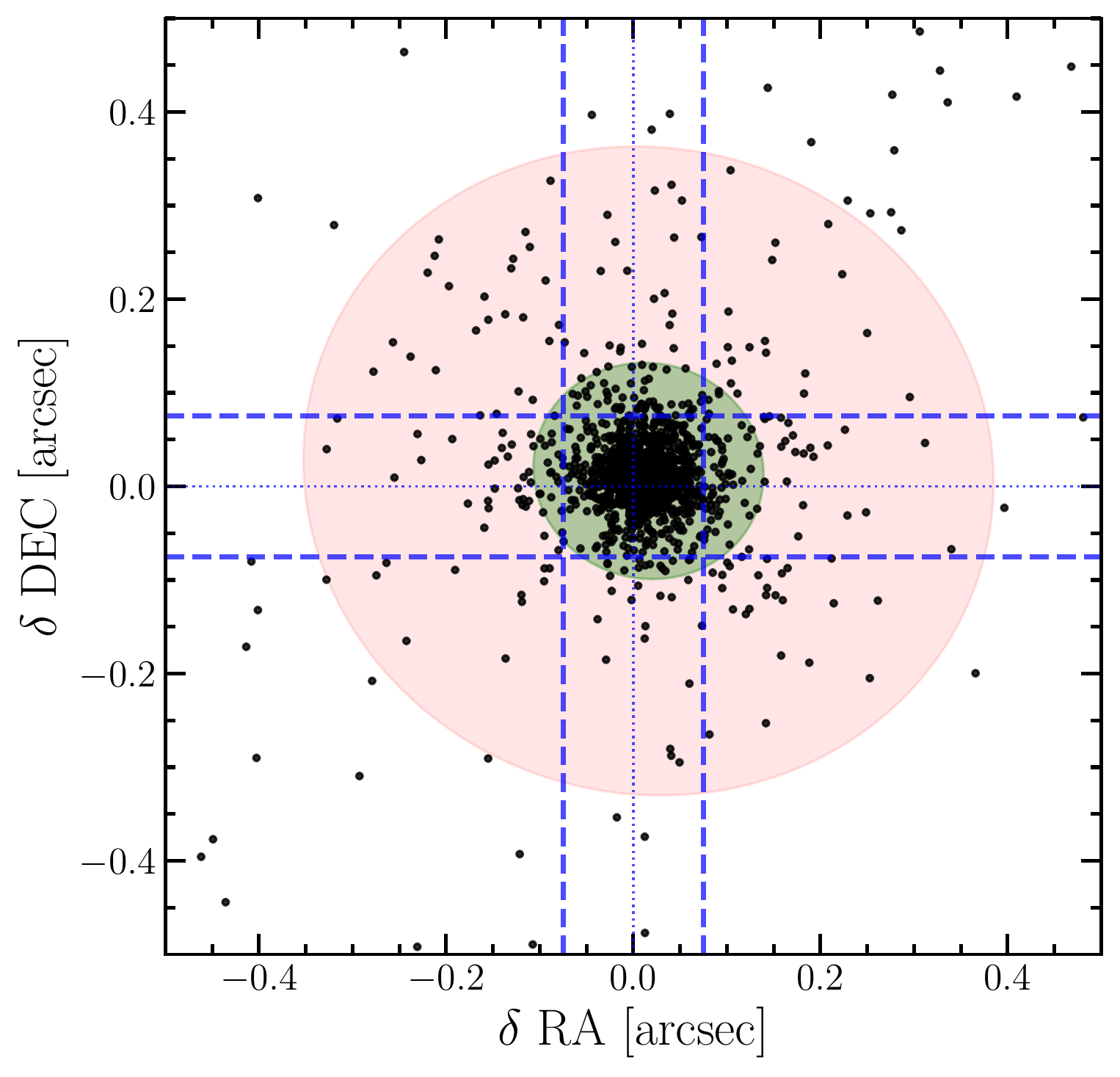}

	\caption{Coordinate offset between sources in the Gaia DR1 catalog and sources extracted in the combined detection image as measured in the aperture-based \classic{} catalog (see Section~\ref{sec:classic_detection}) The spacing between the dashed lines corresponds to the linear dimension of a pixel in the resampled images. Light and dark shaded regions are ellipses containing $68\,\%$ and $99\,\%$ of all sources respectively. For clarity, only one in ten sources are plotted.}
	\label{fig:astrometry2d}
\end{figure}

The astrometry in the previous COSMOS catalogs was based on radio interferometric data. However, with the advent of Gaia, a new, highly precise astrometric reference is available. For COSMOS2020, Astrometric solutions were computed using Gaia data for every data set described here. In the case where data presented in previous papers is included, the astrometric solutions were recomputed and data resampled. The UltraVISTA, HSC, and the reprocessed Suprime-Cam images were calibrated using the Gaia DR1 astrometric reference \citep{gaia_collaboration_gaia_2016}. Figure~\ref{fig:astrometry2d} shows the difference in position between sources in the catalog with HSC $i$ band total magnitudes between 14 and 19 magnitudes and sources in Gaia DR2. The agreement with the reference catalog is excellent, with a standard deviation in both axes of $\sim 10$ mas and an offset of $\sim 1$ mas. This is much better than any previous COSMOS catalog; for example, the size of the residuals shown in Figure 9 of \citet{laigle_cosmos2015_2016} are $\sim 100$ mas. Furthermore, there are no systematic trends of these offsets in either Right Ascension or Declination over the entire field, unlike previous catalogs. Consequently, this improved astrometric precision enables photometric measurements in smaller apertures for faint, unresolved sources.

\subsection{Spectroscopic data}

The spectroscopic data are collected from several spectroscopic surveys, conducted with different target selection criteria and instruments. In this paper, the spectroscopically-confirmed redshifts (\specz{} hereafter) are used to evaluate the accuracy of the \photoz{}. Therefore, this work only includes \specz{} with the highest confidence level. If the observation of one object is duplicated, only the \specz{} associated to the highest confidence level is used.

The spectroscopic surveys presented below share a common system to define the confidence level in the redshift measurement \citep{lilly07_zcosmos, lefevre15, hasinger18_deimos10k, Kashino19_FMOS-COSMOS, masters19_C3R2_2release, Rosani19_catalogue}. They follow a flagging system described in section 6 of \citet{lefevre05}. Each spectrum is inspected visually by two team members, who attribute a flag to the \specz{}, depending on the robustness of the measurement.  A flag 3 or 4 is associated to the \specz{} if several prominent spectral features (e.g. emission and absorption lines, continuum break) support the same \specz{}. While such flagging system is subjective, a posteriori analysis based on duplicated spectroscopic observations indicate that the confidence level of flag 3 and 4 \specz{} is above 95\%.

Two large programs were conducted at ESO-VLT with the VIMOS instrument \citep{2003SPIE.4841.1670L} to cover the COSMOS field. The $z$COSMOS survey \citep{lilly07_zcosmos} gathered 600\,h of observation and is split into a bright and a faint component. The $z$COSMOS-bright surveys targeted 20\,000 galaxies selected at $i^* \le 22.5$, which by construction is highly representative of bright sources. The $z$COSMOS-faint survey (Kashino et al., in prep) targeted star-forming galaxies selected with $B_{\rm J} <25$ and falling within the redshift range $1.5 \lesssim z \lesssim 3$. The VIMOS Ultra Deep Survey \citep[VUDS;][]{lefevre15} includes a randomly selected sample of galaxies at $i<25$, as well as a pre-selected component at $2<z<6$. Included are 8\,280, 739 and 944 galaxies from the $z$COSMOS-bright, $z$COSMOS-faint and VUDS surveys, respectively.

Data from the Complete Calibration of the Color-Redshift Relation Survey \citep[C3R2;][]{masters19_C3R2_2release} are also used. The galaxies were selected to fill the color space using the self-organising map algorithm \citep{kohonen82_som}. Depending on the expected redshift range, various instruments from the Keck telescopes were used, specifically LRIS, DEIMOS, and MOSFIRE. While this sample of 2\,056 galaxies is representative in colors, it is not designed to be representative in brightness. 

A large sample of 4\,353 galaxies taken at Keck with DEIMOS, with various selections over a large range of wavelengths from the X-ray to the far-infrared and radio \citep{hasinger18_deimos10k} are used. Such diversity of selection is crucial to estimate the quality of the \photoz{} for specific populations known to provide less robust results \citep[e.g.,][]{Casey12}.

The FMOS near-infrared spectrograph at Subaru enables tests of the \photoz{} in the redshift range $1.5<z<3$ sometimes referred to as the ``redshift desert'' \citep[e.g][]{lefevre13_VVDS_final}. The sample from \citet{Kashino19_FMOS-COSMOS} contains 832 bright star-forming galaxies at $z\sim 1.6$ with stellar masses $\log_{10}(M_\mathrm{sim}/M_\odot)>9.5$ following the star-forming main sequence.

Also adopted are 447 sources observed with MUSE at ESO/VLT  \citep{Rosani19_paper}. The sample includes faint star-forming galaxies at $z<1.5$ and Lyman alpha emitters at $z>3$, and can be used to test the \photoz{} in a magnitude regime as faint as $i>26$.

Finally, other smaller size samples are added including Darvish et al. (in prep.) and Chu et al. (in prep.) with MOSFIRE, passive galaxies at $z>1.5$ \citep{onodera12_passive}, and star-forming galaxies at $0.8<z<1.6$ from \citet{comparat15}. The full compilation of \specz{} in the COSMOS field, including the contributing survey programs, is described in Salvato et al. (in prep.).

\section{Source Detection and Photometry}
\label{sec:flux}

\subsection{The \classic{} catalog}

\subsubsection{Source Detection}
\label{sec:classic_detection}

The ``chi-squared'' $izYJHK_s$ detection image \citep{1999AJ....117...68S} is created with \texttt{SWarp} from the combined original images without PSF-homogenization using the \texttt{CHI\_MEAN} option. The inclusion of the HSC/$i,z$ band data increases the catalog completeness for bluer objects. In particular, the HSC/$i$ band image is very deep and has excellent seeing of around 0\farcs6. The previous 2015 catalog \citep{laigle_cosmos2015_2016} did not include $i$ band data in their detection image. The inclusion of the deep $i$ band in this detection strategy is the main reason for the higher number of sources detected in the COSMOS2020 catalog compared to COSMOS2015, likely driven by small, blue galaxies at low and intermediate redshift. The increased depth of the near-infrared bands also contributes to the greater number of detected sources.

For the \classic{} catalog, the detection is performed using \texttt{SExtractor} \citep{bertin_sextractor:_1996} with parameters listed in Table~\ref{tab:sextractor_config}. The main difference with respect to COSMOS2015 is \texttt{DETECT\_MINAREA} set to 5\,pix instead of 10\,pix, which is made possible thanks to the lower number of spurious sources in the detection image compared to COSMOS2015, owing to the addition of the $i$ band and deeper imaging in general. The number of detected sources reaches 1\,720\,700 over the whole field, with 790\,579 sources in the UltraVISTA region outside  the HSC bright star masks.

\subsubsection{Point Spread Function Homogenization}
\label{sec:psf_homo}

The procedure to homogenize the PSF in the optical/near-infrared images is similar to the one presented in \citet{laigle_cosmos2015_2016}. 
In the first step, \texttt{SExtractor} is used to build a catalog of bright sources. Stars are identified by cross-matching coordinates with point-like sources from the \textit{HST}/ACS catalog in COSMOS \citep{koekemoer_cosmos_2007,leauthaud_weak_2007}. Saturated stars are removed in the masks (see Section \ref{sec:masking}). Bright, but not saturated stars, are identified by their position in the half-light radius versus apparent magnitude diagram.
The PSF of each image is modeled using \texttt{PSFEx} \citep{bertin_psfex_2013} adopting the polar shapelet basis functions \citep{massey_polar_2005}. The same code also provides a convolution kernel that can modify the image's response into a ``target PSF'', which is modeled as a Moffat profile \citep{moffat_theoretical_1969} with parameters $\theta=0\farcs8$ and $\beta=2.5$ (the former being the FWHM while $\beta$ is the atmospheric scattering coefficient). These two parameters are identical to \citet{laigle_cosmos2015_2016}, whereas the \texttt{PSF\_SAMPLING} parameter is now set to \texttt{1} in order to fix the kernel pixel scale. The core of the homogenization process consists in convolving the entire images with these kernels, so that all of them are affected by the same Moffat-shaped PSF.

\begin{figure}[t]
	\centering
	\includegraphics[width=\hsize]{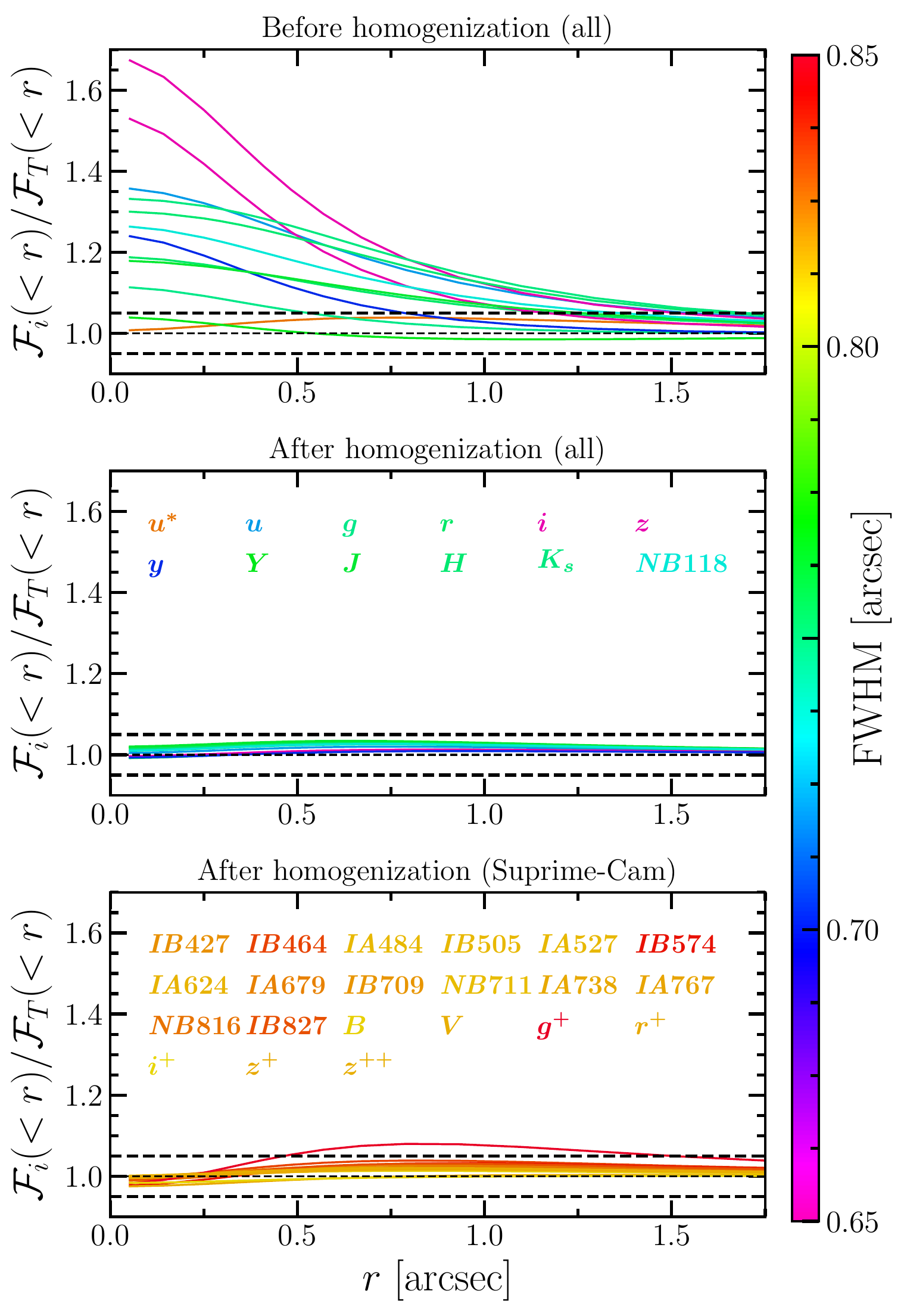}

	\caption{Best-fitting Moffat profile PSF integrated in circular apertures, $\mathcal{F}_{i}$, normalized to the target PSF $\mathcal{F}_{T}$, as a function of the aperture radius for all bands. \textit{Top:} Before PSF-homogenization, for all bands except Suprime-Cam. \textit{Middle:} After PSF-homogenization, for all bands except Suprime-Cam. \textit{Bottom:} After PSF-homogenization, for Suprime-Cam bands. The horizontal dashed lines indicate $\pm5$\,\% relative offset. The color map reflects the PSF FWHM before homogenization for all bands and after homogenization for the Suprime-Cam bands.
	}
	\label{fig:COG_main}
\end{figure}

Figure~\ref{fig:COG_main} illustrates the precision of the PSF homogenization as a function of distance from the center of the source. The integral of the best-fitting PSF within different apertures is plotted for every band, before and after the homogenization; all these functions are normalized by the integral of the target Moffat profile within the same apertures. The ratios of the integrals differ from 1 by less than 5\,\% for all apertures with the exception of Suprime-Cam/$g^{+}$, which has a particularly broad initial PSF. In this case, PSF homogenization kernels can still be consistently computed even when the input PSF is wider than the target PSF, and will give a fraction of the weight to the wings (as opposed to the central region) of the PSF. Although the difference between the Suprime-Cam/$g^{+}$ PSF and the target PSF is below 10\,\% in all apertures, it is poor enough that the band is excluded from SED fitting.

\begin{figure}[t]
	\centering
	\includegraphics[width=\hsize]{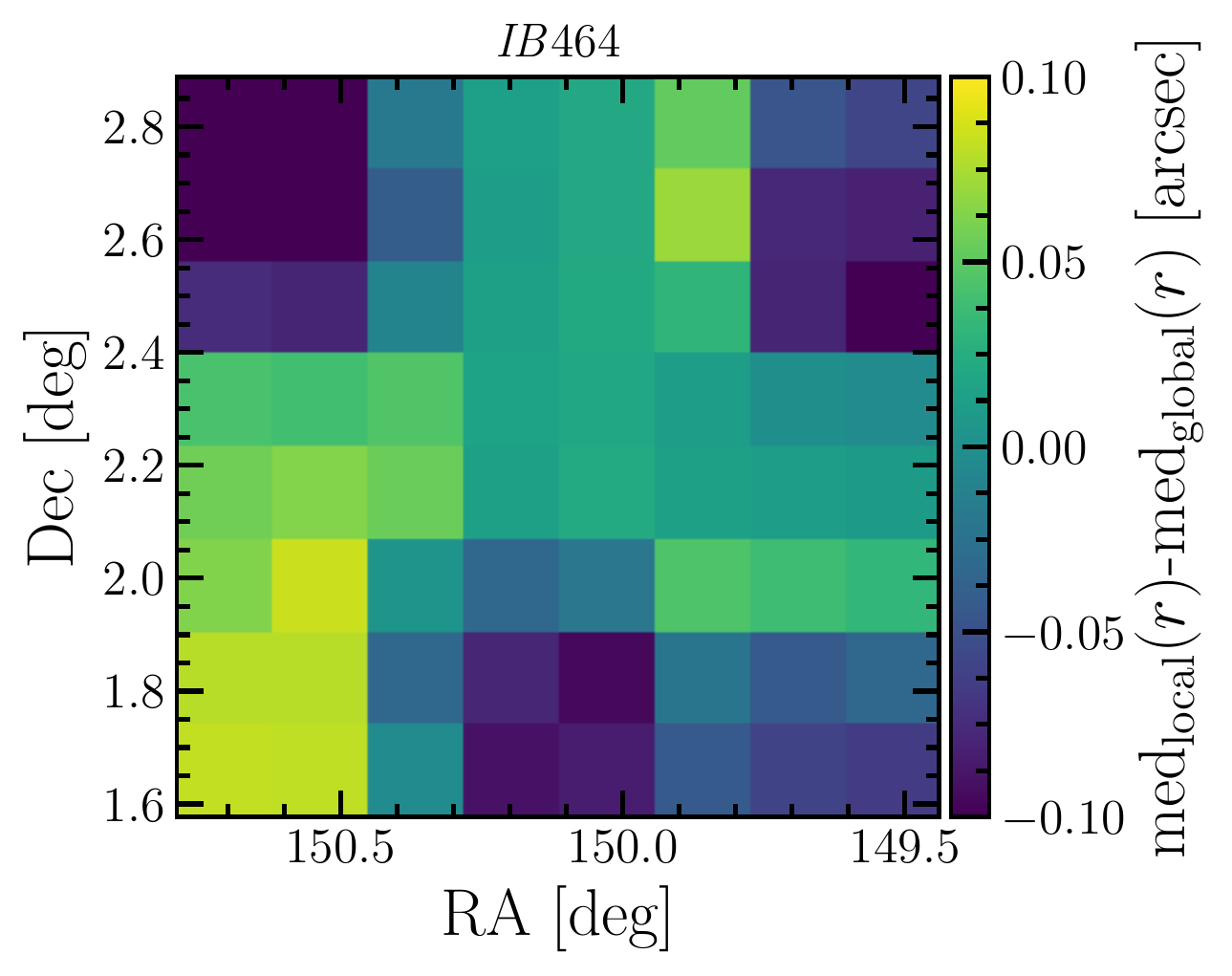}\quad\quad
	\includegraphics[width=\hsize]{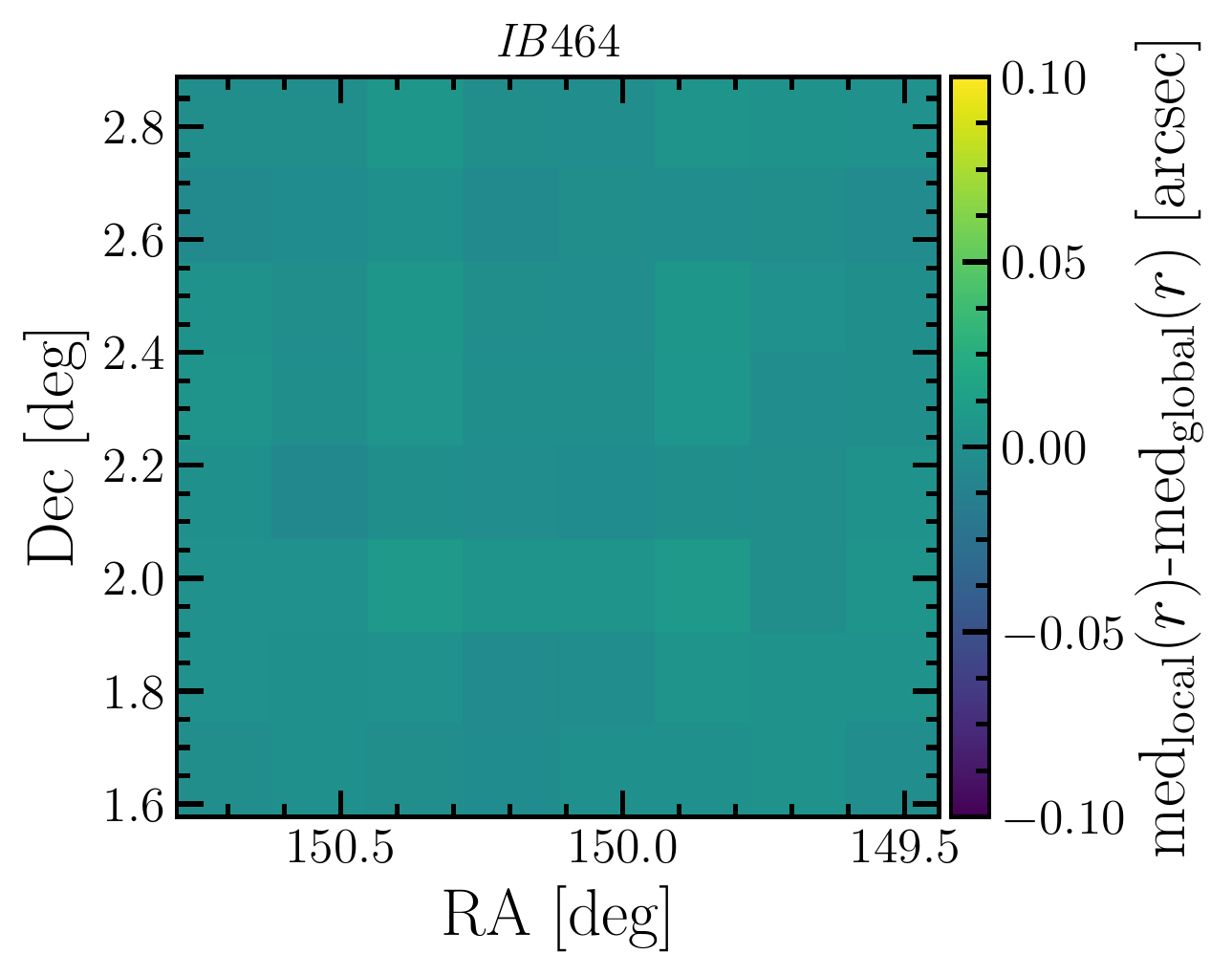}
	\caption{Distribution of the difference between the local and the global median half-light radius for the selected stars in the \textit{IB}$464$ band, as a function of position, before (\textit{top}) and after (\textit{bottom}) PSF-homogenization.
	}
	\label{fig:radmaps}
\end{figure}

In principle, spatial variability of the PSF should be taken into account. For CLAUDS, HSC and UltraVISTA bands this effect is negligible. However, for Suprime-Cam bands the resulting impact of the PSF variability on aperture photometry can be as high as 0.1\,mag (as discussed in \citealt{laigle_cosmos2015_2016}).  As an example, Figure~\ref{fig:radmaps} presents the  variation of the PSF across the sky for the Suprime-Cam/\textit{IB}$464$ band, which has the greatest spatial variability before homogenization among the considered bands.

In this work, the spatially dependent PSF homogenization of Suprime-Cam bands is performed starting from individual exposures, as they cover different patches of the field. First, the single exposure files (SEFs) at the original pixel scale of 0\farcs2 are resampled to the target tangent point with the pixel scale of 0\farcs15, to remove astrometric distortions. Then, the bright object extraction, PSF modeling, and kernel computation are done in the same way as for the other images. Stars are identified in the half-light radius versus apparent magnitude diagram, automatically adjusting the radius threshold using sigma clipping. The PSF-homogenized SEFs are finally coadded to build the final stacks. 
Frames with high sky noise ($>3.5\times$ the median noise) are rejected, representing 1, 5, 28, 16, and 4 images in the $B$, $g^{+}$, $z^{+}$, $z^{++}$, and \textit{NB}$816$ bands, respectively, out of a total of 2219 images. In these  high noise images, only a few objects are detected making it difficult to compute an astrometric solution.

\subsubsection{Aperture Photometry}
\label{sec:aperturephoto} 

Optical and near-infrared fluxes measured in 2\arcsec{} and 3\arcsec{} diameter apertures are extracted using \texttt{SExtractor} in ``dual-image'' mode from PSF-homogenized images, using the \texttt{CHI\_MEAN} as the detection image.
Fixed apertures ensure that the same structures are sampled in different bands for each source, which is necessary for reliable measurement of colors and photometric redshifts.

The photometric errors computed with \texttt{SExtractor} are underestimated in the case of correlated noise in the image \citep[e.g.,][]{leauthaud_weak_2007}. The aperture flux errors and magnitude errors are therefore re-scaled with band-dependent correction factors applied to all sources \citep{bielby_wircam_2012}; see \citet{Mehta2018} for a detailed description. In the PSF-homogenized images, the flux is measured in empty apertures (using the segmentation map estimated in each image) randomly placed over the field. The depths are computed from the standard deviation (3$\sigma$ clipped) of the fluxes in empty apertures inside the UltraVISTA area. The correction factors are then the ratio between the standard deviations of the fluxes measured in empty apertures and the median flux errors in the source catalog, as in \citet{laigle_cosmos2015_2016}. This is performed separately for 2\arcsec{} and 3\arcsec{} diameter apertures, and in the case of UltraVISTA photometry, the deep and ultra-deep regions are treated separately. $3\sigma$ depth estimates for each band computed over the central UltraVISTA area are listed in Table~\ref{tab:band_infos} and illustrated in Figure~\ref{fig:depth_comparison}. Also included in Table~\ref{tab:band_infos} are the photometric uncertainty correction factors used in the \classic{} catalog. The flux and the magnitude errors are already corrected in the \classic{} catalog, as it was done for the COSMOS2015 catalog. The 3$\sigma$ depth of the IRAC bands are computed using the same approach, after tuning the \texttt{SExtractor} configuration to the IRAC images.

Aperture photometry may underestimate the total flux of the sources. Optical and near-infrared aperture fluxes (and flux uncertainties) are converted to total fluxes using a source-dependent correction equivalent to the one adopted by \citet{laigle_cosmos2015_2016}. 
The correction for each object is computed from the pseudo-total flux $f_\text{AUTO}$, provided by \texttt{SExtractor} and defined as the flux contained within the band-independent Kron radius \citep{kron_photometry_1980} as set by \texttt{PHOT\_AUTOPARAMS} (see Table~\ref{tab:sextractor_config}), and the aperture flux $f_\text{APER}$, also provided by SExtractor. The ratio of these two measurements are then averaged over the HSC/$g,r,i,z,y$ and UltraVISTA/$Y,J,H,K_s$ broad-bands and weighted by the inverted quadratic sum of the pseudo-total and the aperture signal-to-noise:
\begin{equation}
o = \dfrac{1}{\sum_i w_i}
\sum_i
\left(\dfrac{f_\text{AUTO}}{f_\text{APER}}\right)_i
w_i,
\end{equation}
where the weights are defined as
\begin{equation}
w_i = \dfrac{1}{
\left(\dfrac{\sigma_\text{AUTO}}{f_\text{AUTO}}\right)_i^2 + 
\left(\dfrac{\sigma_\text{APER}}{f_\text{APER}}\right)_i^2
},
\end{equation}
with $\sigma_\text{AUTO}$ the $f_\text{AUTO}$ uncertainties, and $\sigma_\text{APER}$ the $f_\text{APER}$ uncertainties (corrected for correlated noise). The sum only includes the filters in which both $f_\text{AUTO}$ and $f_\text{APER}$ are positive and unsaturated.
As a result, the optical and near-infrared colors remain unaffected. Since photometry from GALEX and IRAC are measured in total fluxes, this step is required in order to obtain meaningful colors using these bands. Offsets are available (in magnitude units) in the \classic{} catalog for both 2\arcsec{} and 3\arcsec{} diameter apertures.

\subsubsection{IRAC photometry}
\label{sec:iraclean}

Photometry is performed on the \textit{Spitzer}/IRAC channels 1 and 2 images using the \texttt{IRACLEAN} software \citep{Hsieh_2012}. The infrared images of IRAC have a larger PSF (with FWHM between $1\farcs6$ and $2\farcs0$) compared to the optical data and are significantly affected by source confusion which prevents reliable photometric extraction. To tackle this issue, \texttt{IRACLEAN} uses a high-resolution image (and its segmentation map) as a prior to identify the centroid and the boundaries of the source, and iteratively subtract a fraction of its flux (`cleaning') until it reaches some convergence criteria specified by the user. \texttt{IRACLEAN} works in the approximation that an IRAC source can be modeled as a scaled Dirac delta function convolved with the PSF. 

For each source identified in the segmentation map, the software uses a box of fixed size as a filter in the low-resolution image to find the centroid and estimate the flux within a given (square) aperture. The PSF is convolved with a Dirac delta function with an amplitude equal to a fraction of that aperture flux, and then subtracted from the image. Filtering and centroid positioning are executed within the object's boundaries as defined by the prior high-resolution segmentation map. This procedure is repeated on the residual image produced by the previous iteration until the flux of the treated source becomes smaller than a specified threshold. In this case, a minimum signal-to-noise ratio of 2.5 is set so that an object will be considered completed once its aperture flux, compared to the background, becomes smaller than that value. This also implies that not all the sources detected in the prior image will be extracted by \texttt{IRACLEAN}. Moreover, since the global sky background is recomputed at each iteration, the signal of a faint source -- initially disregarded -- may emerge from the background after several passes on the nearby objects. The iterative procedure of centroid positioning within the object’s boundaries allows extended sources to be treated, and the fact that the flux is subtracted by convolving the PSF with a Dirac delta function centered on the centroid controls the contamination by neighbors. For more detail on the workings of \texttt{IRACLEAN}, the reader is referred to Sect.~7 of \citet{Hsieh_2012}, and their Figure~16 for an example of residual images.
 
User-controlled parameters are the threshold below which to stop cleaning, the filtering box size, the square aperture to measure IRAC flux and the fraction of flux to subtract at each iteration. In this configuration, a box of size $7\times7$~pixel is adopted to filter and to find the centroid, and a square aperture of size $9\times9$~pixel to estimate the aperture flux; the fraction of flux subtracted for each cleaning step is 20\,\%. The final flux of each object is the sum of the fluxes subtracted at each step. Since the centroid position is allowed to change at every iteration, the source is eventually modeled by a combination of Dirac delta functions that are not necessarily centered at the same point. The flux error is computed using the residual map by measuring the fluctuations in a local area around the object. 

This implementation adopts the high-resolution $izYJHK_s$ detection image and its segmentation map produced by \texttt{SExtractor}. In order to parallelize the processing of the images, a mosaic of $14\times14$ tiles is made with a $0\farcm3$ overlap in each direction. The PSF is modeled on a grid with spacing of $29\arcsec$ across the full IRAC image in order to take into account its spatial variation using the software \texttt{PRFmap} (A. Faisst, private communication). When modeling the PSF at each grid point, the code takes into account that the final IRAC mosaic is made of multiple overlapping frames that can have different orientations with a PSF that is not rotationally symmetric. \texttt{PRFmap} models the PSF in each of the frames that overlap at a grid point and stacks them to produce the PSF model of the mosaic at that location.
\texttt{IRACLEAN} thus provides photometry in channel 1 and 2 for more than a million sources over the whole field. 

\subsection{\farmer{} catalog}
\label{sec:tractor_catalog}

\subsubsection{Source Detection}
\label{sec:farmer_detection}

The source detection step is entirely equivalent to the procedure adopted for the \classic{} catalog. \farmer{} utilizes the \texttt{SEP} code \citep{barbary_sep_2016} to provide source detection, extraction, and segmentation, as well as background estimation with near identical performance as classical \texttt{SExtractor}. Given their near identical performance, \farmer{} uses \texttt{SEP} as both are written in Python and hence \texttt{SEP} is readily integrated into the existing workflow. 

The detection parameters are configured identically between \texttt{SExtractor} and \texttt{SEP} where possible. Crucially, given that model-based photometry from \farmer{} cannot be readily applied to saturated bright stars and sources contaminated by stellar halos, the HSC PDR2 bright star masks are adopted a priori to ensure the reliability of the derived photometry (see Section~\ref{sec:masking}). Photometric extraction with \farmer{} for COSMOS2020 is limited to the UltraVISTA footprint as this area contains all the bands used in the detection image, which are used by \farmer{} to construct galaxy models. Including areas which lack complete $izYJHK_s$ coverage introduces undesirable inhomogeneities to the model constraints, and hence may adversely change the selection function. Photometry of sources within the HSC bright star masks is also not attempted with \farmer{} as the halo light and the saturated stars are difficult to account for in a model, resulting in poor measurements and exponentially longer computational times. While there are 964\,506 sources in the entire \farmer{} catalog, only 816\,944 sources lie within UltraVISTA footprint but outside the conservative HSC bright star halo masks. This is marginally larger than the number of sources detected in the \classic{} catalog (difference $\sim$3\,\%). Of these, $\sim$95\,\% have counterparts in the \classic{} catalog within 0\farcs6. Conversely, virtually all ($>$99\,\%) \classic{} catalog sources have counterparts in \farmer{} catalog within the same radius over the same area. Generally, sources only included in \farmer{} catalog are concentrated around unmasked bright star halos and their diffraction spikes (further underscoring the need for accurate a priori masking), and which are unlikely to possess well-fit models and so are easily flagged. Some, however, appear to be result of comparably more accurate de-blending of nearby sources by \texttt{SEP}, which given the ability to easily identify nonphysical detections, is advantageous for the important reason that two blended sources will not be well-fit by models unless they are identified as separate objects at detection. This will be further discussed in the context of \farmer{} in Weaver et al., (in prep).

Once sources are detected, \farmer{} identifies crowded regions with multiple nearby sources which although deblended at detection (i.e. have their own centroids), may have some overlapping flux which must be separated by the models. Hence, to avoid double-counting flux and to achieve the most robust modeling possible, these sources are modeled simultaneously. Such crowded regions are identified by dilating the source segmentation map, which assigns pixels to sources, in order to form groups of sources defined by contiguous dilated pixels. Sources which are not in crowded areas are expected to be a group of one source, whereas sources in crowded regions end up as members of larger groups to be modeled together.

\subsubsection{PSF Creation}

In contrast with the PSF-homogenization strategy employed in the \classic{} catalog for all optical and NIR bands, \tractor{} does not operate on images which are PSF-homogenized. Since the models it uses are purely parametric, \tractor{} can simply convolve a given model with the PSF of a given band, which is generally a more tractable operation than PSF homogenization. The approach to generate PSFs for \farmer{} catalog follows similarly with that of \classic{}, using spatially constant PSFs for the broad-bands and spatially varying PSFs for the Subaru medium-bands and IRAC bands.

A spatially constant PSF is computed for $u$, $u^*$, as well as all HSC and UltraVISTA bands with \texttt{PSFEx}. Point-source candidates are selected as described in Section~\ref{sec:psf_homo}. Since models are sensitive to the wings of sources, \farmer{} benefits from particularly large PSF renderings. Typical unsaturated point-sources in optical and NIR images in this work are well-described by PSF stamps generated with 201 pixel diameters ($30\farcs15$).

Another consideration, introduced for the \classic{} catalog in section~\ref{sec:psf_homo}, is the highly variable PSF of the Suprime-Cam medium-bands. Although \farmer{} does not use any kind of PSF-homogenization procedure and hence cannot overcome this variability in the same way as for the \classic{} catalog, it is still possible to overcome highly variable PSFs in model-based photometry by providing a particular PSF to a group of sources, similar to \texttt{PRFMap} which produces a theoretical PSF sampled over a fixed grid. However, this exact approach cannot be readily replicated for other bands, since there is a lack of sufficient theoretical PSFs for the Subaru medium-bands. Instead, a spatial grid is constructed using the PSF FWHM measured from a sample of point-like sources nearest to each grid point. The FWHM distribution is then discretized to form a set of PSFs at a gauge small enough to provide accurate PSFs for each grid point while maintaining the spatial sampling required to describe the variations across the field. Hence, for each medium-band a 20$\times$20 grid consisting of 10 PSFs is built with a typical resolution of less than a tenth of a pixel. Then for a particular group of sources \farmer{} provides the nearest PSF sample to be used in the forced photometry modeling.

Lastly for IRAC, \farmer{} employs \texttt{PRFMap} to provide a spatially-varying PSF to each group of sources based on their nearest PRF sampling point, consistent with the \texttt{IRACLEAN} procedure described in section~\ref{sec:iraclean}. The PSFs are then re-sampled to match the 0\farcs15 pixel scale of the mosaics.

\subsubsection{Model Determination}
Details of the model determination procedure will be found in Weaver et al. (in prep.). This is a brief summary. \farmer{} employs five discrete models to describe resolved and unresolved, stellar and extragalactic sources:

\begin{enumerate}
    \item \textbf{PointSource} models are taken directly from the PSF used. They are parameterized by flux and centroid position and are appropriate for unresolved sources.
    \item \textbf{SimpleGalaxy} models use a circularly symmetric, exponential light profile with a fixed $0\farcs45$ effective radius such that they describe marginally resolved sources and mediate the choice between PointSource and a resolved galaxy model. They are parameterized also by flux and centroid position.
    \item \textbf{ExpGalaxy} models use an exponential light profile. They are parameterized by flux, centroid position, effective radius, axis ratio, and position angle.
    \item \textbf{DevGalaxy} models use a de\,Vaucouleurs light profile. They are parameterized by flux, centroid position, effective radius, axis ratio, and position angle.
    \item \textbf{CompositeGalaxy} models use a combination of ExpGalaxy and DevGalaxy models. They are concentric, and hence share one centroid. There is a total flux parameter as well as a fraction of total flux parameter to distribute the flux between the two components. Components have their own effective radii, axis ratios, and position angles.
\end{enumerate}

\noindent These five models form \farmer{}'s decision tree, whose goal is to both determine the most suitable model for a given source, and provide an optimized set of parameters to describe the shape and position of the source. Unlike some other model-based photometric techniques, the models in \tractor{} are purely parametric and hence do not require a high-resolution image stamp which must undergo PSF kernel convolution when photometering a different band. Although the exact implementation of the modeling can vary (e.g., choice of bands, library of models, etc.), for the present catalog \farmer{} attempts to jointly model a group of nearby sources, using simultaneous constraints from each of the six individual $izYJHK_s$ bands used in the detection image. This ensures that the selection function is preserved by providing a model even for sources detected from one band.

\farmer{} then uses its decision tree to select the most appropriate model type for each source in the group. The decision tree starts with unresolved or marginally resolved models (1,2) and moves towards more complex, resolved ones (3,4,5). Each level of the decision tree assumes the same initial conditions, excepting that some sources may already be assigned a model type in the latter stages. The tree must be tuned according to the data being used. In this work, marginally resolved SimpleGalaxy models must achieve a lower $\chi^2_N$ by a margin of 0.1 compared a unresolved PointSource model, thereby preferring the PointSource model whenever possible. If either model achieves a $\Delta\chi^2_N<1.5$, then the next level is tried. If the ExpGalaxy and DevGalaxy models are not indistinguishable by $|\Delta\chi^2_N|=0.2$ or neither achieves a $\chi^2_N<1.5$, the most complex CompositeGalaxy is tried (see Weaver et al., in prep. for more details). Once a model type has been assigned to each source, the final ensemble of models is re-optimized to ensure that the derived model parameters reflect the actual model ensemble. If instead the parameters were adopted during the initial stages of the decision tree, then it would be possible for one source which has not yet been fit with the appropriate model type to influence the parameters of another nearby source. By re-computing the model parameters at the very end, when all the model types have been assigned, this case is avoided.

\begin{figure}[t]
	\centering
	\includegraphics[width=\hsize]{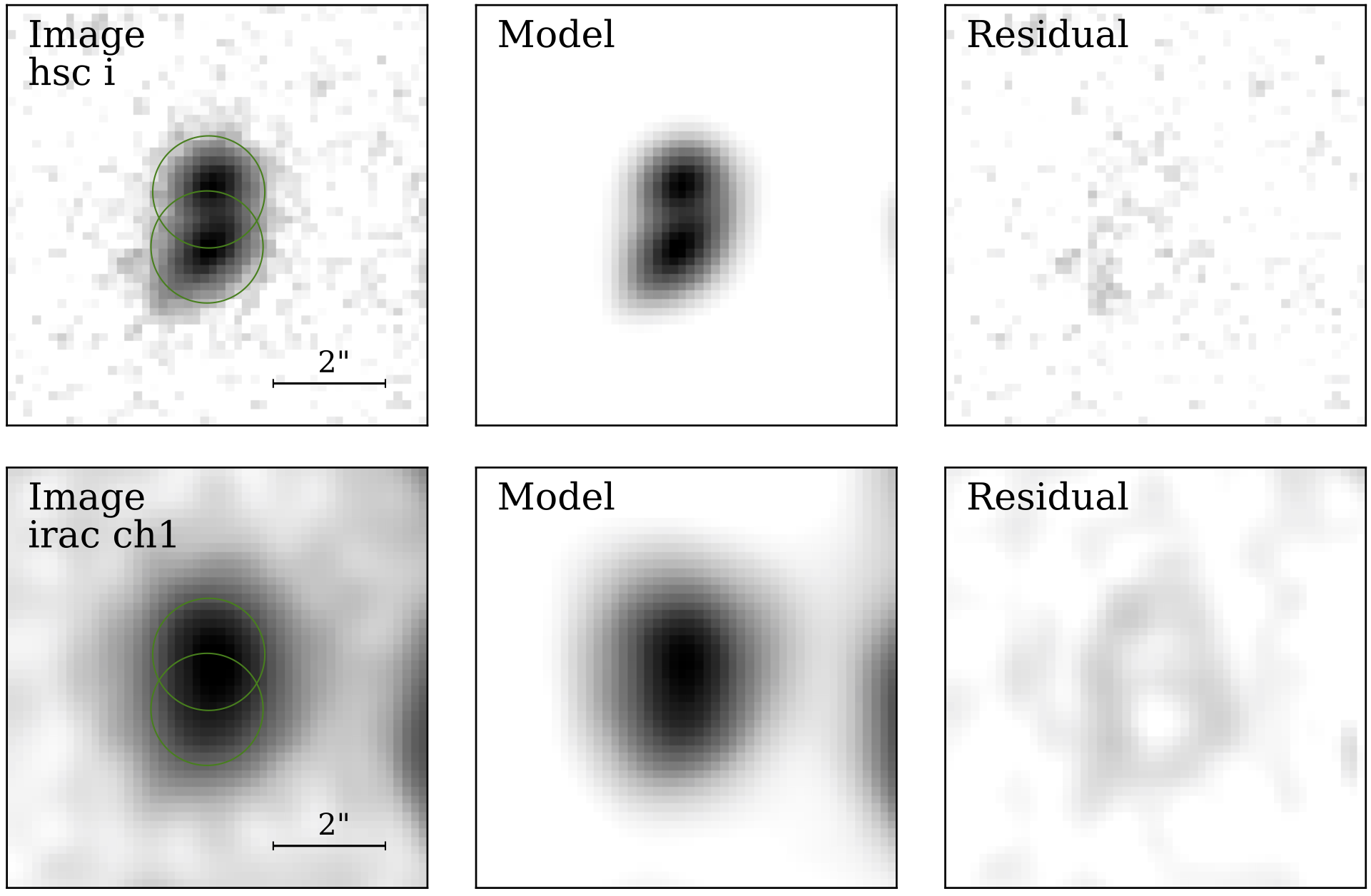}

	\caption{Demonstration of the model-fitting method from \tractor{}. A pair of detected but overlapping sources is shown in the HSC $i$ band (\textit{top}). They are jointly modeled using \farmer{} with constraints from each of the $izYJHK_s$ images in order to provide a parameterized solution which is suitably optimized, and from which the total flux is measured. The same pair of sources are shown in the less resolved IRAC channel~1 (\textit{bottom}), where the two models are convolved with the channel~1 PSF and re-optimized using the channel~1 image to measure the flux contributed by each source. The extremely blended nature of this pair is underscored by the overlapping 2\arcsec apertures, consistent with the methodology of the \classic{} catalog. Pixel values are logarithmically scaled between the rms level and 95\,\% of the peak flux per pixel.
	}
	\label{fig:demo_tractor}
\end{figure}

An example of the modeling procedure is shown in Figure~\ref{fig:demo_tractor}, whereby two models are jointly determined for two nearby sources using each of the individual $izYJHK_s$ bands, simultaneously. It is stressed that the models are not constructed on the detection image itself, which suffers from PSF inhomogeniety which makes it not suitable for deriving morphologically-sensitive model constraints. The $i$ band is shown as it is the deepest high-resolution band in the detection image and hence provides the greatest constraints on the morphology. Forced photometry on IRAC channel~1 (See Section~\ref{subsec:farmer_fphot}) is shown to demonstrate the extent to which the prior information derived jointly from $izYJHK_s$ can adequately model IRAC flux, even for the most severely blended sources which apertures cannot accurately photometer.

\subsubsection{Forced Photometry}
\label{subsec:farmer_fphot}
With the model catalog complete for all detected sources, \farmer{} can measure total model fluxes for every band of interest. \farmer{} does this in a ``forced photometry'' mode, similar to the ``dual-image'' mode in \texttt{SExtractor}. In brief, the model catalog of a given group is initialized with the optimized parameters from the preceding stage. For each band, model centroids are allowed to vary with a strict Gaussian prior of 0.3\,pix to prevent catastrophic failures. By doing so, \farmer{} can overcome subtle offsets in astrometric frames between different images, and this can be done on an object-by-object basis to even overcome spatially varying offsets which may arise due to bulk flows in the astrometry. The optimization of these models produces total fluxes and flux uncertainties for each band of interest, keeping the shape parameters fixed. The flux measurement is obtained directly from the scaling factor required to match the models, which are normalized to unity, to the source in question. However, the flux uncertainties are derived by computing a quadrature sum over the weight map, weighted by the unit profile of the model, producing a similar result as traditional aperture methods but where the model profile is used in place of a fixed aperture. The weight maps are the same as those used by \classic{}. Importantly, the flux uncertainties reported in \farmer{} catalog are not corrected with empty apertures, in contrast with the \classic{} catalog (see Section~\ref{sec:aperturephoto}). The aperture-derived procedure used in \classic{} is inappropriate for model-based photometry, and although it may be expected that model-based methods would produce more precise measurements, they may still underestimate the true extent of correlated noise in the images and hence underestimate the uncertainty. This will be further discussed in Weaver et al. (in prep), and briefly evaluated later in Section~\ref{sec:photoz_valid} in terms of photometric redshift precision.

Photometry is performed with \farmer{} for all CFHT, HSC, VISTA, and IRAC bands, as well as the Suprime-Cam intermediate bands. As such, there are two main differences with respect to \classic. Firstly, the older Suprime-Cam broad bands suffer from high spatial PSF variability, which is resolved in the \classic\, catalog by PSF-homogenizing each tile (see Section~\ref{sec:aperturephoto}). However, this cannot be done for profile-fitting methods like \farmer{} that do not operate on psf-homogenized images. Combined with the fact that these broad bands are eclipsed by deeper imaging from Hyper Suprime-Cam in almost all cases, they contribute very little to improving \photoz{} precision and can indeed even decrease accuracy if the PSF variability is not properly controlled. For these reasons the Suprime-Cam broad bands are only used when deriving \photoz{}s from the \classic{} photometry using \lephare{}, as described in Section~\ref{sec:photoz}. Secondly, photometry for IRAC channels 3 and 4 are performed with \farmer{} to extend the wavelength baseline. This is largely due to the significantly cheaper computational power required for \farmer{} relative to \texttt{IRACLEAN}. Although relatively shallow, in limited cases they can help place constraints on the rest-frame optical emission of potentially high-$z$ sources. Details as to precisely which bands are available with each catalog can be found in associated \texttt{README} files.

\subsubsection{Advantages and Caveats}
\label{subsec:adv}
An important distinction between the two catalogs is that \farmer{} provides total fluxes natively, without the need to correct for aperture sizes or perform PSF-homogenization. Since this advantage can be leveraged over different resolution regimes, \farmer{} computes photometric measurements which are self-consistent. Additional metrics are also readily available from \farmer{}. This includes the goodness-of-fit reduced $\chi^{2}_{N}$ estimate computed for the best-fit model of each source on a per-band basis, obtained by dividing the $\chi^{2}$ value by the number of degrees of freedom, i.e. the pixels belonging to the segment for each source minus the number of fitted parameters. Measurements of source shape are provided for resolved sources, and as such they yield estimates of effective radii, axis-ratios, and position angles. These measurements are directly fitted in \farmer{}, unlike in \texttt{SExtractor} where they are estimated from moments of the flux distribution. Uncertainties on shape parameters are deliverable as well, in the sense that they are a fitted parameter which is the result of a likelihood maximization and not a directly calculated quantity. Likewise, centroids for both the modeling and forced photometry stages are also fitted parameters, and are delivered with associated uncertainties.

Another important consideration is that given the diversity of galaxy shapes and source crowding across ultra-deep imaging, it is inevitable that a model, or group of models, will fail to converge. Often it is due to either a bright, resolved source not being well described by smooth light profiles, an extremely dense group of sources, or a failure at detection to separate nearby sources (and hence assign the correct number of models to use), or a combination of all three. This problem is endemic to these methods and one which cannot be practically solved by manually tuning each fit; nor at this time by selecting tuning parameter based on statistics, which are unlikely to be effective in the most ill-conditioned cases. Thankfully, like in \texttt{SExtractor}  which indicates failures by a combination of boolean flags, model based photometry can also be accompanied by a flag to indicate a failure to converge. Importantly, for those which do converge, however, model-based methods can provide more information about untrustworthy measurements than any aperture-based method by leveraging the statistical properties of the residual pixel distribution (e.g. $\chi^2$ and other $\chi$-pixel statistics) to precisely indicate the extent of these failures, and hence convey in comparably greater detail the extent to which the user can rely on any given measurement.

\section{Photometry Comparison}
\label{sec:comparison}

With the photometry from the two independent methods in hand, this section presents a comparison of the photometric catalogs as measured by differences in magnitudes, colors, and photometric uncertainties. In addition, a comparison is made with literature results of galaxy number counts. The primary motivation for these tests is to validate the two catalogs, in particular the performance of the relatively newer photometry from \tractor{} generated with \farmer{}. The performance of \tractor{} code has been demonstrated previously (see \citealt{lang_tractor_2016}), hence this work focuses on additional validation of the performance particular to \farmer{} configuration used here. Additional validation of \farmer{} where its performance is benchmarked against simulated galaxy images is provided in Weaver et al (in prep.). 

A matched sample of sources common to both \farmer{} and \classic{} is constructed consisting of 854\,734 sources matched within 0\farcs6, for which \farmer{} obtained a valid model and hence has extracted photometry. The sample contains 95.8\,\% of valid \farmer{} sources, most of which are matched well below 0\farcs6. As explained in Section~\ref{sec:farmer_detection}, those which are unmatched are typically marginally detected sources, or blends which are de-blended by only one of the detection procedures.

\subsection{Magnitudes}

\begin{figure*}[t]
	\centering
	\includegraphics[width=\hsize]{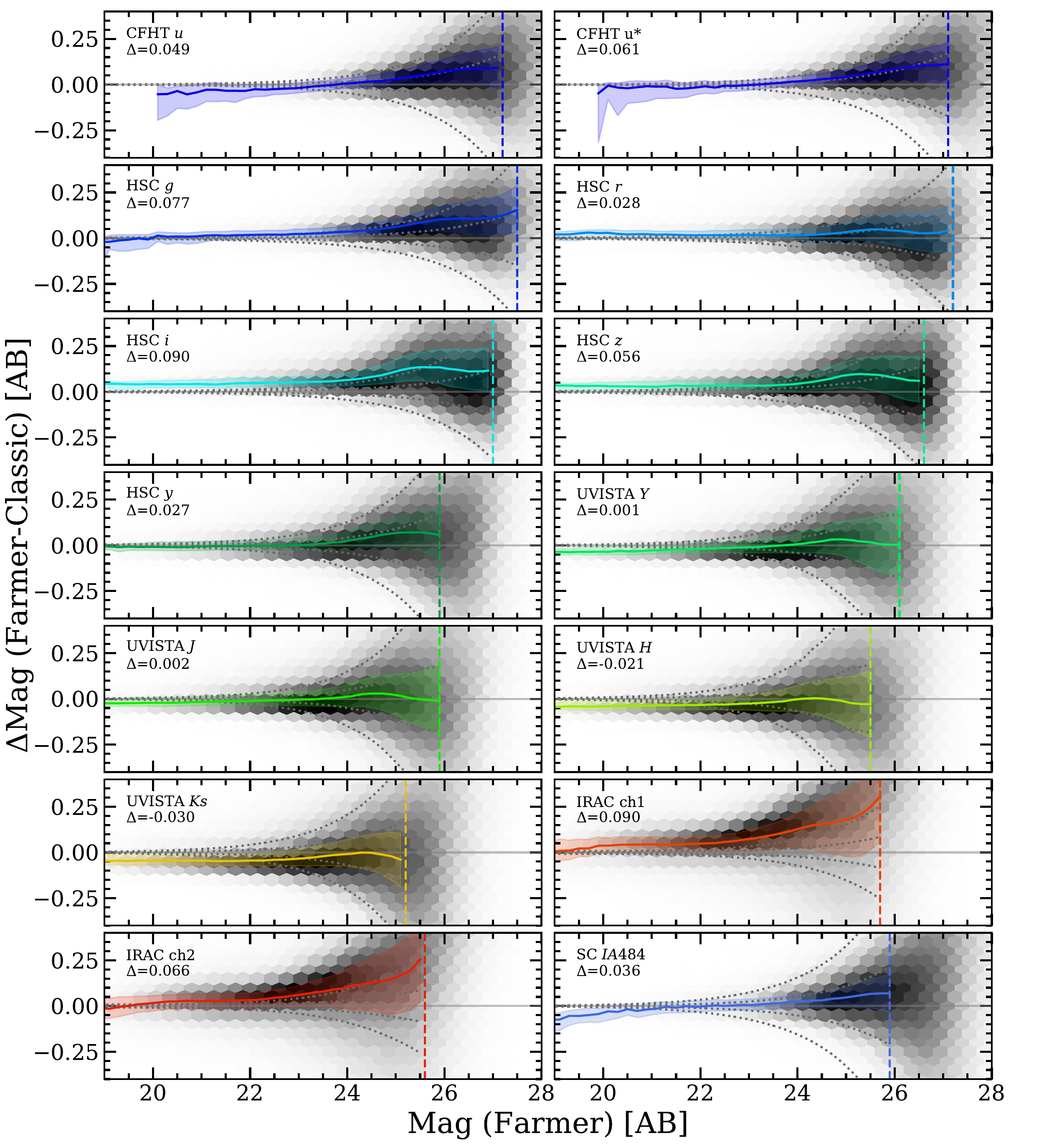}
	\caption{Summary of the difference between broad-band magnitudes measured by \farmer{} and \classic{} catalogs, $\Delta\mathrm{Mag}$. Magnitudes for \classic{} are the re-scaled 2\arcsec total magnitudes. For UltraVISTA, sources in both the ultra-deep and deep regions are shown.
	Agreement for individual sources is shown by the underlying density histogram which is described by the overlaid median binned by 0.2\,AB with an envelope containing 68\,\% of points per bin (solid line and shaded area). 1$\sigma$ and 3$\sigma$ photometric uncertainty estimates on $\Delta\mathrm{Mag}$ are indicated by the grey dotted curves. The 3$\sigma$ depths measured with 3\arcsec{} diameter apertures as reported in Table~\ref{tab:band_infos} are shown by vertical dashed lines. The median $\Delta$ magnitudes for sources brighter than the depth limit are reported in each panel.
	}
	\label{fig:mag_comparison}
\end{figure*}

A comparison of broad-band magnitudes derived independently with the two methods is shown in Figure~\ref{fig:mag_comparison}. One medium-band is included for reference. Here the re-scaled 2\arcsec{} total aperture magnitudes are used to compare with the model magnitudes from \farmer{}. The comparison is limited only to sources brighter than the 3$\sigma$ depth as reported in Table~\ref{tab:band_infos} and indicated by the vertical dashed lines. For bands not included in the detection \texttt{CHI\_MEAN}, these depths are upper bounds. The quadrature combined $\pm$3$\sigma$ and $\pm$1$\sigma$ uncertainty envelopes on $\Delta\mathrm{Mag}$, computed by quadrature addition of the photometric uncertainties from both catalogs, are shown for reference by the grey dotted curves.

In general, there is excellent agreement between the photometric measurements from the two methods. As shown in Figure~\ref{fig:mag_comparison}, the median systematic difference taken over all magnitudes is typically below 0.1\,mag in all bands, and in some cases is noticeably smaller. If one were to remove this systematic median difference, then the remaining median differences in each magnitude bin would, for all bands, lie within the 3$\sigma$ uncertainty threshold expected given the stated photometric uncertainties. In other words, the two sets of photometry are consistent within the expected uncertainties. The largest median differences occur for the faintest sources, but in most cases this is found to be $\lesssim0.25\,\mathrm{mag}$, which is on the order of the expected uncertainty at these magnitudes. There is also noticeably low scatter between the measurements, as illustrated by the tight 68\,\% range envelopes about the medians. In most cases, the 68\,\% range envelope on the median spans the same range as the expected $\pm$1$\sigma$ uncertainty envelope, the coincidence of which provides the first evidence validating the photometric uncertainties, discussed in full later in this section. Hence, it is established by multiple quantitative means that the two photometric measurements are broadly consistent.

A closer inspection, however, reveals a minor second-order curvature observed in all comparisons (including \textit{IA}$484$) at the threshold where sources become unresolved in our ground-based NIR detection images, around $\sim24.5$\,mag. At these magnitudes, photometry from \farmer{} tends to be slightly fainter than that reported by \texttt{SExtractor} (or \texttt{IRACLEAN} for channel~1 and channel~2). However, these differences are generally very small and by median estimate are within the 3$\sigma$ uncertainties for all bands. The fact that these features occur around the magnitude of each band where increasingly fainter sources are more likely to be point-sources may suggest that these sources are inadequately modeled because \farmer{} chose a resolved model for a point-source, or conversely an unresolved model for a resolved source. If a resolved source is fitted with an unresolved model then the flux may be underestimated. Differences (in bands other than IRAC) may also arise from imperfections in rescaling the 2\arcsec{} apertures to total fluxes, compared to the native total fluxes obtained with \tractor{}. This is particularly relevant given the high density of sources which can led to inaccurate estimates of object size, consequently producing inaccurate total flux measurements.

Regarding the IRAC photometry, which was obtained in both instances by profile-fitting techniques, discrepancies for faint sources cannot arise from aperture corrections. However, whereas \texttt{IRACLEAN} performs iterative subtraction of the PSF until convergence and sums all of the flux which has been subtracted, \farmer{} solves for the flux as a model parameter without iterative subtraction. Yet, there is no evidence that any residual flux remaining from \farmer{} fitting is significant enough to explain the observed discrepancy. Another potential difference which might explain the trend with brightness is that \texttt{IRACLEAN} performs iterative local background subtraction whereas \farmer{} performs a static background subtraction before performing photometry. However, it remains unclear as to exactly which methodology is most accurate. Definitively elucidating the cause of this observed discrepancy can only be obtained through simulation and is hence included in detail in Weaver et al. (in prep).

\subsection{Colors}

\begin{figure*}[t]
	\centering
	\includegraphics[width=\hsize]{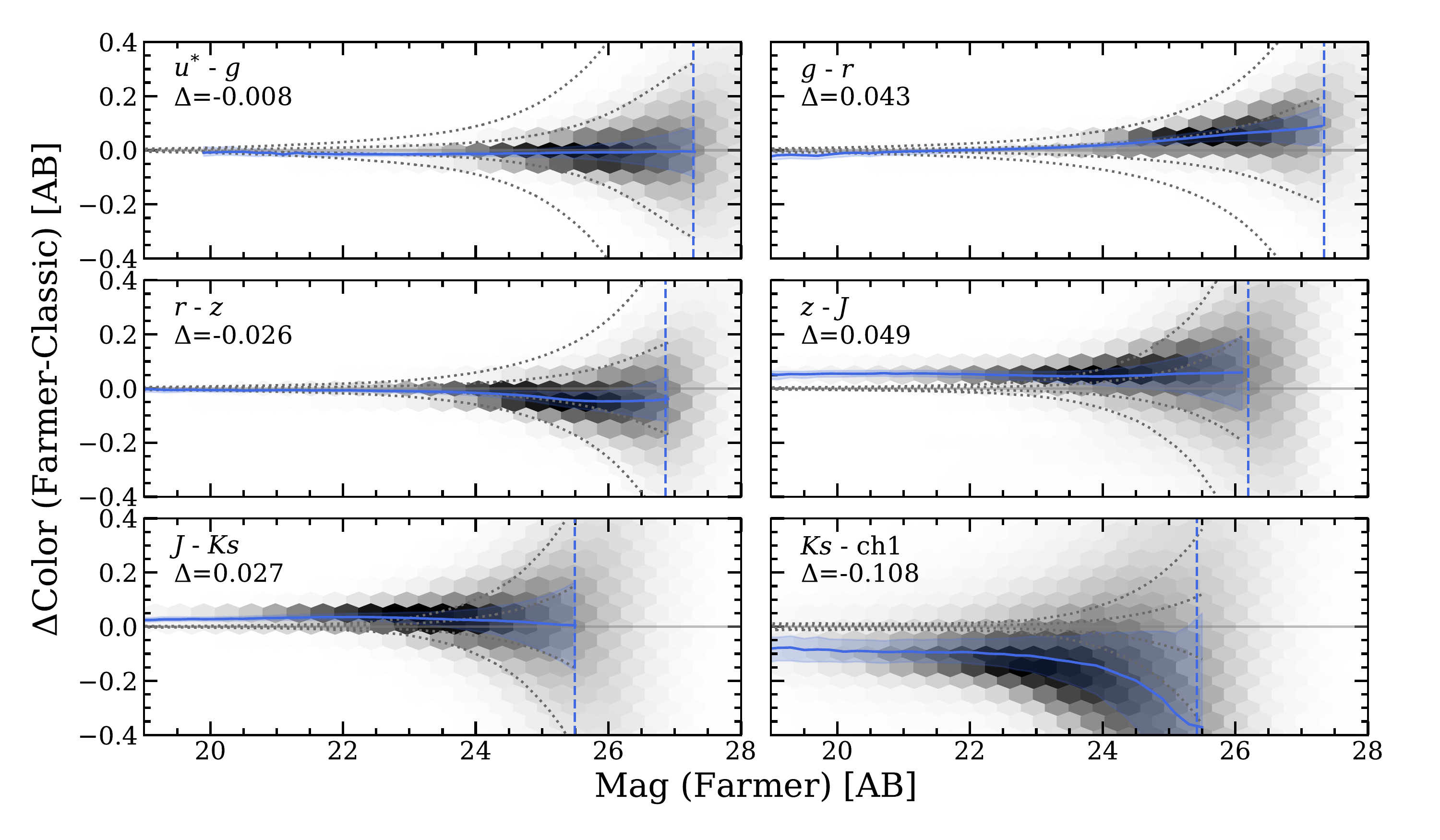}
	\caption{Comparison of broad-band colors between the \farmer{} and \classic{} catalogs, $\Delta\mathrm{color}$. \farmer{} magnitudes of the first color term in each panel are shown on the x-axis. Colors for individual sources are shown by the underlying density histogram which is described by the overlaid median binned by 0.2\,AB with a 68\,\% confidence interval. 1$\sigma$ and 3$\sigma$ photometric uncertainty estimates on the colors are indicated by the grey dotted curves and the mean 3$\sigma$ depth computed from both bands of interest and measured with 3\arcsec{} diameter apertures as reported in Table~\ref{tab:band_infos} are shown by vertical dashed lines, brighter than the median $\Delta$ are reported.
	}
	\label{fig:color_comparison}
\end{figure*}

A comparison of six colors which contribute significantly to constraining a SED is shown in Figure~\ref{fig:color_comparison}. In similar fashion to the previous comparison, the distributions are described with a running median and 68\,\% range up to the nominal 3$\sigma$ depth which is averaged for the two bands of interest. The expected $\pm$3$\sigma$ and $\pm$1$\sigma$ uncertainty thresholds on $\Delta\mathrm{Color}$, computed by the quadrature addition of the color uncertainties for each catalog, are shown by the grey dotted curves.

There is excellent agreement in colors, in some cases well-beyond the level of agreement achieved between individual bands. The median difference in color $\Delta$ is below 0.1\,mag for all colors, with the best agreement seen by $u^*-g$, $g-r$, and $r-z$. Indeed, there is a lack of systematic difference in color and the observed scatter is well below the 1$\sigma$ uncertainty expected for the color difference. The remaining panels show some level of systematic disagreement which is significant for bright sources. However, colors for faint sources are statistically consistent as they lie within the $\pm$1$\sigma$ thresholds on the color uncertainty. This may be helped by the fact that the \classic{} catalog does not require aperture-to-total rescaling to compute colors, thereby eliminating any relevant uncertainties present when comparing magnitudes only. In general, there is no evidence for a significant systematic difference in colors obtained by the two methods. Second-order curvatures are only visible at the faintest magnitudes, and are not significant even at the 1$\sigma$ level after correcting for median shifts. The most significant deviation in color shown here is $K_{s}-\text{ch1}$, which features a relatively large systematic offset for bright sources and a strong second-order curvature for faint sources whereby \farmer{} obtains systematically bluer colors. Given that $K_{s}$ magnitudes are well-matched between the two catalogs, this discrepancy in color must originate from the disagreement in faint IRAC channel~1 fluxes demonstrated in Figure~\ref{fig:mag_comparison}. However, after correcting for the systematic median offset, the median curvature of the $K_{s}-\text{ch1}$ lies between the 3$\sigma$ color uncertainty thresholds.

\subsection{Photometric uncertainties}
\label{subsec:photoUncertainties}
\begin{figure*}[t]
	\centering
	\includegraphics[width=\hsize]{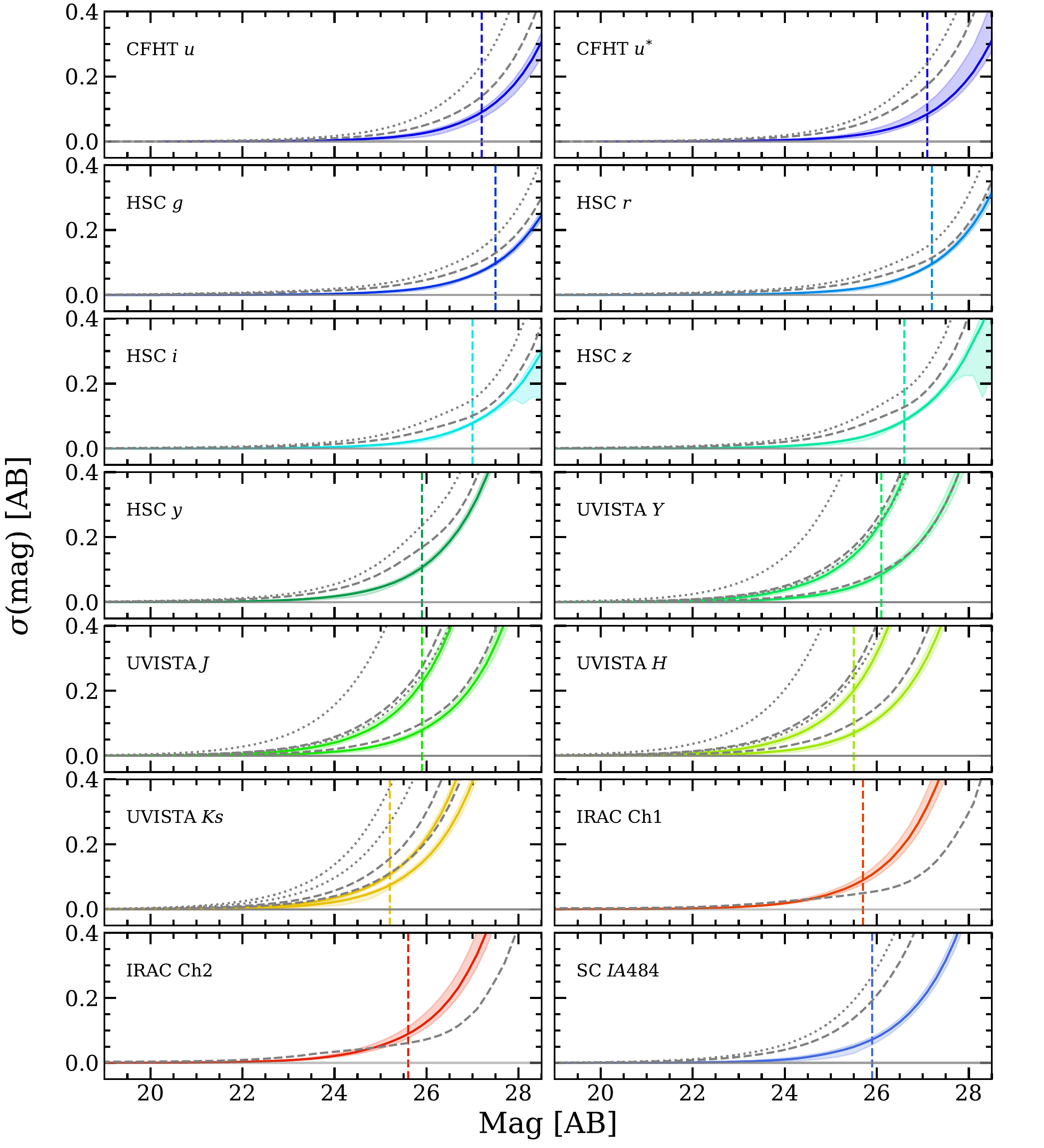}
	\caption{Growth of photometric uncertainties as a function of magnitude. The colored curves indicate the distributions for individual sources in \farmer{} catalog, described by the running median and a tight envelope containing 68\,\% of sources. The grey curves represent the median growth of uncertainty for the total magnitudes in the \classic{} catalog derived from 2\arcsec{} aperture photometry, shown by the dashed and dotted curves for the uncorrected and corrected uncertainties, respectively. The 3$\sigma$ depths measured with 3\arcsec{} diameter apertures as reported in Table~\ref{tab:band_infos} are shown by vertical dashed lines. The two curves shown for each band in $YJHK_{s}$ are due to different depths of the deep and ultra-deep regions. 
	}
	\label{fig:unc_comparison}
\end{figure*}

One critically important aspect to compare is photometric uncertainties. The uncertainties from \texttt{SExtractor} are measured by quadrature summation of the $1/\sigma^{2}$ inverse-variance per-pixel (i.e. weight) map corresponding to the aperture on the source in the image. In contrast, \tractor{} reports minimum variance estimates on the photometric uncertainty, although still using the same weight map. \tractor{} computes flux uncertainties by a quadrature summation of weight map pixels, weighted by the unit-normalized model profile, which for point-sources is simply the PSF. This thereby prioritizes the per-pixel uncertainty directly under the peak of the model profile and places less weight on the per-pixel uncertainty near the edges of the model.

Figure~\ref{fig:unc_comparison} shows a comparison of magnitude uncertainties between \farmer{} and the \classic{} catalogs. Unlike the magnitude and color comparisons, the sources which constitute this particular comparison are not matched between catalogs. They are however restricted to sources within the UltraVISTA area and clear of stellar halos indicated by the HSC bright star masks.

The distributions of magnitude uncertainties as a function of magnitude as measured by \farmer{} for the primary broad-bands, as well as a medium-band for reference, are shown by colored binned medians with an envelope enclosing 68\,\% of sources per bin. The uncertainties in the UltraVISTA bands grow more quickly for the deep region compared to the ultra-deep region, and hence they are visualized here separately. The greatest differences between the rate of growth of uncertainties can be seen most noticeably for the $Y$, $J$, and $H$ bands which feature the greatest difference in depth (see Table~\ref{tab:band_infos}). $K_s$ does not feature a significantly different growth rate between the deep and ultra-deep regions due to the near homogeneous coverage in DR4, a fact which will be useful when determining the mass completeness of the catalog.

For comparison, binned medians on the uncorrected magnitude uncertainties from the \classic{} catalog are indicated by the grey dashed curves. As described in Section~\ref{sec:aperturephoto}, the uncertainties for most bands were then corrected using empty apertures, and are indicated by the grey dotted curves. The exception is IRAC, where the uncertainties for \classic{} are computed with \texttt{IRACLEAN} (see Section~\ref{sec:iraclean}). Like with \farmer{}, the magnitude uncertainties for UltraVISTA bands are split by depth. The faster growing curve is from the deep region, and the slower is from the ultra-deep region.

Photometric uncertainties smoothly and monotonically increase for fainter sources. For \farmer{}, there is no evidence for discontinuities related to the transition between the resolved and unresolved regimes. There is, however, a difference between the magnitude uncertainties in that those measured with \texttt{SExtractor} and corrected are always larger than those from \farmer{} for all bands except IRAC, where \texttt{IRACLEAN} was used. Yet in the case of the initial, uncorrected \texttt{SExtractor} uncertainties, this difference is much smaller. Moreover the two sets of uncertainties are in better agreement in the bluest bands (e.g., $u$, $u^*$, and HSC) where the spatial resolution is generally better than in the UltraVISTA bands. The opposite is true when comparing IRAC photometry, whereby \farmer{} reports larger uncertainties than \texttt{IRACLEAN}. However, a noticeable level of consistency is achieved by \farmer{} in that uncertainties from IRAC are similar to those from UltraVISTA, which should be expected given the similarity in the depths reported in Table~\ref{tab:band_infos}. This consistency is not present in the \classic{} catalog, due to the difference between the methods of extraction from UltraVISTA and IRAC images.

Given that the photometric uncertainties measured with \farmer{} are intrinsically linked to the underlying weight map, it is possible to quantify the internal consistency of these uncertainties using the reduced $\chi^{2}_{N}$ statistic, described in Section~\ref{subsec:adv}. In general, $\chi^{2}_{N}$ values are roughly unity for all bands. While this provides one measure of internal consistency, both the uncertainties reported by \farmer{} and the $\chi^{2}$ statistics fail to take into account pixel co-variance, which may be quite large, particularly in the lower resolution UltraVISTA mosaics which have been upsampled from their native 0\farcs34 per pixel to 0\farcs15 per pixel. It is then reasonable to conclude on this basis that although the uncertainties provided by \farmer{} may be underestimated, they are indeed internally consistent with measurements which likewise ignore correlated noise, such as $\chi^{2}$, and are in general suitable for use in SED-fitting. Additional correction of the photometric uncertainties from both \farmer{} and \classic{} catalogs appropriate for SED-fitting is discussed further in Section~\ref{sec:photoz}.

\subsection{Galaxy number counts}

\begin{figure*}[t]
	\centering
	\includegraphics[width=\hsize]{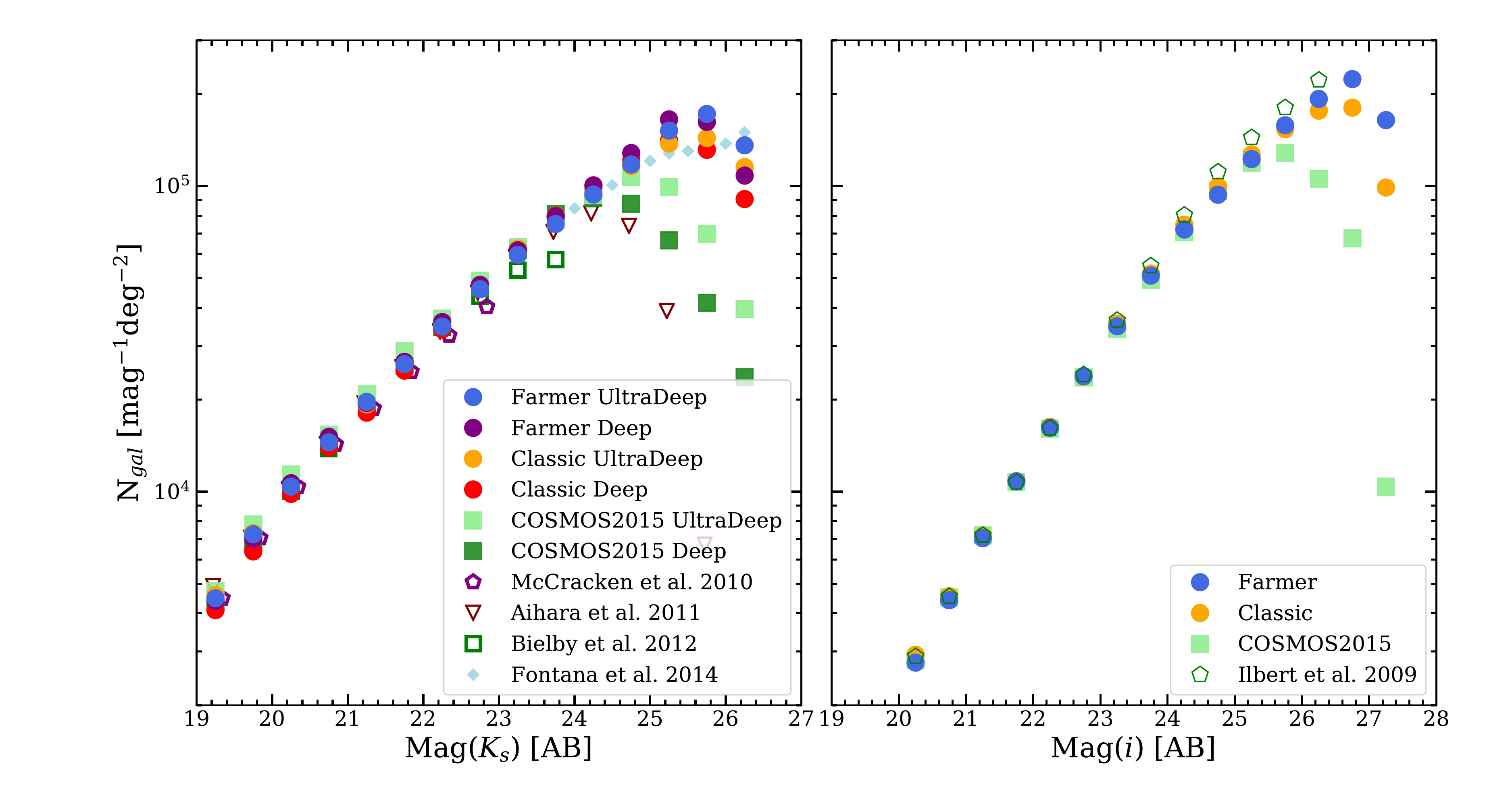}
	\caption{$i$- and $K_{s}$ band galaxy number counts of the $izYJHK_s$-detected galaxies in the UltraVISTA ultra-deep and deep regions, compared to a selection of literature measurements, including previous COSMOS catalogs.
	The bins follow increments of 0.5\,mag, with the exception of \citet{2014A&A...570A..11F} which uses 0.25\,mag.
	}
	\label{fig:gal_counts}
\end{figure*}

\begin{table}
\caption{Bin centers and values of the $izYJHK_s$-selected logarithmic galaxy number counts shown in Figure~\ref{fig:gal_counts} for both the \farmer{} and \classic{} catalogs, in units of mag$^{-1}$\,deg$^{-2}$ with bin widths of 0.5\,mag.}
\centering\footnotesize
\begin{tabular*}{\columnwidth}{lcccccc}
\hline\hline
 & \multicolumn{2}{c}{$K_s$-Deep} & \multicolumn{2}{c}{$K_s$-UltraDeep} & \multicolumn{2}{c}{$i$} \\
Mag & Farmer & Classic & Farmer & Classic & Farmer & Classic \\
\hline
19.25 & 3.64 & 3.61 & 3.65 & 3.66 & 3.01 & 3.04  \\
19.75 & 3.85 & 3.80 & 3.86 & 3.86 & 3.23 & 3.27  \\
20.25 & 4.03 & 3.99 & 4.02 & 4.02 & 3.44 & 3.47  \\
20.75 & 4.18 & 4.14 & 4.16 & 4.16 & 3.64 & 3.66  \\
21.25 & 4.29 & 4.26 & 4.29 & 4.29 & 3.85 & 3.86  \\
21.75 & 4.42 & 4.40 & 4.42 & 4.43 & 4.03 & 4.03  \\
22.25 & 4.56 & 4.53 & 4.54 & 4.54 & 4.21 & 4.21  \\
22.75 & 4.68 & 4.66 & 4.66 & 4.68 & 4.38 & 4.38  \\
23.25 & 4.79 & 4.78 & 4.78 & 4.80 & 4.54 & 4.55  \\
23.75 & 4.90 & 4.90 & 4.88 & 4.90 & 4.71 & 4.71  \\
24.25 & 5.00 & 5.00 & 4.97 & 4.99 & 4.86 & 4.87  \\
24.75 & 5.11 & 5.08 & 5.07 & 5.07 & 4.97 & 5.00  \\
25.25 & 5.22 & 5.15 & 5.18 & 5.14 & 5.08 & 5.10  \\
25.75 & 5.21 & 4.12 & 5.24 & 5.16 & 5.20 & 5.19  \\
26.25 & 5.03 & 4.96 & 5.13 & 5.06 & 5.29 & 5.25  \\
26.75 & - & - & - & - & 5.35 & 5.26  \\
27.25 & - & - & - & - & 5.22 & 5.00  \\
\hline
\end{tabular*}
\label{tab:gal_counts}
\end{table}

The galaxy number counts measured in COSMOS2020 are now compared to measurements in the literature. Figure~\ref{fig:gal_counts} shows the galaxy number counts measured for bands on the bluest and reddest ends of the \texttt{CHI\_MEAN} detection image, namely $K_{s}$ (left panel) and $i$ (right panel). The star-galaxy classification is adopted from the photometric redshift code \lephare{}, as described in Section~\ref{sec:lephare}, and is carried out similarly for both catalogs.

The effective area of \farmer{} catalog is smaller than that of the \classic{} as photometry is not returned in the case of model failure with \tractor{}, most often due to  the presence of unexpected bright stars or large resolved galaxies which cannot be adequately modeled with one of the assumed smooth galaxy profiles (see Section~\ref{subsec:adv}). In this case the effective survey area is corrected by subtracting the area occupied by sources for which a model is not available. Galaxy counts from COSMOS2015 are included for the deep and ultra-deep regions as the detection and photometry are equivalent to the \classic{} approach.
The left panel of Figure~\ref{fig:gal_counts} shows the $izYJHK_s$-detected $K_{s}$ band galaxy number counts computed over the $0.812$/$0.757$\,deg$^2$ of the HSC-masked ultra-deep region of UltraVISTA and over the $0.592$/$0.536$\,deg$^2$ of the deep region as measured by photometry from both the \classic{} and \farmer{} catalogs, respectively (see the corresponding \texttt{README} file for most up-to-date areas). There is good agreement with previous studies both within COSMOS \citep{mccracken_ultravista_2012, laigle_cosmos2015_2016} and from other surveys \citep{Aihara_2011,bielby_wircam_2012, 2014A&A...570A..11F} over the regime where comparison is possible. The counts from both COSMOS2020 catalogs are in excellent agreement. \farmer{} counts have slightly better completeness which may be due to the larger number of deblended sources at faint magnitudes. Notably, the COSMOS2020 completeness limit is $\sim1$\,mag deeper compared to COSMOS2015, which is due to a combination of both deeper infrared data and a much deeper detection image.

Similarly, the right panel of Figure~\ref{fig:gal_counts} shows the $izYJHK_s$-detected $i$ band galaxy number counts computed over the entire $1.403$/$1.234$\,deg$^2$ of the HSC-masked UltraVISTA region for the \classic{} and \farmer{} catalogs, respectively.
Literature results from the $i$-selected counts of \citet{ilbert_photoz_2009} are included for reference. At the bright end, these counts are in excellent agreement with our measurements. At the faint end, however, \citeauthor{ilbert_photoz_2009} are above our COSMOS2020 measurements. To identify the cause of this disagreement, a representative sample of $24 < i < 25$ objects detected only in \citeauthor{ilbert_photoz_2009} were visually inspected in the detection \texttt{CHI\_MEAN}, $i$, and $K_s$ images, finding virtually all to be within the halos of bright foreground objects and stars. This is especially true for the $K_s$ image, whose halos are significantly more extended relative to $i$, which in a \texttt{CHI\_MEAN} construction can lead to noise structures resembling real sources even at $i\sim25$. A reasonable explanation, therefore, is that the higher counts of \citeauthor{ilbert_photoz_2009} are due to spurious sources created by an overly-aggressive deblending threshold.

\section{Photometric redshifts}
\label{sec:photoz}

Photometric redshifts are computed using both \classic{} and \farmer{} catalogs. First, photometric measurements are corrected for Galactic extinction at each object position using the \citet{schlafly_measuring_2011} dust map\footnote{\citet{schlafly_measuring_2011} re-scaled the entire \citet{schlegel_maps_1998} dust map by a factor of 0.86.}. In the next sections, photometric redshifts are computed using both \lephare{} \citep{arnouts_measuring_2002, ilbert_accurate_2006} and \eazy{} \citep{brammer08_eazy}, followed by a comparison between the two methods.

\subsection{\lephare{}}
\label{sec:lephare}

The first set of \photoz{} is computed following the same method as in \citet{laigle_cosmos2015_2016}. Both galaxy and stellar templates are fitted to the observed photometry using the  code \lephare{}\footnote{\url{https://www.cfht.hawaii.edu/~arnouts/LEPHARE/lephare.html}} \citep{arnouts_measuring_2002,ilbert_accurate_2006} with the same configuration as \citet{ilbert_mass_2013}.

Before fitting, $0.02$\,mag is added in quadrature to the photometric errors of the data in the optical, $0.05$\,mag for $J$, $H$, $K_s$, ch1, and the three narrow-bands, and $0.1$\,mag for ch2. Such an approach is common in numerous surveys \citep[e.g.][]{arnouts07}, i.e., to include uncertainties in the color-modeling (more important near-infrared and in the narrow-bands due to the emission lines). Fluxes are used to perform the fit (as opposed to magnitudes), with the clear advantage of not introducing upper-limits. Given the uncertainties in the calibration of the Suprime-Cam/$g^+$, and the availability of deeper HSC images covering the same wavelength, this band is not included. Similarly, the shallow $z^+$ photometry is not used, since the Suprime-Cam/$z^{++}$ and HSC/$z$ images are deeper and already cover this wavelength range. IRAC channel~3 and 4 are not included given the difficulty to model the emission from polycyclic aromatic hydrocarbon (PAH) in the mid-infrared\footnote{The 6.2\,$\mu$m and 7.7\,$\mu$m PAH lines contribute to the IRAC channel~4 photometry at $z<0.3$, and the 3.3\,$\mu$m line to both channels 3 and 4 with a lower contribution.} and their shallower depth \citep{Sanders_2007}.

Stellar templates include the library from \citet{1998PASP..110..863P}, the white dwarf templates of \citet{1995AJ....110.1316B}, and the brown dwarf templates from \citet{2000ApJ...542..464C}, \citeauthor{2015A&A...577A..42B} (\citeyear{2015A&A...577A..42B}, BT-Settl/CIFIST2011\_2015) and \citet{2012ApJ...756..172M,2014ApJ...787...78M}. 
All the brown dwarf templates extend to at least 10\,$\mu$m in the infrared. The blue limit of these templates is between 0.3 and 0.6\,$\mu$m, and the flux density at bluer wavelengths is set to zero. Indeed, cool brown dwarfs belong to the very faint population of sources, and are expected to not be detected in the optical. 
Stellar templates with an effective temperature $T_{\rm eff}<4000$\,K are rejected in the case that the physical parameters do not satisfy the constraints from \citet{2008ApJ...689.1327S}.

Regarding galaxy templates, the original library \citep{ilbert_photoz_2009} includes elliptical and spiral galaxy models from \citet{polletta_spectral_2007} interpolated into 19 templates to increase the resolution, and 12 blue star-forming galaxy models from \citeauthor{bruzual_stellar_2003} (\citeyear{bruzual_stellar_2003}, hereafter BC03). Two additional BC03 templates with exponentially declining star-formation rate (SFR) were added to improve the \photoz{} of quiescent galaxies \citep{onodera12_passive}. Extinction is a free parameter with reddening ${\rm E}(B-V)\leq0.5$, and the considered attenuation curves are those of \citet{calzetti_dust_2000}, \citet{prevot_typical_1984}, and two modifications of the Calzetti law including the bump at 2175\,\AA{} \citep{fitzpatrick_analysis_1986} with two different amplitudes. Emission lines are added using the relation between the UV luminosity and [O{\sc ii}] emission line flux, as well as fixed ratios between dust-corrected emission lines following \citet{ilbert_photoz_2009}. It is imposed that the absolute magnitude in the rest-frame Suprime-Cam/$B$ band is $M_B\ge-24$\,mag which acts as a unique prior. The predicted fluxes for the templates are computed using a redshift grid with a step of $0.01$ and a maximum redshift of $10$.

Also included are a set of templates to account for active galactic nuclei (AGN) as well as quasars (see Table~3 of \citealt{Salvato2009ApJ}, and \citealt{Salvato2011} for details). A measure of the goodness of fit and \photoz{} are provided for the best-fit AGN template which can be readily compared with that of the galaxy template to identify cases where the SED can be explained by emission from an AGN. This is especially important when considering stellar mass estimates, which can be inflated in the case of an undiagnosed AGN where the stellar continuum emission is unknowingly contaminated.

An initial run of \lephare{} fitting galaxies with spectroscopic redshifts provides a method to optimize the absolute calibration in each band. The method is the same as \citet{ilbert_accurate_2006}:  after having fixed the redshift to the \specz{} value, the photometric offset of each band is derived by minimizing the difference between the predicted and observed fluxes. This procedure is applied iteratively until the offsets converge. The offset values are given in Table~\ref{tab:zp}.

A key output of the \photoz{} code is the likelihood of the observed photometry given the redshift, ${\mathcal L}({\rm data}|z)$, after having marginalized over the template set. The official \photoz{} estimate included in the catalog, noted $z_{\rm phot}$ hereafter, is defined as the median of the likelihood distribution. The $z_{\rm phot}$ error bar  is comprised between  $z_{\rm phot}^{\rm min}$  and $z_{\rm phot}^{\rm max}$, which are defined as 34\,\% of the likelihood surface below and above the median, respectively. The galaxy spectroscopic sample can be used to verify that these error bars actually represent 68\,\% confidence level intervals (see Section~\ref{sec:photoz_valid} for more details).

Galaxies are separated from stars and AGN in \lephare{} by combining morphological and SED criteria. The stellar sequence is isolated by comparing half-light radii and magnitude for bright sources in the \textit{HST}/ACS and Subaru/HSC images. All the point-like sources falling on this sequence are classified as stars at $i<23$ and $i<21.5$ for ACS and HSC images, respectively. Point-like AGN sources are also removed by this criterion. Sources with $\chi^2_{star}<\chi^2_{gal}$ are also classified as a star, with $\chi^2_{star}$ and $\chi^2_{gal}$ being the best $\chi^2$ obtained using the stellar and galaxy templates, respectively. This criterion is applied only for sources detected at 3$\sigma$ in the $K_s$ band or the IRAC channel~1, since the lack of near-infrared data could increase the risk of stellar contamination in the galaxy sample \citep[][]{daddi04,coupon09}. We do not apply the criteria based on the $\chi^2$ if the source is resolved, to avoid creating incompleteness in the galaxy sample.
 
 \begin{figure}[t]
	\includegraphics[width=\columnwidth]{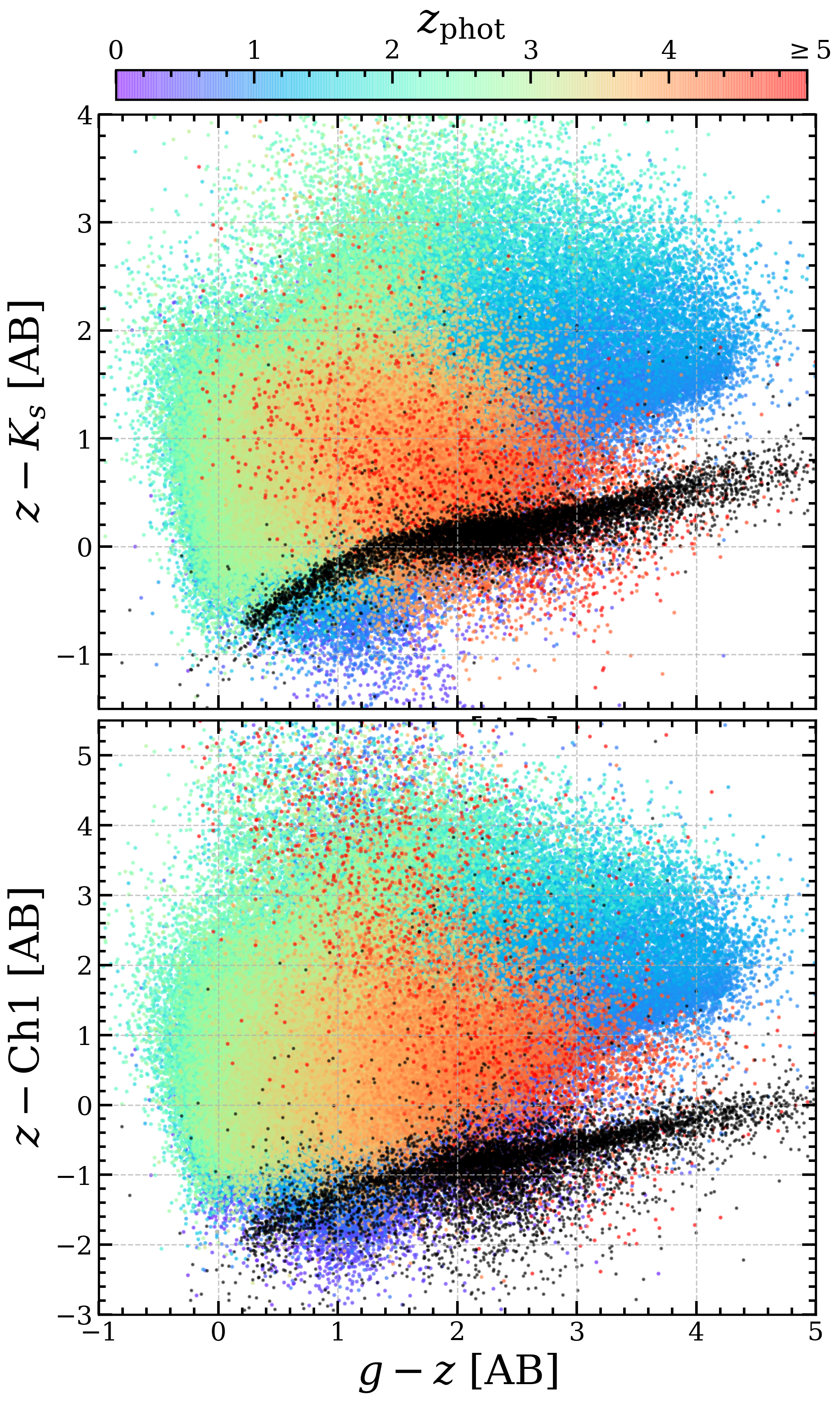}
	\caption{Color-color diagrams showing stars (black) and galaxies (colored by $z_{\rm phot}$) classified by \lephare{} for \farmer{} photometry, shown in $gzK_s$ (\textit{top}) and $gz{\rm ch1}$ (\textit{bottom}) color-color diagrams. For simplicity, galaxies with $z_{\rm phot}>5$ are shown also by red points. Only sources with $\text{S/N}>3$ in $g,\,z,\,K_s\,$and$\, {\rm ch1}$ in the UltraVISTA area outside the HSC bright star halos are shown.
	}
	\label{fig:stargal}
\end{figure}

The result of this star-galaxy separation is shown in Figure~\ref{fig:stargal}. Here again \farmer{} photometry is used, and the result is unchanged with \classic{}. Most of the sources classified as stars fall on the expected stellar locus of the two color-color diagrams \citep[e.g., Figure 2 of][]{arcilaosejo2013}.

Although these classifications are made available in the catalogs (and explained in detail in the accompanying release documentation), it should be cautioned that this precise classification scheme may be sub-optimal for certain science investigations (e.g. where galaxies with stellar-like SEDs are science targets). Hence, this star-galaxy separation method is aimed at providing a baseline, conservative galaxy population from which to demonstrate the overall effectiveness of these catalogs, for instance with the galaxy number counts in Figure~\ref{fig:gal_counts}.


\subsection{\eazy{}}
\label{sec:eazy}

Photometric redshifts are computed along with physical parameters using an updated version of the \eazy{} code\footnote{\url{https://www.github.com/gbrammer/eazy-py}} \citep[]{brammer08_eazy} rewritten in Python. \eazy{} shares much of the strategy outlined for \lephare{} in the previous section, with the primary difference being the source of the population synthesis templates and how they are fit to the observed photometry. This computation uses a set of 17 templates derived from the Flexible Stellar Population Synthesis models \citep{2009ApJ...699..486C,2010ApJ...712..833C} with a variety of dust attenuation and ages from log-normal star formation histories that are chosen to broadly span the rest-frame $UVJ$ color-space populated by galaxies over $0<z<3$. For each galaxy in the catalog, \eazy{} fits a non-negative linear combination of these templates integrated through the redshifted filter bandpasses to the observed flux densities and associated uncertainties. In this way, \eazy{} fits combinations of dust attenuation and star-formation histories to efficiently span the continuous color space populated by the majority of galaxies across the survey. For the \eazy{} \photoz{} estimates, the Subaru Suprime-Cam broad-band photometric measurements are not used, as these are generally significantly shallower than other nearby filters. Furthermore, the GALEX FUV and NUV are ignored, as these bands are relatively shallow and have broad PSFs that are difficult to combine with the other deeper filters.

As with \lephare{}, \eazy{} iteratively derives multiplicative corrections to the individual photometric bands (Table~\ref{tab:zp}). For such a task, galaxies without a spec-$z$ are also used, to mitigate the possible bias due to selection effects in the spectroscopic sample.
At each step of the iteration, the median fractional residual is computed both for all bands individually and for all measurements in all bands sorted as a function of rest-frame wavelength. With many filters that overlap in the observed frame and galaxies across a broad range of redshifts, the catalog can largely break the degeneracy between systematic offsets in individual filters (e.g., from poor photometric calibration) and systematic effects resulting from the properties of the template set (e.g., continuum shape and emission line strengths). The correction routine is stopped after five iterations, where the updates are generally less than 1\,\%. For the final photometric redshift estimates, \eazy{} uses the ``template error function'' and apparent magnitude prior as described by \cite{brammer08_eazy}.

Regarding star-galaxy separation, the current Python implementation of \eazy{} provides functionality for fitting stellar templates to the observed photometry, similar to \lephare{}. By default, \eazy{} uses a set of theoretical \texttt{PHOENIX} BT-Settl stellar templates \citep{Allard_2012} spanning a range of effective temperatures and calculates the $\chi^{2}$ goodness of fit for each template individually (i.e., not as linear combinations). Included in the catalog is the minimum $\chi^{2}$ of the fits to the stellar templates, as well as the effective temperature of the best-fit stellar model, which together may be used to separate stars from galaxies, possibly with the addition of morphological information to determine point-like sources.


\begin{table}
\caption{Values of the magnitude offsets used to optimize the absolute calibration in each band, derived with \lephare{} and \eazy{} for both photometric catalogs. When no value is indicated, the band was not used in the fit. The relative calibrations are normalized in $K_s$. Although included in \farmer{} catalog, IRAC channels 3 and 4 are not used during the zeropoint calibration by \lephare{}. Observed photometry may be corrected by adding the appropriate values.}
\centering\footnotesize
\begin{tabular}{lcccc}
\hline\hline
Band & \lephare{} & \lephare{} & \eazy{} & \eazy{} \\
 & \farmer{}  & \classic{} & \farmer{} & \classic{} \\
\hline
NUV & 				     $-$0.145  &  0.005 &  ...  &  ... \\  
 \hline
$u$      &			     $-$0.092 & 0.001  &  $-$0.128  &  $-$0.097 \\   		 
$u^{*}$ & 		         $-$0.002  &  0.058  &  $-$0.182  &  $-$0.151 \\              		 
\hline
$g$ & 			          0.058  &  0.133  &  $-$0.010  &  0.020 \\         	
$r$ & 			          0.081  &  0.133  &  0.046  &  0.057 \\  			 
$i$ & 			          0.018  &  0.102  &  0.006  &  0.054 \\ 			 
$z$ & 			          0.019  &  0.090  &  0.038  &  0.078 \\ 			 
$y$ & 			          0.070  &  0.105  &  0.091  &  0.103 \\ 			 
\hline
$B$       & 	            ...  &  $-$0.069  &  ...  &  ... \\	  
$V$ & 		    	        ...  &   0.128  &  ...  &  ... \\	 	 
$r^{+}$ & 	    	        ...  &   0.044  &  ...  &  ... \\	 	 
$i^{+}$ & 	    	        ...  &   0.058  &  ...  &  ... \\	 	 
$z^{++}$ &	      	        ...  &   0.101  &  ...  &  ... \\     	 
\textit{IB}$427$ &       $-$0.111  &  $-$0.007  &  $-$0.187  &  $-$0.135 \\			   
\textit{IB}$464$ &       $-$0.057  &  0.014  &  $-$0.119  &  $-$0.094 \\
\textit{IA}$484$ &       $-$0.036  &   0.027  &  $-$0.086  &  $-$0.066 \\
\textit{IB}$505$ &       $-$0.035  &   0.031  &  $-$0.074  &  $-$0.051 \\			 
\textit{IA}$527$ &       $-$0.062  &   0.009  &  $-$0.092  &  $-$0.066 \\
\textit{IB}$574$ &       $-$0.104  &  $-$0.027  &  $-$0.120  &  $-$0.089 \\
\textit{IA}$624$ &       $-$0.015  &   0.037  &    $-$0.027  &  $-$0.012 \\
\textit{IA}$679$ &          0.145  &   0.213  &    0.146  &  0.174 \\
\textit{IB}$709$ &       $-$0.043  &  0.015  &  $-$0.036  &  $-$0.017 \\
\textit{IA}$738$ &       $-$0.054  &   0.009  &  $-$0.047  &  $-$0.021 \\
\textit{IA}$767$ &       $-$0.052  &  $-$0.009  &  $-$0.038  &  $-$0.032 \\
\textit{IB}$827$ &       $-$0.087  &  0.007  &  $-$0.060  &  $-$0.008 \\
\textit{NB}$711$ &       $-$0.030  &   0.028  &  ...  &  ... \\   
\textit{NB}$816$ &       $-$0.082  &  $-$0.016  &  ...  &  ... \\                 
\hline
$Y$ & 	   	     	      0.039  &   0.055  &    0.065  &  0.058 \\ 		
$J$ &		    	      0.005  &   0.028  &    0.037  &  0.050 \\ 			
$H$ &		             $-$0.049  &  $-$0.043  &  $-$0.029  &  $-$0.023 \\ 			
$K_s$ &			          0.000  &   0.000  &    0.000  &  0.000 \\
\textit{NB}$118$ & 	     $-$0.034  &  $-$0.013  &  ...  &  ... \\      
\hline
 ch1 &                   $-$0.184  &  $-$0.067  &  $-$0.127  &  $-$0.119 \\ 
 ch2 &                   $-$0.186  &  $-$0.091  &  $-$0.200  &  $-$0.174 \\  
 ch3 &                   ...  &  ...  &  $-$0.168  &  ... \\  
 ch4 &                   ...  &  ...  &  $-$0.265  & ... \\  
 \hline
\end{tabular}
\label{tab:zp}
\end{table}

\subsection{Photometric Redshift Validation}
\label{sec:photoz_valid}

\begin{figure*}[t]
	\centering
	\includegraphics[width=\hsize]{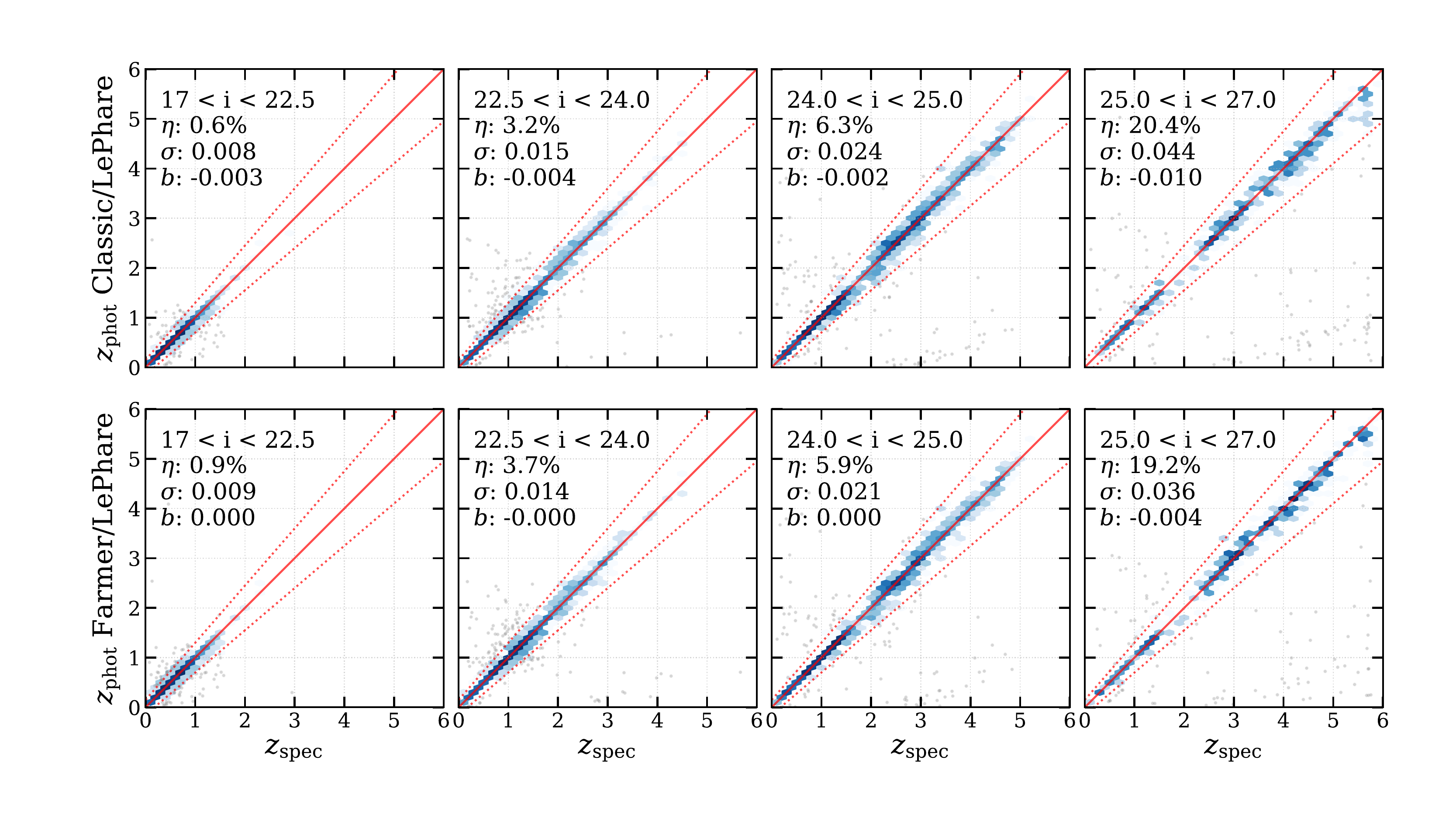}\\
	\vspace{0.5em}
	\includegraphics[width=\hsize]{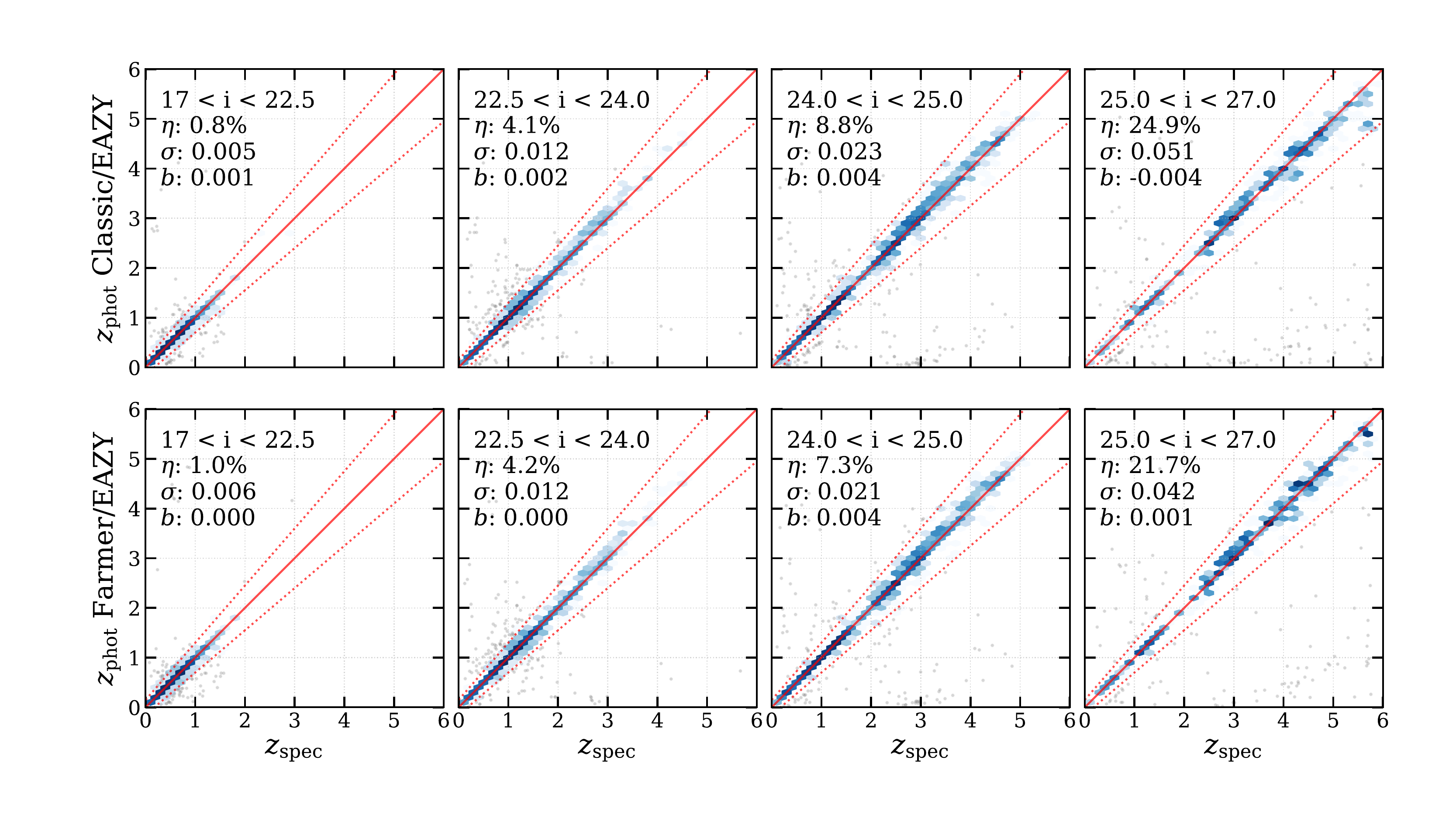}\\
	\caption{Photometric redshifts computed with \lephare{} and \eazy{}, split by apparent magnitude bin (from $i<22.5$ on the left to $25<i<27$ on the right). \textit{Top:} \photoz{} versus \specz{} for the \classic{} and \farmer{} photometric catalogs computed with \lephare{}. \textit{Bottom:} \photoz{} versus \specz{} for the \classic{} and \farmer{} photometric catalogs computed with \eazy{}. The red solid line corresponds to the one-to-one relation, and the dashed lines correspond to the \photoz{} at $\pm 0.15(1+z_{\rm spec})$. The fraction of sources outside the dashed lines (noted $\eta$), the precision measured with the normalized absolute deviation (noted $\sigma$), and the overall bias (noted $b$) are indicated in each panel. The nature of the off-diagonal points, shown individually, are discussed in the text. Bin color increases on a log$_{10}$ scale. \specz{} of $i>26$ comprise 18\% of sources shown in the rightmost $25<i<27$ panels.}
	\label{fig:zp_zs_both}
\end{figure*}

\begin{figure*}[t]
	\centering
	\[\begin{array}{l}
	\includegraphics[width=\hsize]{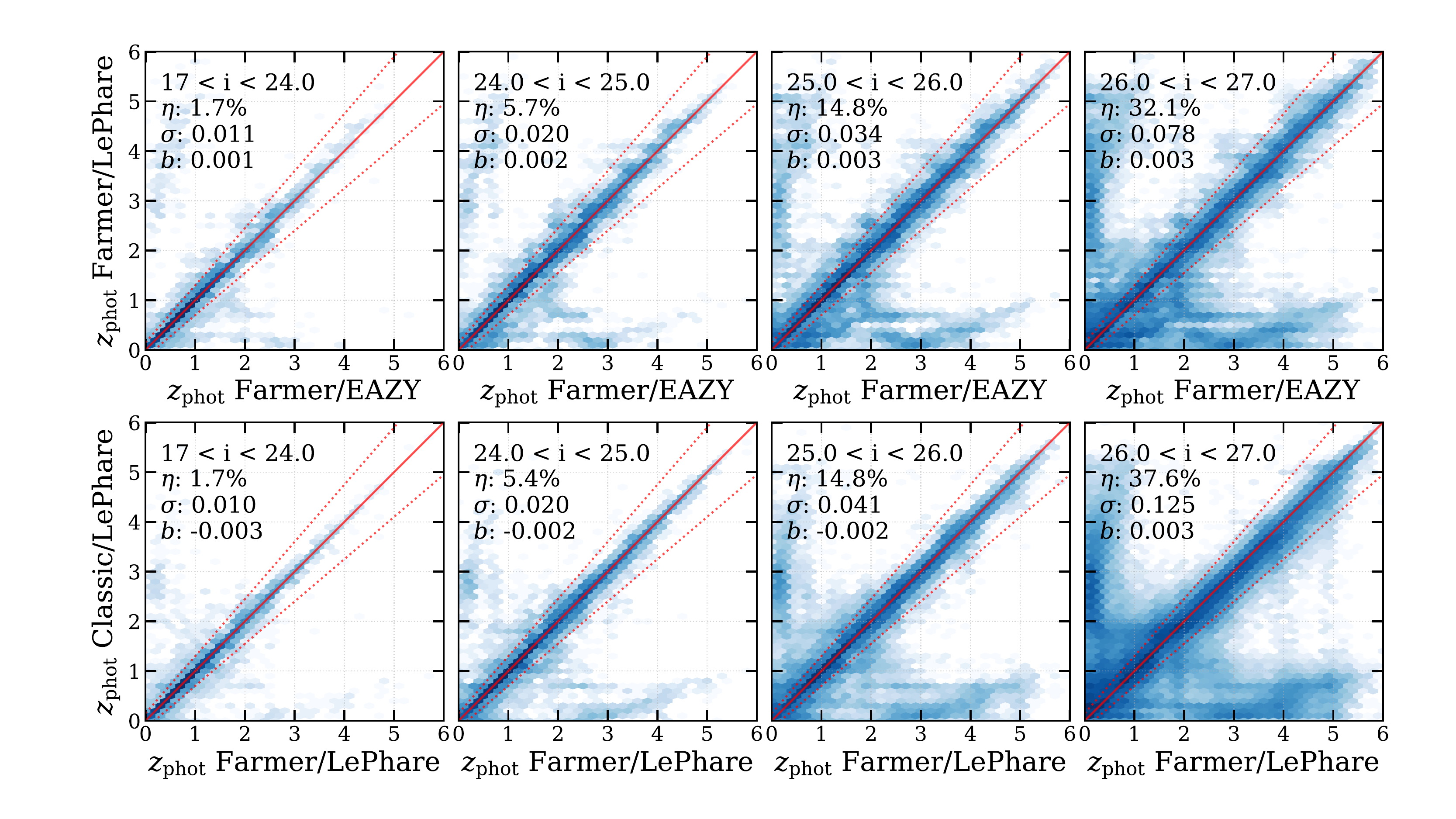}\\
    \end{array}\]
	\caption{Photometric redshifts computed with \lephare{} and \eazy{} for the \classic{} and \farmer{} photometric catalogs, split by apparent magnitude bin (from $i<22.5$ on the left to $25<i<27$ on the right). \textit{Top:} Comparison between the photometric redshifts computed with \lephare{} and \eazy{} for the full \farmer{} photometric catalog. \textit{Bottom:} Comparison between the \photoz{} derived from the \classic{} and \farmer{} full catalogs computed with \lephare{} (excluding masked regions). The nature of the two groups of off-diagonal points is discussed in the text. Bin color increases on a log$_{10}$ scale. Note that the magnitude bins are different than in Figure~\ref{fig:zp_zs_both}, to illustrate the behavior at faint magnitudes.}
	\label{fig:zp_comparison}
\end{figure*}

\begin{figure}
    \centering
    \includegraphics[width=\columnwidth]{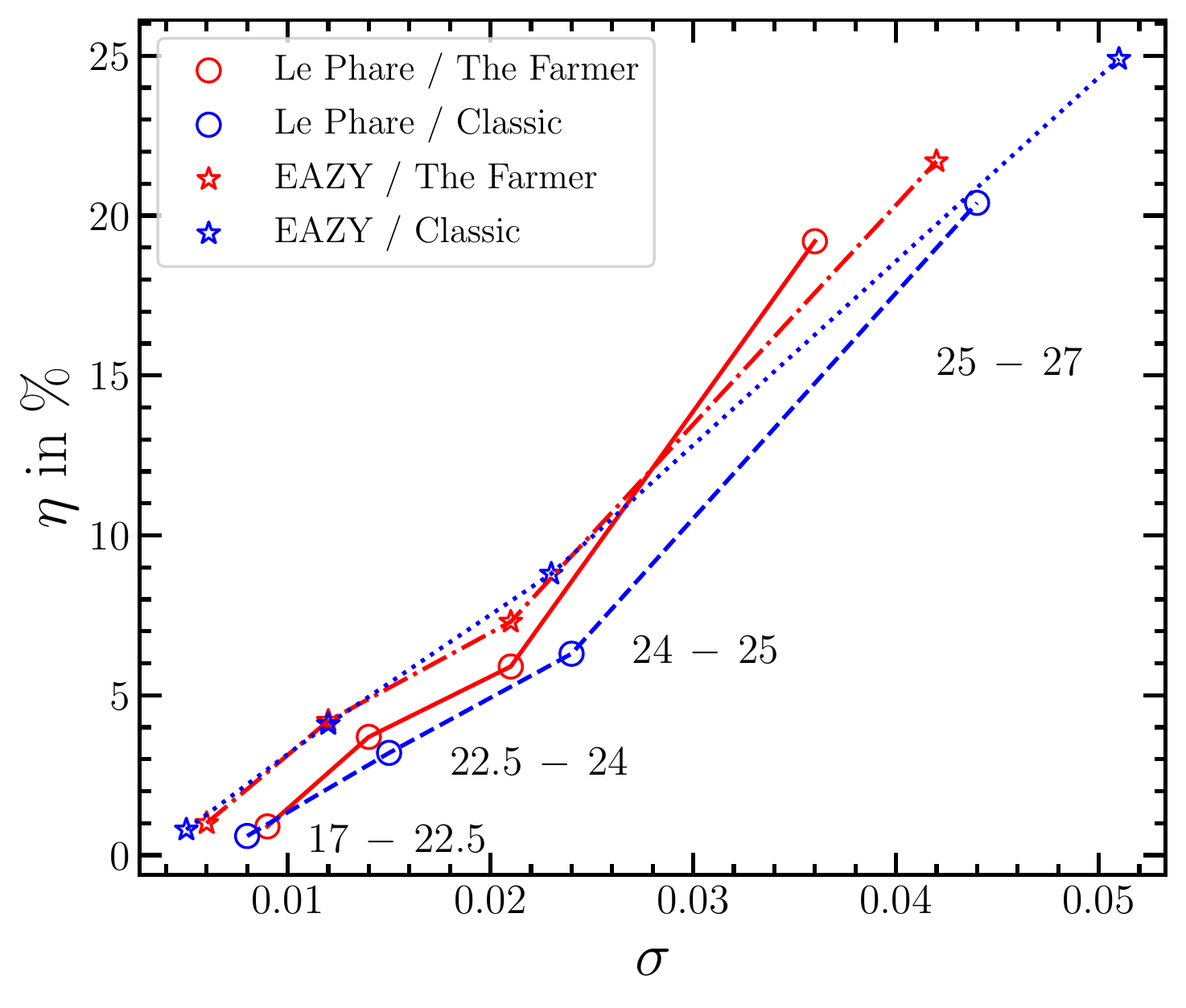}

    \caption{Comparison between the precision ($\sigma_{\rm NMAD}$) and the outlier fraction for the two catalogs (the \classic{} in blue and \farmer{} in red), and for the two \photoz{} codes (\lephare{} with circles and \eazy{} with stars). The statistics are computed per $i$ band apparent magnitude bin, as indicated on the side of the points.
    }
    \label{fig:final_comp}
\end{figure}

\begin{figure}[t]
	\centering
	\includegraphics[width=\hsize]{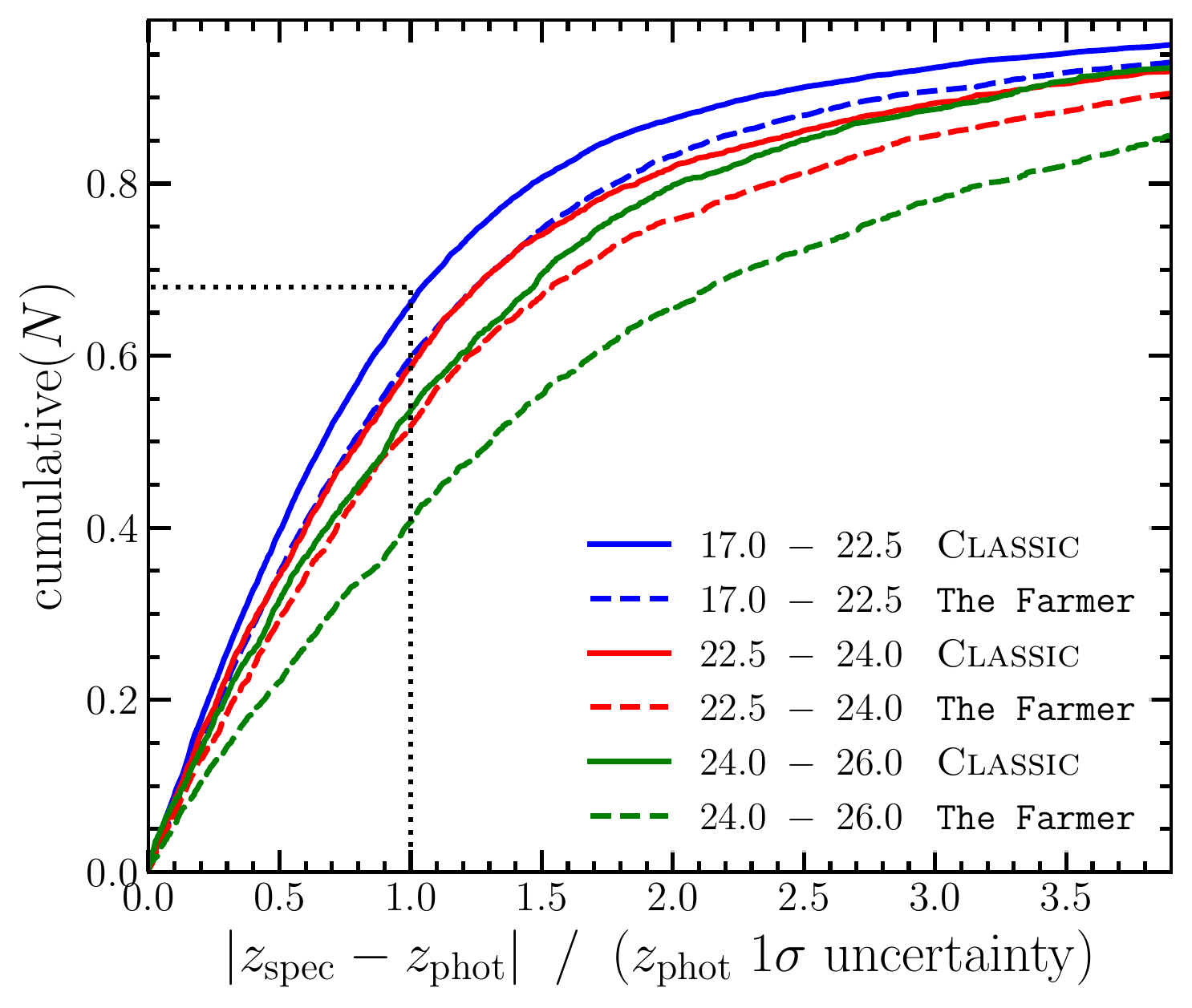}

	\caption{Cumulative distribution of the ratio between $\vert z_{\rm phot}-z_{\rm spec} \vert$ and the \photoz{} 1$\sigma$ uncertainty, for both photometric catalogs and using \lephare{}. The \photoz{} 1$\sigma$ uncertainty is taken as the maximum between $(z_{\rm phot}-z_{\rm phot}^{\rm min})$ and $(z_{\rm phot}^{\rm max}-z_{\rm phot})$. The solid and dashed lines correspond to the uncertainties from the \classic{} and \farmer{} catalogs, respectively. For an unbiased estimate of the \photoz{} 1$\sigma$ uncertainties, the cumulative number should reach 0.68 when the ratio equals 1 (black dotted line). The distributions are shown per bin of $i$ band magnitude. }
	\label{fig:cumulative_aper}
\end{figure}

\begin{figure}[t]
	\centering
	\includegraphics[width=\hsize]{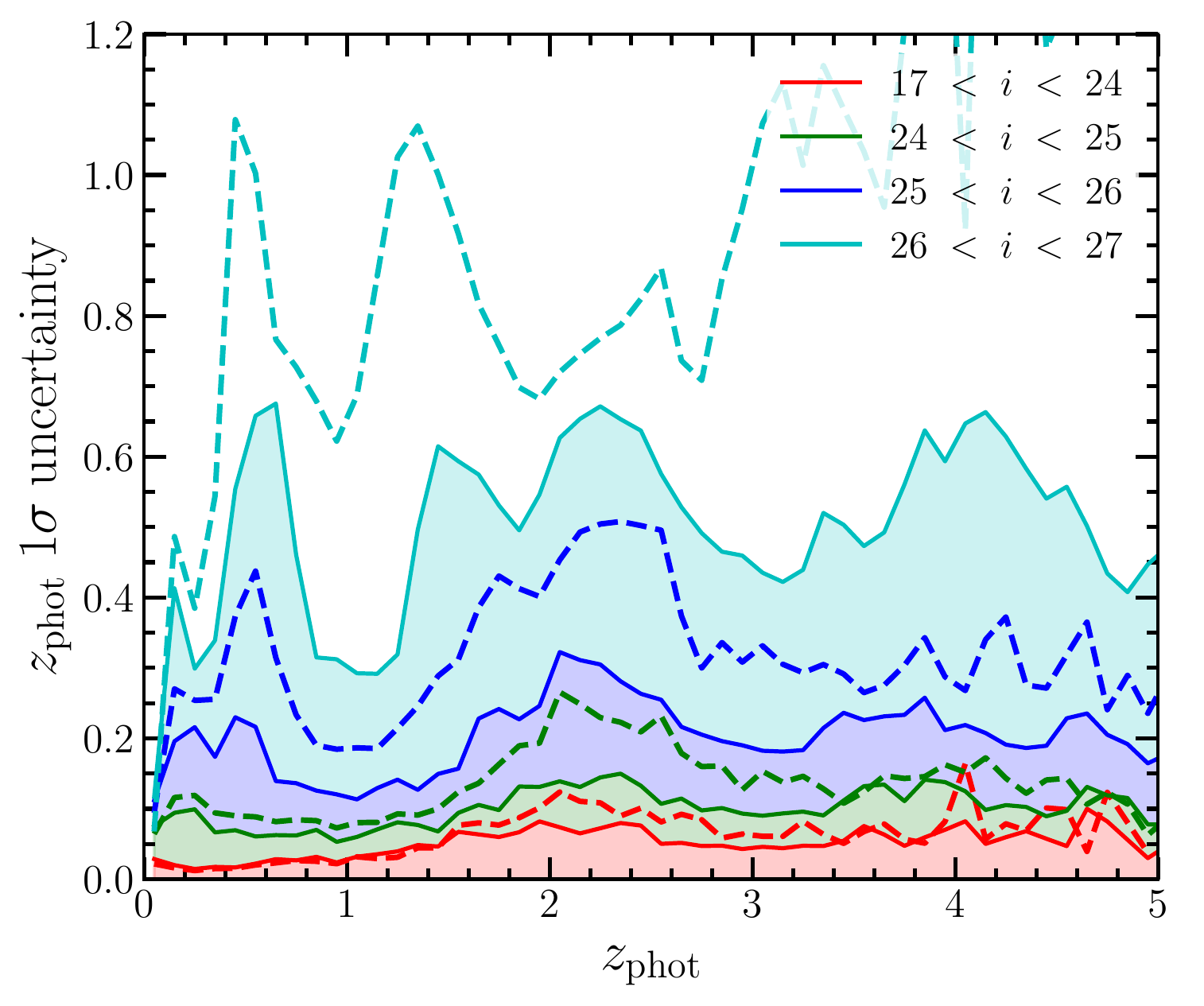}
	\caption{Median of the \photoz{} $1\sigma$ uncertainties (defined as in Section~\ref{sec:photoz_valid}) shown as a function of redshift. The shaded areas correspond to the COSMOS2020 \classic{} catalog computed with \lephare{} and the dashed lines correspond to the COSMOS2015 catalog. The distributions are shown per bin of $i$ band magnitude. } 
	\label{fig:comp2015}
\end{figure}

\begin{figure}[t]
	\centering
	\includegraphics[width=\hsize]{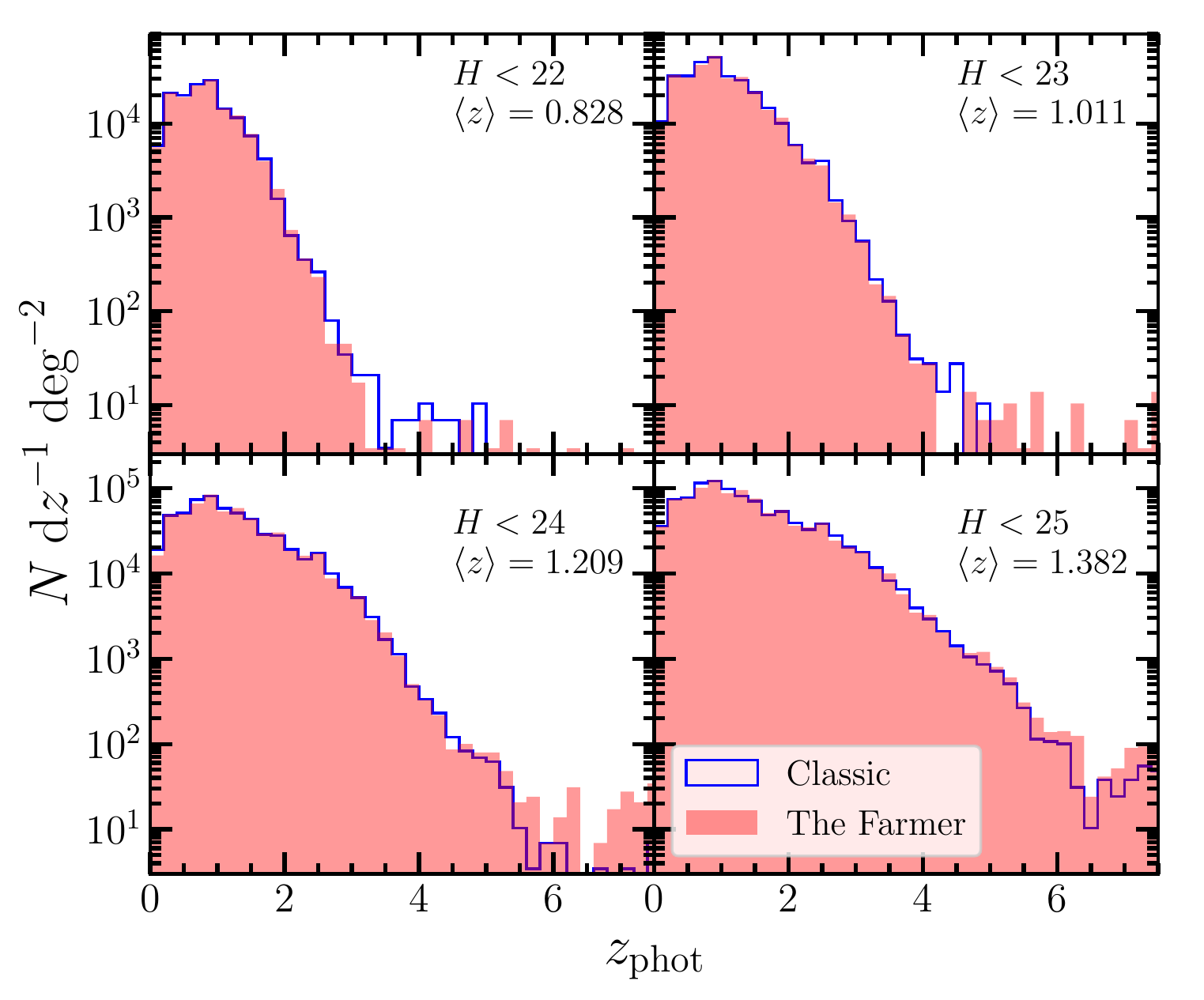}
	\caption{Redshift distribution for the \classic{} (blue) and \farmer{} (red) full catalogs computed with \lephare{}. Each panel corresponds to a different magnitude limit in $H$ band from \farmer{}.} 
	\label{fig:distz}
\end{figure}

One unique aspect of this work different from \citet{laigle_cosmos2015_2016} is the availability of two photometric catalogs created with different photometric extraction methods (see Section~\ref{sec:flux}). By applying the same \photoz{} code to the \classic{} and \farmer{} catalogs, it is possible to assess if one method to extract the photometry produces better results than the other. This is done by quantifying the precision of the \photoz{} using the normalized median absolute deviation \citep[NMAD,][]{hoaglin83_MAD}, defined as 
\begin{equation}
\sigma_{\rm NMAD}=1.48\times\mathrm{median}\left(\frac{\vert \Delta z-\mathrm{median}(\Delta z)\vert}{1+z_\mathrm{spec}}\right),
\label{eq:MAD}
\end{equation}
following \citet{brammer08_eazy} as it is less sensitive to outliers compared to the normal definition \citet[e.g.][]{ilbert_accurate_2006}. The fraction of outliers is noted $\eta$ and defined, following \citet{hildebrandt12_cfhtlens}, as galaxies whose \photoz{} deviate from their spec-$z$ by $\vert \Delta z\vert>0.15\,(1+z_\mathrm{spec})$. Lastly, the bias $b$ and is computed as the median difference between \photoz{} and \specz{}.

Comparisons between \photoz{} and \specz{} are shown for both \classic{} and \farmer{} catalogs in combination with \lephare{} and \eazy{} in Figure~\ref{fig:zp_zs_both} and summarized in Figure~\ref{fig:final_comp}. In general, the \photoz{} precision (given by $\sigma_{\rm NMAD}$) is on the order of $0.01\,(1+z)$ at $i<22.5$, and the precision is degraded at fainter magnitudes, but is still better than $0.025\,(1+z)$ at $i<25$. For both catalogs, there is a population of galaxies with $z_{\rm spec}\, > \, 2$ and $z_{\rm phot}\, < \, 1$. This population is explained by the mis-identification between the Lyman and Balmer breaks in the observed SED. This degeneracy appears clearly when comparing the \photoz{} derived for the full catalogs in Figure~\ref{fig:zp_comparison}, especially for fainter objects where the lower signal-to-noise is not sufficient to constrain the identity of the break. The figure provides a straightforward demonstration of the remarkable similarity between the catalogs computed using the same \photoz{} code (\lephare{}) and the \photoz{} codes with the same catalog (\farmer{}). The \photoz{} quality is similar between both catalogs, with a slight trend of having better results at $i<22.5$ for the \classic{} catalog, while \farmer{} catalog provides better results at fainter magnitudes. 

The \photoz{} uncertainties are also an important aspect of the \photoz{} quality. If correctly estimated  (i.e., representing the 1$\sigma$ uncertainty) the fraction of \specz{} which belong to the interval $[\zmin{} ,\,\zmax{}]$ should be 0.68. Initially, this fraction was significantly smaller due to the photometric uncertainties being underestimated; therefore the error bars associated with the observed fluxes have been multiplied by a factor of 2$\times$ for the SED fitting.   Figure~\ref{fig:cumulative_aper} shows the cumulative distribution of the ratio between $|z_{\rm phot}-z_{\rm spec}|$ and the 1$\sigma$ uncertainty derived for the \lephare{} \photoz{} solutions after boosting the flux error bars. The 1$\sigma$ uncertainty is defined as the maximum between $(z_{\rm phot}-z_{\rm phot}^{\rm min})$ and $(z_{\rm phot}^{\rm max}-z_{\rm phot})$. The cumulative distribution of the bright sample ($i<22.5$) now reaches 0.68 as expected, while the \photoz{} uncertainties of objects at $i>22.5$ are still underestimated. This effect was already discussed in \citet{laigle_cosmos2015_2016} and is seen also in \eazy{}. Since it is limited to faint galaxies, it may be due to a selection bias in the spectroscopic sample rather than a problem in the \photoz{} uncertainties    \citep[see][]{laigle19_horizonAGN}. For this reason, no further correction is applied to the uncertainties of $i>22.5$ objects.
The effect is more pronounced in the \farmer{} catalog since its photometric uncertainties are typically smaller, as they are not re-scaled to the same extent as in the \classic{} catalog (see Section~\ref{subsec:photoUncertainties}). These larger uncertainties explain the more realistic \photoz{} errors in \classic{}, and may also help to explain the lower precision for faint sources as the \photoz{} are more uncertain. 

Figure~\ref{fig:comp2015} illustrates the evolution with redshift of the 1$\sigma$ \photoz{} uncertainties in several $i$ band magnitude bins, as derived from the \lephare{} \photoz{}. There is an increase of the $1\sigma$ uncertainty between $z<1$ and $1.5<z<2.5$. This increase is explained by the Balmer break being shifted out of the medium-band coverage, as well as blue galaxies at high redshift with low signal-to-noise in the near-infrared bands. Since the \photoz{} based on the \classic{} catalog are estimated using similar techniques as \citet{laigle_cosmos2015_2016}, the \photoz{} uncertainties computed with both catalogs can be compared. For this comparison, the \photoz{}\,uncertainties in both catalogs are re-scaled in order to make them consistent with 68\,\% of the \specz{} falling into the $1\sigma$ error\footnote{The COSMOS2020 \photoz{} uncertainties are re-scaled by a factor $1+0.1\,(i-21)$ for the galaxies fainter than $i>21$. Applying the same method and using the new \specz{} sample, the COSMOS2015 \photoz{} uncertainties are re-scaled by a factor 1.3.}. The result is that the \photoz{} are improved at $1.4<z<3$ at all magnitudes owing to the gain in UltraVISTA depth, and at faint magnitudes ($i>25$) over the full redshift range thanks to the new HSC and CFHT data. While COSMOS2015 \photoz{} were unreliable at $i>26$, the new catalog can be used also at fainter magnitudes, depending on the scientific application. In summary, \photoz{} uncertainties reported in COSMOS2020 match those found 0.7 magnitudes brighter in COSMOS2015, a considerable gain.

Figure~\ref{fig:distz} shows the \photoz{} distribution of sources common to both the \classic{} and \farmer{} catalogs in four selections of $H$ band magnitude. As expected, the mean redshift increases toward faint magnitude from $z\sim 0.82$ at $H<22$ to $z\sim 1.37$ at $H<25$. There is an excellent agreement between the mean redshifts of both catalogs, within $\sim 0.01-0.02$. The mainly near-infrared selection in $izYJHK_s$ allows for the detection of a significant sample of galaxies above $z>6$ ($100-300$ at $H<25$ depending on the catalog). \farmer{} catalog includes a higher density of $z>6$ sources (by a factor almost two in the faintest bin). This is discussed in detail in Kauffmann et al. (in prep.). 

\begin{figure*}
    \centering
    \includegraphics[width=\hsize]{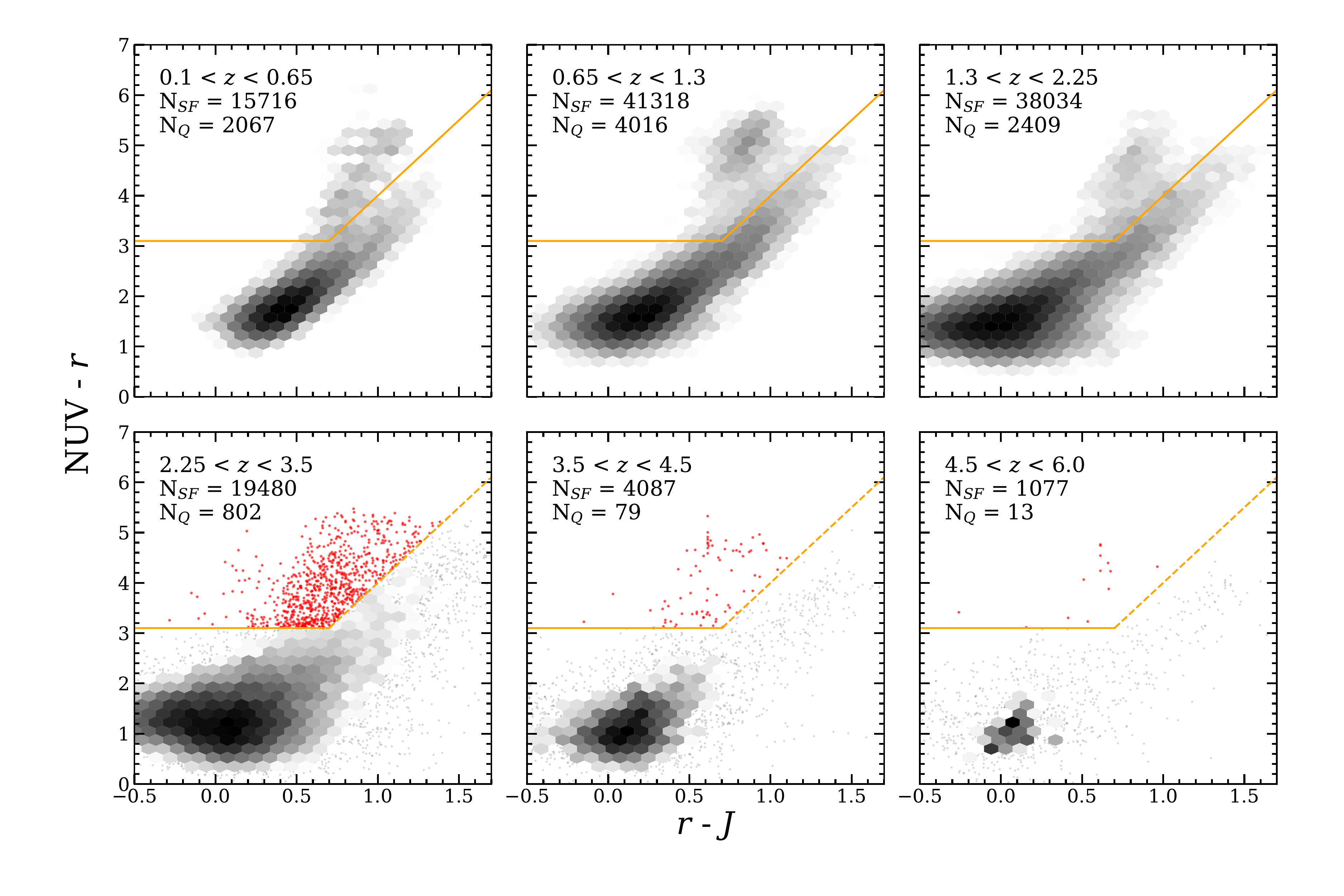}
    \caption{Identification of quiescent galaxies in bins of redshift by selection in rest frame $\textrm{NUV}-r$ and $r-J$ colors using the \lephare{} results, computed with \farmer{} for sources which lie above their respective mass completeness limit. The selection is made using the prescription of \citet{ilbert_mass_2013} shown in orange. For clarity, quiescent galaxies at $z>2.25$ are shown by individual red points. $r-J$ colors are highly uncertain at $z>2.6$ where the rest-frame $J$ band is extrapolated redward of the available photometry, and hence have an uncertain classification marked by an orange dashed line.}
    \label{fig:qgsel}
\end{figure*}

\begin{figure}
    \centering
    \includegraphics[width=\hsize]{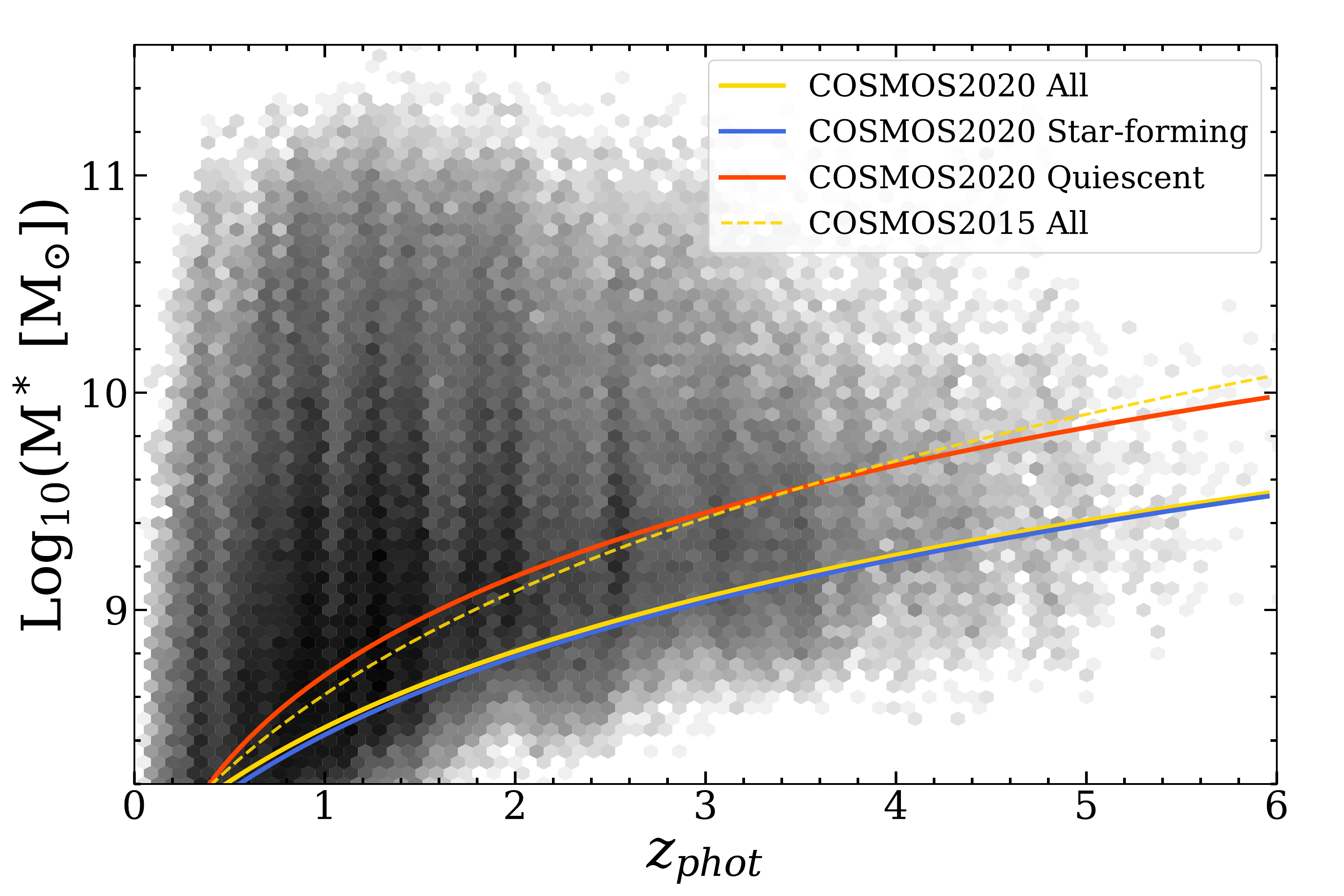}
    \caption{Mass completeness for the total sample (yellow), as well as the star-forming (blue) and quiescent (red) populations using quantities derived from \farmer{} and \lephare{} considering magnitude limits of IRAC channel~1. Limits are calculated based on the method introduced in \citet{Pozzetti2010} in a manner consistent with COSMOS2015 \citep[yellow dashed]{Davidzon17_mass}. For clarity, the total sample limit has been raised by 0.02\,dex so that both it and the star-forming limit are visible.
    }
    \label{fig:mass_completeness}
\end{figure}

\section{Physical properties of COSMOS galaxies}
\label{sec:pp}

Now a first characterization of the sources classified as galaxies in Section~\ref{sec:lephare} can be presented. Physical properties such as absolute magnitudes and stellar mass are computed using \lephare{} with the same configuration as COSMOS2015: a template library generated by BC03 models is fit to the observed photometry after fixing the redshift of each target to the \photoz{} estimated in the previous \lephare{} run \citep[for more details, see][]{laigle_cosmos2015_2016}. It should be noted that this standard configuration has been selected to be consistent with previous SED fitting results, even though recent work shows that the resulting stellar masses could be underestimated. For example, \citet{Leja2019} finds that $M^*$ estimates are $0.1-0.3$\,dex larger when using complex SFHs to build their library, instead of the standard templates of \textsc{FAST} \citep{Kriek2018}. However, integrated fluxes (as provided by these catalogs) merges together the light of young stellar populations outshining the older ones. \citet{Sorba2018} show that when these different stellar components can be resolved (e.g. in the \textit{Hubble} eXtreme Deep Field), a pixel-by-pixel SED fitting results in a galaxy stellar mass a factor $2-5\times$ larger \citep[see also][]{Abdurrouf2018, Mosleh2020}. On the other hand, tests with mock galaxy catalogs in \citet{laigle19_horizonAGN} did not find such a significant bias, with an underestimation $<20$\%.

The present analysis is limited to a classification of COSMOS2020 galaxies between star forming and quiescent, and a subsequent determination of their stellar mass completeness as a function of redshift; further investigation is deferred to future studies. Moreover, the following illustrates only the results generated with \farmer{} and \lephare{} to provide the most direct comparison to \citet{laigle_cosmos2015_2016} template fitting while demonstrating the effectiveness of the new \farmer{} photometry. There are no significant differences when repeating the analysis with either \classic{} photometry or with \eazy{}. 

\subsection{Galaxy classification}

Previous studies have devised a variety of techniques to identify quiescent galaxies using broad-band photometry. \citet{Williams2009} provides a prescription utilizing $U-V$ and $V-J$ rest-frame colors which has been broadly adopted in the literature \citep[e.g.,][]{muzzin13_mass,Tomczak14}. \citet{ilbert_mass_2013} and \citet{2013A&A...558A..67A} proposed improving the selection by replacing $U-V$ with $\textrm{NUV}-r$, since the latter can better separate galaxies with different star formation histories \citep[see also][]{leja_uvj_2019}. 

This analysis adopts the rest-frame $\textrm{NUV}-r$ vs. $r-J$ diagram described in \citet{ilbert_mass_2013}, where quiescent galaxies are defined to be those with $M_{\text{NUV}}-M_{r}>3\,(M_{r}-M_{J})+1$ and $M_{\text{NUV}}-M_{r}>3.1$. Measurements are provided by \lephare{} by convolving the best-fit template with the appropriate pass-band in the observed frame. Figure~\ref{fig:qgsel} shows the rest-frame $NUVrJ$ color-color diagram in six redshift bins from $z=0.1$ to 6. The assembly of the quiescent population at late cosmic times is evident. Quiescent galaxies are rare at $z>2$ \citep[e.g.,][]{ilbert_mass_2013, muzzin13_mass, Tomczak14, Davidzon17_mass} but the large cosmic volume probed by COSMOS allows us to identify a significant number of candidates. However, a portion of them are expected to be star-forming galaxies that contaminate the high-$z$ quiescent locus due to large uncertainties in their rest-frame colors (especially at $z>2.6$ where $M_J$ corresponds to observed wavelengths redder than channel~2).

\subsection{Stellar Mass Completeness}

The stellar mass completeness of our galaxy sample is empirically computed following the method described in \citet{Pozzetti2010}, discriminating between star-forming and quiescent populations. 
This method is commonly used in the literature \citep[e.g.,][]{ ilbert_mass_2013, Moustakas2013}. It converts the detection limit of a given survey, given by the apparent magnitude $m_\mathrm{lim}$, into a redshift-dependent threshold in stellar mass $M_\mathrm{lim}$ computed using the mass-to-light ratio of galaxies brighter than $m_\mathrm{lim}$. Their stellar masses, estimated via template fitting, are re-scaled by a factor $10^{-0.4(m_i-m_\mathrm{lim})}$, where $m_i$ is the magnitude of the $i$-th galaxy. One can determine $M_\mathrm{lim}$ in a given redshift bin from the distribution of such re-scaled masses: e.g., their 95$^\mathrm{th}$ percentile can define the smallest mass at which most of the objects would still be observable. 

The case of COSMOS2020 is more complicated because it is now possible to quantify $m_\mathrm{lim}$ not in a single band but for the \texttt{CHI\_MEAN} $izYJHK_s$ detection image itself. Adopting the sensitivity limit in the $K_s$ band (Table~\ref{tab:band_infos}) is a conservative choice that disregards the numerous NIR-faint objects detected thanks to the deep HSC photometry. This bias has already been discussed for COSMOS2015 \citep[see][]{Davidzon17_mass} and it is now more relevant after the addition of the $i$ band in the \texttt{CHI\_MEAN} image which was not considered in 2015. Therefore the analysis proceeds as in \citet{Davidzon17_mass} by computing $m_\mathrm{lim}$ in IRAC channel~1, using the CANDELS-COSMOS catalog \citep{2017ApJS..228....7N} as a reference parent catalog\footnote{In the COSMOS field, CANDELS detection image \textit{HST}/F160W has a 5$\sigma$ limit at 27.56\,mag within $0\farcs34$ diameter apertures, corresponding to twice the PSF FWHM.}. Source completeness in channel~1 is related not only to the properties of the IRAC mosaic itself, but also to the depth of the $izYJHK_s$ image, which is used as a prior for source extraction (Section~\ref{sec:iraclean} and \ref{sec:farmer_detection}). The choice to use channel~1 over $K_{s}$ is motivated by the fact that channel~1 probes the bulk of stellar mass at $z>2.5$, where the Balmer break is shifted beyond the optical-NIR bands. While \lephare{} and \eazy{} estimate broadly similar masses, this particular mass completeness is computed with masses reported by \lephare{} with \farmer{}. Other combinations may produce a marginally different mass completeness limit, and should be re-derived for specific science applications.

A common sample is constructed by cross-matching IRAC channel~1 sources of COSMOS2020 to the deeper CANDELS catalog in the $\sim200$\,arcmin$^2$ where the two overlap. At $m_\mathrm{lim}=26$\,mag, about 75\,\% of the CANDELS sources are also recovered by \farmer{}\footnote{The fraction of recovered CANDELS sources is the same with the \classic{} catalog.}; the completeness at that magnitude was $<50$\,\% in COSMOS2015. 
With $m_\mathrm{lim}$ in hand, galaxy masses are re-scaled to compute $M_\mathrm{lim}$ in bins of redshift (see Figure~\ref{fig:mass_completeness}), to which a polynomial function in $1+z$ is fitted The result is: 
\begin{equation*}
M_\mathrm{lim}(z)= -1.51\times10^{6}(1+z) +6.81\times10^{7}(1+z)^2
\end{equation*}

\noindent for $z<6$, which is more complete by $\sim$0.5\,dex compared to \citet{Davidzon17_mass}. Since the boundary used here is the 95th percentile of the re-scaled mass distribution and the choice of $m_\mathrm{lim}$ already implied that about 25\,\% of the objects are missing, it is expected that $M_\mathrm{lim}$ corresponds to a 70\,\% completeness threshold.

The procedure is repeated separately for the star-forming and the quiescent sample, both shown in Figure~\ref{fig:mass_completeness}. Quiescent galaxies start to be incomplete at stellar masses $\sim0.4$\,dex higher than the total sample since they have larger mass-to-light ratios.
$M_\mathrm{lim}$ at $z<2.5$ is additionally computed starting from the $K_s$ limit (Table~\ref{tab:band_infos}) and following precisely the procedure of \citet{laigle_cosmos2015_2016}. However, due to the nearly uniform coverage of the new data set, there is not a significant difference between the completeness limits of the ultra-deep and deep regions. The $K_s$-based completeness is well-described by the function:
\begin{equation*}
M_\mathrm{lim}(z)= -3.55\times10^{8}(1+z)+2.70\times10^{8}(1+z)^2
\end{equation*}

\noindent for $z<2.5$ and is more complete by $\sim0.5$\,dex compared to the same threshold found in COSMOS2015 \citep{laigle_cosmos2015_2016}.

\section{Conclusions}
\label{sec:conclu}

This paper describes the creation and validation of COSMOS2020, a new set of two multi-wavelength catalogs of the distant Universe, each of which includes photometric redshifts and other physical parameters computed from two independent codes. COSMOS2020 builds on more than a decade of panchromatic observations on the COSMOS field. Compared to previous releases, COSMOS2020 features significantly deeper optical, infrared, and near-infrared data all tied to a highly precise astrometric reference frame, Gaia.

Starting from a very deep multi band detection image and using two different photometric extraction codes, one based on aperture photometry and one based on a profile-fitting technique, two photometric catalogs have been extracted. These photometric catalogs were then used to estimate photometric redshifts and stellar masses using two different codes, \lephare{} and \eazy{}. This enables us, for the first time, to make a robust estimate of the systematic errors introduced by photometric extraction and photometric redshift estimation over a large redshift baseline with an unprecedented number of objects over 2\,deg$^{2}$. 
Our results show that all methods are in remarkable agreement. COSMOS2020 gains almost one order of magnitude in photometric redshift precision compared to COSMOS2015 \citep{laigle_cosmos2015_2016}. In the brightest bin, $i<22.5$, the catalogs reach redshift precision and outlier fraction are both below 1\%. Even in the faintest $25<i<27$ bins, photometric redshift precision is still $\sim 4\,\%$ with an outlier fraction of $\sim 20\,\%$. A detailed comparison in Section~\ref{sec:photoz} shows that at bright magnitudes the classic aperture catalog is marginally superior whereas at faint magnitudes the trend is reversed with the profile fitting technique providing a better result. This close agreement provides a unique validation of our measurement and photometric redshift techniques. Superseding our previous catalogs, COSMOS2020 represents an unparalleled deep and wide picture of the distant Universe. It will be of invaluable assistance in preparing for the next generation of large telescopes and surveys. 

One can already start to imagine what COSMOS2025 might contain. After fifteen years of observations, the UltraVISTA survey will have been completed, providing an unparalleled near-infrared view of COSMOS. These data, combined with the \textit{Spitzer} data presented here, will lay the foundation for a next-generation catalog combining deep high-resolution optical and infrared imaging data from \textit{Euclid} and the \textit{James Webb} Space Telescope with ultra-deep optical data from \textit{Rubin}. Such a catalog will be an important step towards producing a mass-complete survey comprising every single galaxy in a representative volume from the present day to the epoch of reionization.




\begin{acknowledgements}
\subsection*{Acknowledgements}
This paper is dedicated to Olivier Le F\`evre. Spectroscopic redshifts from his VIMOS instrument (often collected in surveys that he designed and led) played an invaluable role in preparing this catalog. 

The authors thank 
Nathaniel Strickley,
Dustin Lang,
Clara Gim\'enez Arteaga,
Istvan Sz\'apudi,
Andrew Repp,
and
Emmanuel Bertin
for helpful discussions. We are also grateful for the many helpful and constructive comments from the anonymous referee. We gratefully acknowledge the contributions of the entire COSMOS collaboration consisting of more than 100 scientists. The HST COSMOS program was supported through NASA grant HST-GO-09822. More information on the COSMOS survey is available at \url{http://www.astro.caltech.edu/cosmos}. 

The Cosmic Dawn Center (DAWN) is funded by the Danish National Research Foundation under grant No. 140. ST, GB and JW acknowledge support from the European Research Council (ERC) Consolidator Grant funding scheme (project ConTExt, grant No. 648179). OI acknowledges the funding of the French Agence Nationale de la Recherche for the project ``SAGACE''. HJMcC acknowledges support from the PNCG. ID has received funding from the European Union’s Horizon 2020 research and innovation programme under the Marie Sk\l{}odowska-Curie grant agreement No. 896225.
This work used the CANDIDE computer system at the IAP supported by grants from the PNCG and the DIM-ACAV and maintained by S. Rouberol. BMJ is supported in part by Independent Research Fund Denmark grant DFF - 7014-00017. CMC thanks the National Science Foundation for support through grants AST-1714528, AST-1814034 and AST-2009577, and additionally the University of Texas at Austin College of Natural Sciences, and the Research Corporation for Science Advancement from a 2019 Cottrell Scholar Award sponsored by IF/THEN, an initiative of Lydia Hill Philanthropies. The work of DS was carried out at the Jet Propulsion Laboratory, California Institute of Technology, under a contract with NASA. GEM acknowledges the Villum Fonden research grant 13160 “Gas to stars, stars to dust: tracing star formation across cosmic time”. DR acknowledges support from the National Science Foundation under grant numbers AST-1614213 and AST-1910107. D.R. also acknowledges support from the Alexander von Humboldt Foundation through a Humboldt Research Fellowship for Experienced Researchers. MS acknowledges the support of the Natural Sciences and Engineering Research Council of Canada (NSERC)

The authors wish to recognize and acknowledge the very significant cultural role and reverence that the summit of Mauna Kea has always had within the indigenous Hawaiian community.  We are most fortunate to have the opportunity to conduct observations from this mountain.
This work is based on data products from observations made with ESO Telescopes at the La Silla Paranal Observatory under ESO program ID 179.A-2005 and on data products produced by CALET and the Cambridge Astronomy Survey Unit on behalf of the UltraVISTA consortium. This work is based in part on observations made with the NASA/ESA \textit{Hubble} Space Telescope, obtained from the Data Archive at the Space Telescope Science Institute, which is operated by the Association of Universities for Research in Astronomy, Inc., under NASA contract NAS 5-26555. Some of the data presented herein were obtained at the W.M. Keck Observatory, which is operated as a scientific partnership among the California Institute of Technology, the University of California and the National Aeronautics and Space Administration. The Observatory was made possible by the generous financial support of the W.M. Keck Foundation.
This research is also partly supported by the Centre National d'Etudes Spatiales (CNES). These data were obtained and processed as part of the CFHT Large Area U-band Deep Survey (CLAUDS), which is a collaboration between astronomers from Canada, France, and China described in Sawicki et al. (2019, [MNRAS 489, 5202]).  CLAUDS is based on observations obtained with MegaPrime/ MegaCam, a joint project of CFHT and CEA/DAPNIA, at the CFHT which is operated by the National Research Council (NRC) of Canada, the Institut National des Science de l’Univers of the Centre National de la Recherche Scientifique (CNRS) of France, and the University of Hawaii. CLAUDS uses data obtained in part through the Telescope Access Program (TAP), which has been funded by the National Astronomical Observatories, Chinese Academy of Sciences, and the Special Fund for Astronomy from the Ministry of Finance of China. CLAUDS uses data products from TERAPIX and the Canadian Astronomy Data Centre (CADC) and was carried out using resources from Compute Canada and Canadian Advanced Network For Astrophysical Research (CANFAR).

Authors contributed to the paper as follows: AM, HJMcC, PC, SG processed the imaging data; JW, OK, ID, MSh, BCH produced the photometric catalogs; JW, OI, GB produced the photometric redshifts and physical parameters catalogs; HJMcC, ST supervised this study. All these authors contributed to the validation and testing of the catalogs. The second group of authors (CL to ZL) covers those who have either made a significant contribution to assemble the data products or to the scientific analysis. The remaining authors (SA to GZ) contributed in a some way to the data products, conceptualization, validation, and/or analysis of this work. 

\facilities{ESO:VISTA, Subaru(HSC), Spitzer(IRAC), CFHT, GALEX, HST, Gaia} \\
\software{\texttt{numpy} \citep{numpy2011},
\texttt{matplotlib} \citep{matplotlib2007},
\texttt{astropy} \citep{2013A&A...558A..33A, 2018AJ....156..123A},
\texttt{SExtractor} \citep{bertin_sextractor:_1996},
\texttt{PSFEx} \citep{bertin_psfex_2013},
\texttt{The Tractor} \citep{2016ascl.soft04008L}, and
\textsc{The Farmer} (Weaver et al., in prep.)
}
\end{acknowledgements}


\bibliography{references.bib}

\appendix

\section{Data Release}
\label{sec:release}
Both the \classic{} and \farmer{} catalogs detailed in this work is publicly available in FITS format through the ESO Phase 3 System (\url{https://www.eso.org/qi/}) and through servers at the Institut d'Astrophysique de Paris (\url{https://cosmos2020.calet.org}). Each catalog includes object positions, region mask flags, photometry, limited ancillary data (e.g., \textit{HST}/ACS, GALEX), as well as photometric redshifts and physical parameters measured by both \lephare{} and \eazy{}, for each set of photometry. Four additional files will contain the redshift probability distributions for the two photometric catalogs in combination with both photometric redshift codes. Corresponding documentation will include information about the use of mask flags, and their respective regions. This dataset will also be made available through the IPAC-IRSA and CDS VizieR systems. Each catalog is a distinct item in the Digital Object Identifier (DOI) system: in work relying on COSMOS2020 data, the DOI name(s) should be cited, in addition to a reference to the present article, to keep track of which file(s) are actually used.

Science investigators who publish software analyzing these catalogs are encouraged to link their (e.g., \texttt{github}) repository to the website \url{https://paperswithcode.com/}; in this way the code will be also visible in the \texttt{arXiv} entry of the present publication under the section ``Code \& Data -- Community Code''.

\section{Source Detection Parameters}
\begin{table}[h]
\centering
\renewcommand{\arraystretch}{0.9}
\begin{threeparttable}
\caption{\texttt{SExtractor} parameters used for the aperture detection and photometry.}

\begin{tabular}{ll}
 \hline \hline
Name & Value \\
 \hline
\texttt{ANALYSIS\_THRESH} & \texttt{1.5} \\
\texttt{BACKPHOTO\_THICK} & \texttt{30} \\
\texttt{BACKPHOTO\_TYPE} & \texttt{LOCAL} \\
\texttt{BACK\_FILTERSIZE} & \texttt{3} \\
\texttt{BACK\_SIZE} & \texttt{128} \\
\texttt{BACK\_TYPE} & \texttt{AUTO} \\
\texttt{CLEAN} & \texttt{Y} \\
\texttt{CLEAN\_PARAM} & \texttt{1.0} \\
\texttt{DEBLEND\_MINCONT} & \texttt{0.00001} \\
\texttt{DEBLEND\_NTHRESH} & \texttt{32} \\
\texttt{DETECT\_MAXAREA} & \texttt{100000} \\
\texttt{DETECT\_MINAREA} & \texttt{5} \\
\texttt{DETECT\_THRESH} & \texttt{1.5} \\
\texttt{DETECT\_TYPE} & \texttt{CCD} \\
\texttt{FILTER} & \texttt{Y} \\
\texttt{FILTER\_NAME} & \texttt{gauss\_4.0\_7x7.conv} \\
\texttt{GAIN} & \texttt{band-dependent} \\
\texttt{MAG\_ZEROPOINT} & \texttt{band-dependent} \\
\texttt{MASK\_TYPE} & \texttt{CORRECT} \\
\texttt{PHOT\_APERTURES} & \texttt{13.33,20.00,47.33} \\
\texttt{PHOT\_AUTOAPERS} & \texttt{13.3,13.3} \\
\texttt{PHOT\_AUTOPARAMS} & \texttt{2.5,3.5} \\
\texttt{PHOT\_FLUXFRAC} & \texttt{0.2,0.5,0.8} \\
\texttt{RESCALE\_WEIGHTS} & \texttt{N} \\
\texttt{SATUR\_LEVEL} & \texttt{30000} \\
\texttt{THRESH\_TYPE} & \texttt{ABSOLUTE} \\
\texttt{WEIGHT\_GAIN} & \texttt{N} \\
\texttt{WEIGHT\_TYPE} & \texttt{MAP\_WEIGHT,MAP\_WEIGHT} \\
 \hline
\end{tabular}

\label{tab:sextractor_config}
\end{threeparttable}
\end{table}

\pagebreak 

\section{Comparison with Reference Photometry}

\begin{figure}
    \centering
    \includegraphics[width=0.9\hsize]{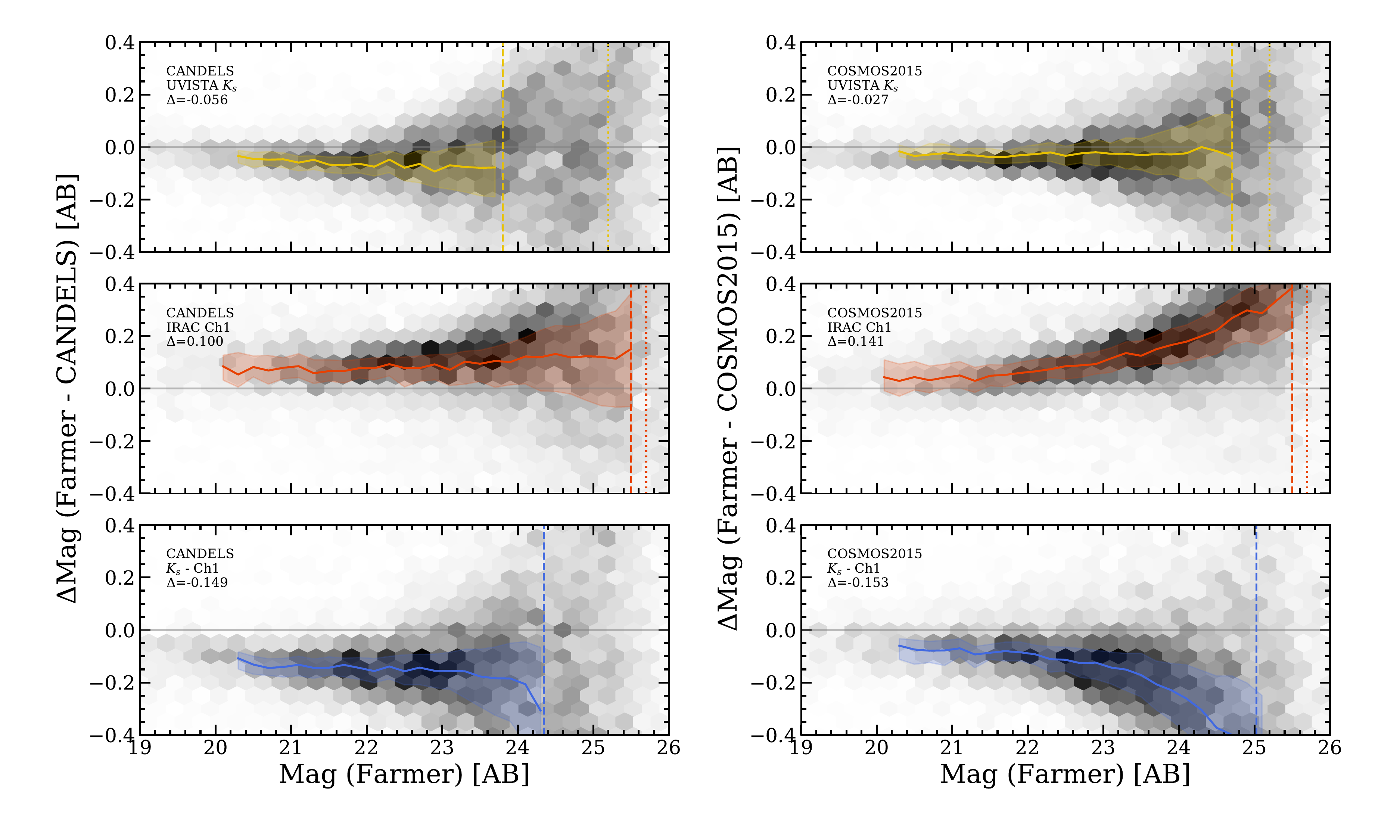}
    \caption{Comparison of broad-band $K_s$ and IRAC channel~1 magnitudes and color between the  \farmer{} catalog of this work with those of CANDELS~\citep{2017ApJS..228....7N} and COSMOS2015~\citep{laigle_cosmos2015_2016}. Individual sources are shown by the underlying density histogram which is described by the overlaid median binned by 0.2\,AB with an envelope containing 68\,\% of sources per bin. For the magnitudes, depths are shown for the comparison sample (dashed) and for COSMOS2020 (dotted), corresponding to 3$\sigma$ depths measured with 3\arcsec{} diameter apertures. For colors, averaged 3$\sigma$ depth computed from both bands of interest measured with 3\arcsec{} diameter apertures. The median $\Delta$ magnitude offsets are reported for sources below the dashed magnitude limit.
    }
    \label{fig:lit_comparison}
\end{figure}

The comparisons shown in Section~\ref{sec:comparison} are here supplemented by comparing selected bands in this work to two well-known COSMOS-field literature catalogs for which this work is readily comparable: CANDELS~\citep[][using UltraVISTA DR1 and IRAC/SPLASH]{2017ApJS..228....7N} and COSMOS2015~\citep[][using UltraVISTA DR2 and IRAC/SPLASH]{laigle_cosmos2015_2016}. As shown in Figure~\ref{fig:lit_comparison}, broad-band $K_s$ and IRAC channel~1 magnitudes and their colors are compared up to the depth limit of the shallower literature data set indicated by the vertical dashed line. For fairness, the sample includes only the $\sim$18\,000 sources which are common to all three catalogs with $0\farcs6$.

A brief analysis reveals three main points. Firstly, the COSMOS2020 depths in the bands considered exceed both those in CANDELS and COSMOS2015, as indicated by the vertical dashed and dotted lines, which manifests in the high scatter beyond the brightest magnitude limit. This restricts a meaningful comparison to sources below this limit. Secondly, the comparison with COSMOS2015 looks identical to the comparison of those bands between \farmer{} and \classic{}, both in terms of offset and any trends with magnitude. This suggests that the \classic{} photometry is highly consistent with COSMOS2015, as verified directly during the catalog preparation process. Finally, the comparison of the \farmer{} photometry with CANDELS is broadly similar. Although the $K_s$ offset is larger than in comparisons with COSMOS2015 and \classic{}, the trend with magnitude in channel~1 is more constant than with either COSMSO2015 or \classic{}. The differences in $K_s$ and channel~1 are similarly reflected in the colors, being more constant when comparing with CANDELS but not COSMOS2015.
The similarity in the comparison with COSMOS2015 and \classic{} is expected, since both employed the same methodologies, by design. Similarly, the model-fitting employed in the IRAC photometry in CANDELS is more similar to that used by \farmer{} and hence their agreement is unsurprising.

\end{document}